\documentclass[prd,preprint,superscriptaddress,preprintnumbers,eqsecnum,showpacs,nofootinbib,nobibnotes,longbibliography]{revtex4-1}
\usepackage{amsfonts,amsmath,amssymb,bm,natbib}
\usepackage{graphicx} 
\usepackage{notoccite}

\newcommand{\be}{\begin{equation}}
\newcommand{\bea}{\begin{eqnarray}}
\newcommand{\ee}{\end{equation}}
\newcommand{\eea}{\end{eqnarray}}

\def\Y#1{K_#1}

\def\s#1{{\scriptscriptstyle #1}}

\def\noeq#1{(\ref{#1})}
\def\1eq#1{Eq.~(\ref{#1})}

\def\2eqs#1#2{Eqs.~(\ref{#1}) and~(\ref{#2})}
\def\3eqs#1#2#3{Eqs.~(\ref{#1}),~(\ref{#2}), and~(\ref{#3})}

\def\fig#1{Fig.~\ref{#1}}

\def\gA{g^2 C_A}

\def\ie{{\it i.e.}, }

\def\n#1{({\it #1}\,)}

\def\gtree{\Gamma^{(0)}}
\def\bcj{J}
\def\gfullb{\widetilde{\Gamma}}
\def\bqqm{\gfullb_{m}}
\def\NV{\gfullb'}
\def\Jm{J_m}
\def\Deltam{\Delta_{m}}
\def\Vbqq{\widetilde{V}}
\def\NP{\widetilde{V}}

\def\Y{Y}

\def\chic#1{{\scriptscriptstyle #1}}

\def\m2L{m^2_\s{\mathrm{L}}}
\def\gAsq{g^4 C^2_A}

\begin{document}

\title{The gluon mass generation mechanism: a concise primer}

\author{A.~C. Aguilar}
%\email{aguilar@ifi.unicamp.br}
\affiliation{University of Campinas (UNICAMP), 
``Gleb Wataghin'' Institute of Physics\\
13083-859 Campinas, SP, Brazil}

\author{D. Binosi}
%\email{binosi@ectstar.eu}
\affiliation{European Centre for Theoretical Studies in Nuclear
Physics and Related Areas (ECT*) and Fondazione Bruno Kessler, \\Villa Tambosi, Strada delle
Tabarelle 286, 
I-38123 Villazzano (TN)  Italy}

\author{J. Papavassiliou}
%\email{Joannis.Papavassiliou@uv.es}
\affiliation{\mbox{Department of Theoretical Physics and IFIC, 
University of Valencia and CSIC},
E-46100, Valencia, Spain}

\begin{abstract}

We present a pedagogical overview of the nonperturbative mechanism that endows gluons with a dynamical mass. This analysis is performed based on pure Yang-Mills theories in the Landau gauge, within the theoretical framework that emerges from the combination of the pinch technique with the background field method. In particular, we concentrate on the Schwinger-Dyson equation satisfied by the gluon propagator and examine the necessary conditions for obtaining finite solutions within the infrared region. The role of seagull diagrams receives particular attention, as do the identities that enforce the cancellation of all potential quadratic divergences. We stress the necessity of introducing nonperturbative massless poles in the fully dressed vertices of the theory in order to trigger the Schwinger mechanism, and explain in detail the instrumental role of these poles in maintaining the Becchi-Rouet-Stora-Tyutin symmetry at every step of the mass-generating procedure. The dynamical equation governing the evolution of the gluon mass is derived, and its solutions are determined numerically following implementation of a set of simplifying assumptions. The obtained mass function is positive definite, and exhibits a power law running that is consistent with general arguments based on the operator product expansion in the ultraviolet region. A possible connection between confinement and the presence of an inflection point in the gluon propagator is briefly discussed.

%The general framework employed is the one 

\end{abstract}


\maketitle

\section{Introduction}

The assertion that quantum chromodynamic (QCD) interactions endow the gluon with an effective mass 
through a subtle mechanism that respects gauge invariance~\cite{Cornwall:1981zr} is conceptually 
intriguing  and has far-reaching theoretical and phenomenological implications~\cite{Aguilar:2002tc,Aguilar:2004td,Binosi:2014aea,Cloet:2013jya,Roberts:2015dea}.
Although the necessity for resolution of the infrared divergences appearing in the theory through production of such a mass seems more than evident, establishing a specific, self-consistent 
realization of this scenario is a notoriously complex task~\cite{Jackiw:1973tr,Jackiw:1973ha,Cornwall:1973ts,Eichten:1974et,Poggio:1974qs}. 
In fact, the purely nonperturbative character of the problem is 
compounded by the need to demonstrate, at every step,     
the compatibility of any proposed mechanism 
with the crucial concepts of gauge invariance and renormalizability.

\begin{figure}[!t]
\centerline{\includegraphics[scale=1.0]{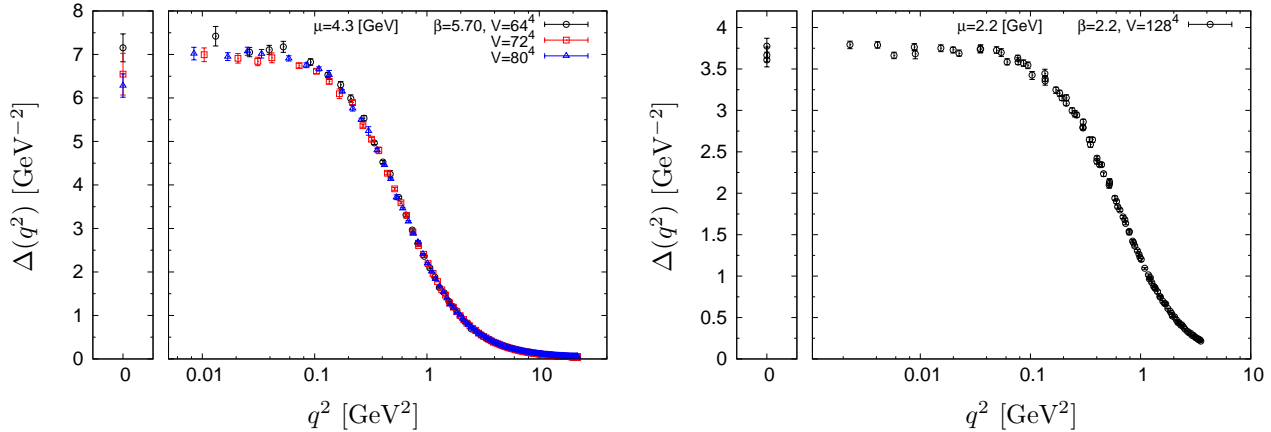}}
\caption{\label{fig:lattice-gluon}The SU(3) (left) and SU(2) (right) gluon propagator $\Delta$ measured on the lattice. Lattice data are from~\cite{Bogolubsky:2009dc,Bogolubsky:2007ud} [SU(3)] and~\cite{Cucchieri:2010xr} [SU(2)].}
\end{figure}

The notion that gluons acquire 
a dynamical, momentum-dependent mass due to their self-interactions 
was originally put forth in the early 1980s~\cite{Cornwall:1981zr,Bernard:1982my,Donoghue:1983fy}, 
but has only gained particular impetus relatively recently; this is primarily 
the result of the continuous accumulation of indisputable   
evidence from large-volume  
lattice simulations, both for SU(3)~\cite{Bogolubsky:2009dc,Bogolubsky:2007ud,Bowman:2007du,Oliveira:2009eh}
and SU(2)~\cite{Cucchieri:2007md,Cucchieri:2007rg,Cucchieri:2009zt,Cucchieri:2010xr}.
As shown in~\fig{fig:lattice-gluon}, according to these high-quality simulations, the Landau gauge gluon propagator saturates at a nonvanishing value in the deep infrared range, a feature that corresponds to an unequivocal signal of gluon mass generation~\cite{Aguilar:2008xm} (for related but somewhat different approaches to this issue, see~\cite{Szczepaniak:2001rg,Maris:2003vk,Szczepaniak:2003ve,Aguilar:2004sw,Kondo:2006ih,Braun:2007bx,Epple:2007ut,Boucaud:2008ky,Dudal:2008sp,Fischer:2008uz,Szczepaniak:2010fe,Watson:2010cn,RodriguezQuintero:2010wy,Campagnari:2010wc,Tissier:2010ts,Pennington:2011xs,Watson:2011kv,Kondo:2011ab,Siringo:2014lva}).

The primary theoretical concept underlying this entire topic 
is none other than 
Schwinger's fundamental observation~\cite{Schwinger:1962tn,Schwinger:1962tp}. That is,   
a gauge boson may acquire mass even if the gauge symmetry forbids a mass term 
at the level of the fundamental Lagrangian, 
provided that its vacuum polarization function develops a pole at 
zero momentum transfer.  In this paper, which is based upon a brief series of lectures~\cite{lectures}, we outline the 
implementation of this fascinating concept in QCD, using the general 
formalism of the Schwinger-Dyson equations (SDEs)~\cite{Roberts:1994dr,Maris:2003vk}.   
In particular, we focus on a variety 
of subtle conceptual issues, and explain how they can be self-consistently
addressed within a particularly suitable framework that has been developed in recent years. 

The present work is organized as follows. 
In Sect.~I, we present the main characteristics and advantages of the 
new SDE framework 
that emerges from the 
combination of the pinch technique (PT)~\cite{Cornwall:1981zr,Cornwall:1989gv,Binosi:2002ft,Binosi:2003rr,Binosi:2009qm}   with the background field method (BFM)~\cite{Abbott:1980hw,Abbott:1981ke}, 
which is simply referred to as ``PT-BFM''~\cite{Aguilar:2006gr,Binosi:2007pi,Binosi:2008qk}. In Sect.~II, we conduct 
a detailed study of the special 
identity that enforces the masslessness of the gluon propagator when the 
Schwinger mechanism is non-operational, and demonstrate conclusively 
that the seagull graph is not responsible 
for the mass generation, nor does it give rise to quadratic divergences   
once such a mass has been generated~\cite{Aguilar:2009ke}. 
In Sect.~III, we explain how the massless poles required for the 
implementation of the Schwinger mechanism enter the treatment of the 
gluon SDE, and why their inclusion is crucial for maintaining the 
Becchi-Rouet-Stora-Tyutin (BRST) symmetry of the theory in the presence of a dynamical gluon mass~\cite{Aguilar:2011ux}. 
Then, in Sect.~IV, we derive the ``gluon gap equation''~\cite{Binosi:2012sj}, namely, 
the homogeneous integral equation
that governs the dependence of the gluon mass function on the momentum.
In Sect.~V, we proceed to the numerical treatment of this equation, 
and discuss its compatibility with some basic field-theoretic criteria.
Finally, we present our conclusions in Sect.~VI.

\section{General considerations}

In this section, we present a general overview of the conceptual 
and technical tools necessary for the analysis that follows. 

\subsection{Preliminaries}

The Lagrangian density of the SU($N$) Yang-Mills theory can be expressed as the sum of three terms:
\begin{align}
	{\cal L}={\cal L}_{\mathrm{YM}}+{\cal L}_{\mathrm{GF}}+{\cal L}_{\mathrm{FPG}}.
	\label{lagden}
\end{align}
The first term represents the gauge covariant action, which is usually expressed in terms of the field strength of the gluon field $A$
\begin{align}
	{\cal L}_{\mathrm{YM}}&=-\frac14F^a_{\mu\nu}F_a^{\mu\nu};&
	F^a_{\mu\nu}=\partial_\mu A^a_\nu-\partial_\nu A^a_\mu+gf^{abc}A^b_\mu A^c_\nu,
\end{align}
with $g$ being the strong coupling constant, $a=1,\dots,N^2-1$  the color indexes and $f^{abc}$ the totally antisymmetric SU($N$) structure constants.

The last two terms in~\1eq{lagden} represent the gauge-fixing  and 
Faddev-Popov ghost terms, respectively. The most general means of expressing these terms is by introducing a gauge-fixing function ${\cal F}^a$ and coupling it to a set of Lagrange multipliers $b^a$ (the so-called Nakanishi-Lautrup multipliers~\cite{Nakanishi:1966cq,Lautrup:1966cq}); one then obtains
\begin{align}
	{\cal L}_{\mathrm{GF}}+{\cal L}_{\mathrm{FPG}}=s\left[\overline c^a{\cal F}^a-\frac\xi2\overline c^a b^a\right].
	\label{GF+FPG}
\end{align}
In the equation above, $\overline c^a$ (and, respectively, $c^a$ appearing below) are the antighost (ghost) fields, whereas $\xi$ is a non-negative gauge-fixing parameter. Finally, $s$ is the BRST operator~\cite{Becchi:1974md,Tyutin:1975qk}, which acts on the various fields according to
\begin{align}
	sA^a_\mu={\cal D}^{ab}_\mu c^b;\qquad
	sc^a=-\frac12f^{abc}c^bc^c;\qquad
	s\bar c^a=b^a;\qquad
	sb^a=0,
	\label{BRST}
\end{align}
with the adjoint covariant derivative ${\cal D}$ defined as
\begin{align}
	{\cal D}^{ab}_\mu=\partial_\mu\delta^{ab}+gf^{acb}A^c_\mu.
\end{align}
Note that the $b^a$ fields have no dynamical content and can be eliminated through their trivial equations of motion.

There are two gauge classes that have been found to be particularly relevant for what follows. In the so-called renormalizable $\xi$ (abbreviated as $R_\xi$) gauges, one chooses~\cite{Fujikawa:1972fe}
\begin{align}
	{\cal F}^a=\partial^\mu A^a_\mu.
	\label{lincov}
\end{align}
The Landau gauge, which is almost exclusively used in this analysis, is a particular case of this gauge class and corresponds to $\xi=0$.

BFM $R_\xi$ gauges~\cite{Abbott:1981ke,Abbott:1980hw} are also central to the 
methodology described here. The conventional means of obtaining these gauges is to split the gauge field into background ($B$) and quantum fluctuation ($Q$) components according to
\begin{align}
	A^a_\mu=B^a_\mu+Q^a_\mu.
	\label{split}
\end{align} 
Next, one imposes a residual gauge invariance with respect to $B$ on the gauge-fixed Lagrangian; this can be achieved by choosing a gauge-fixing function transforming in the adjoint representation of SU($N$), in particular through the replacements 
\begin{align}
	\partial_\mu\delta^{ab}\to\widehat{\cal D}^{ab}_\mu\equiv\partial_\mu\delta^{ab}+f^{acb}\widehat{B}^c_\mu;\qquad
A^a_\mu\to Q^a_\mu,
\end{align}
which, once implemented in~\1eq{lincov}, lead to the BFM $R_\xi$ gauge-fixing function
\begin{align}
	\widehat{\cal F}^a=\widehat{\cal D}^{ab}_\mu Q^\mu_b.
	\label{BFM-gf}
\end{align}
Inserting~\1eq{BFM-gf} into~\1eq{GF+FPG}, one obtains the Feynman rules characteristic of the BFM, namely, a symmetric $Bc\overline{c}$ trilinear vertex and the four-particle vertex $BQc\overline{c}$. Finally, inserting~\1eq{split} back into the original invariant Lagrangian, one obtains the conventional Feynman rules, together with those involving $B$; however, to lowest order, only vertices containing exactly two $Q$ differ from the conventional vertices. We encounter one of these vertices in Sect.~II, namely, the $BQ^2$ vertex. 

As a result of the residual gauge invariance, the contraction of the Green's functions with the momentum corresponding to a $B$ gluon leads to Abelian-like Slavnov-Taylor identities (STIs), that is, linear identities that preserve  
their tree-level form to all orders. The divergence of $Q$ instead yields the non-Abelian STIs, akin to 
those of the conventional $R_\xi$ gauges.

It has been found that the conventional and BFM $R_\xi$ gauges are related by symmetry transformations. In fact, as has been shown in~\cite{Binosi:2013cea}, Yang-Mills theories quantized in the BFM emerge in a natural manner 
from Yang-Mills theories quantized in the $R_\xi$ gauges, if one renders the latter also invariant under anti-BRST symmetry. 
This is a crucial construction, because it clarifies the origin  of a plethora of identities, including the so-called background-quantum identities (BQIs)~\cite{Binosi:2002ez,Grassi:1999tp}. The BQIs relate Green's functions evaluated in the conventional $R_\xi$ gauge to the same functions evaluated in the BFM $R_\xi$ gauge. The simplest of these identities, {\it i.e.}, that connecting the corresponding gluon propagators, has been found to be of paramount importance for the self-consistency of the proposed formalism. 

\subsection{Notation and definitions}

%%%%%%%%%%%%%%%%%%%%%%
\begin{figure}[!t]
\begin{center}
\includegraphics[scale=1.0]{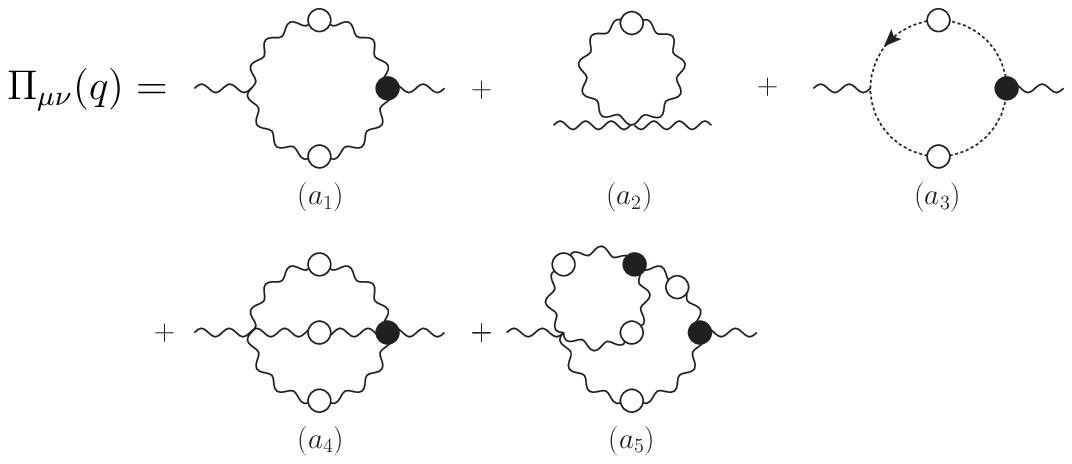}
\caption{\label{glSDEQQ} The conventional SDE of the standard gluon propagator ($QQ$). Black blobs represent fully dressed one-particle irreducible  vertices, whereas the white ones denote fully dressed propagators.}
\end{center}
\end{figure}
%%%%%%%%%%%%%%%%%%%%%%

In the general renormalizable $R_\xi$ gauge defined by means of~\1eq{lincov}, the gluon propagator is given by (we suppress the color factor $\delta^{ab}$) 
\be 
i\Delta_{\mu\nu}(q) = -i\bigg[\Delta(q^2)P_{\mu\nu}(q) + \xi\frac{q_\mu q_\nu}{q^4}\bigg]; \quad P_{\mu\nu}(q) = g_{\mu\nu}-\frac{q_\mu q_\nu}{q^2},
\label{QQprop}
\ee
with inverse
\be \label{invQQprop}
-i\Delta_{\mu\nu}^{-1}(q) = \Delta^{-1}(q^2)P_{\mu\nu}(q) + \xi^{-1}q_\mu q_\nu.
\ee
The function $\Delta(q^2)$, which at tree-level is simply given by $1/q^2$, contains all the dynamics of the gluon propagator, and  is  related to  the  corresponding scalar co-factor of the standard gluon self-energy, $\Pi_{\mu\nu}(q)$ (\fig{glSDEQQ}). Specifically, as $\Pi_{\mu\nu}(q)$  is  both perturbatively and nonperturbatively transverse as a consequence of the BRST symmetry, one obtains   
\be
q^\nu\Pi_{\mu\nu}(q)=0;\qquad \Pi_{\mu\nu}(q)=\Pi(q^2)P_{\mu\nu}(q),
\label{trangen}
\ee
such that 
\be
\Delta^{-1}({q^2}) = q^2+i\Pi(q^2).
\label{defPi}
\ee
Furthermore, it is advantageous for the discussion that follows 
to define the dimensionless function $J(q^2)$ as~\cite{Ball:1980ax}
\be
\Delta^{-1}({q^2})=q^2 J(q^2).
\label{defJ}
\ee
Evidently, $J(q^2)$ corresponds to the {\it inverse} of the gluon dressing function, 
which is frequently employed in the literature.

\begin{figure}[!t]
\centerline{\includegraphics[scale=1.0]{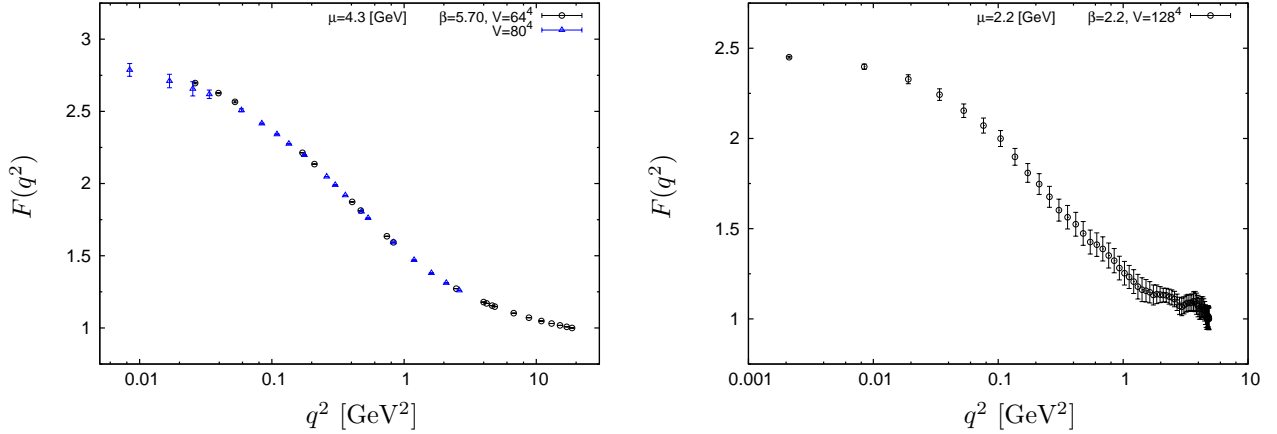}}
\caption{\label{fig:lattice-ghost}The SU(3) (left) and SU(2) (right) ghost dressing function $F$ measured on the lattice. As before, lattice data are from~\cite{Bogolubsky:2009dc,Bogolubsky:2007ud} [SU(3)] and~\cite{Cucchieri:2010xr} [SU(2)].}
\end{figure}

An additional fundamental Green's function, which is extremely relevant for our considerations, is the full ghost propagator denoted by $D(q^2)$. This is usually expressed in terms of the corresponding ghost dressing function $F(q^2)$, according to  
\be
D(q^2) = \frac{F(q^2)}{q^2}.
\ee 
It is important to emphasize that the large-volume lattice simulation mentioned earlier has established beyond any reasonable doubt that, while the ghost remains massless, $F(q^2)$ saturates at a non-vanishing value in the deep infrared region (see~\fig{fig:lattice-ghost}).  This particular feature may be conclusively explained from the SDE that governs $F(q^2)$, as a direct consequence of the fact that the gluon propagator entering the SDE is effectively massive~\cite{Boucaud:2008ky,Aguilar:2008xm}.

\begin{figure}[!t]
\includegraphics[scale=0.5]{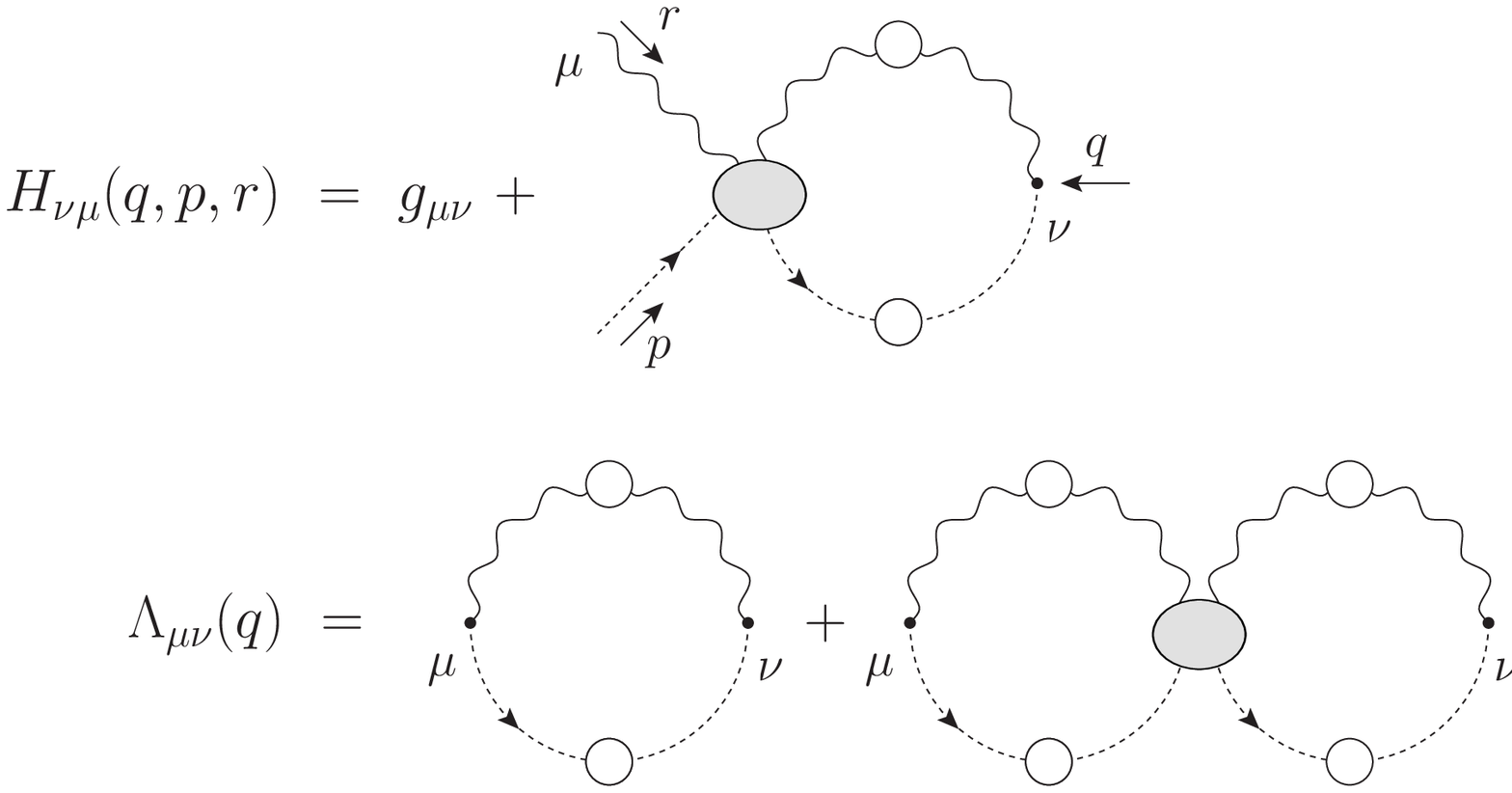}
\caption{\label{fig:H-Lambda-new} The diagrammatic representation of the gluon-ghost functions $H$ (top) and $\Lambda$ (bottom).}
\end{figure}

The $Q^3$ three-gluon vertex at tree-level is given by the 
standard expression
\be   
\Gamma^{(0)}_{\alpha\mu\nu}(q,r,p) = (r-p)_\alpha g_{\mu\nu} + (p - q)_\mu g_{\nu\alpha} + (q - r)_\nu g_{\alpha\mu}\,,
\label{Q3tree}
\ee
and satisfies the simple identity 
\be 
q^{\mu}\gtree_{\mu\alpha\beta}(q,k,-k-q) = (k+q)^2 P_{\alpha\beta}(k+q) - k^2 P_{\alpha\beta}(k).
\label{Q3treeWI}
\ee
The fully dressed version of this vertex (which is the subject of a very active investigation, see, {\it e.g.},~\cite{Pelaez:2013cpa,Aguilar:2013vaa,Eichmann:2014xya,Blum:2014gna}), denoted by $\Gamma_{\alpha\mu\nu}(q,r,p)$,  satisfies instead a rather complicated STI  
\be
q^\alpha \Gamma_{\alpha\mu\nu}(q,r,p) =  F(q)[ \Delta^{-1}(p^2) P^\alpha_\nu(p)H_{\alpha\mu}(p,q,r) - \Delta^{-1}(r^2)P^\alpha_\mu(r)H_{\alpha\nu}(r,q,p)], 
\label{STI3g}
\ee
along with cyclic permutations~\cite{Ball:1980ax}. The function $H$ appearing in~\1eq{STI3g} is the gluon-ghost kernel appearing in the top panel of~\fig{fig:H-Lambda-new}. 

The tree-level value of the $Q^4$ four-gluon vertex is given by
\begin{eqnarray}
{\Gamma}_{\mu\nu\rho\sigma}^{(0)mnrs} &=& -ig^2[f^{msx}f^{xrn}(g_{\mu\rho}g_{\nu\sigma}-
g_{\mu\nu}g_{\rho\sigma}) + f^{mnx}f^{xsr}(g_{\mu\sigma}g_{\nu\rho}-g_{\mu\rho}g_{\nu\sigma}) \nonumber \\
&+& f^{mrx}f^{xsn}(g_{\mu\sigma}g_{\nu\rho}-g_{\mu\nu}g_{\rho\sigma})].
\label{tree-levelfour}
\end{eqnarray}
and its divergence satisfies the identity 
\bea 
\label{WIBQ3}
q^\mu {\Gamma}^{(0) mnrs}_{\mu\nu\rho\sigma}(q,r,p,t) &=& f^{mse}f^{ern} \Gamma^{(0)}_{\nu\rho\sigma}(r,p,q+t) + f^{mne}f^{esr}\Gamma^{(0)}_{\rho\sigma\nu}(p,t,q+r) \nonumber \\
&+& f^{mre}f^{ens} \Gamma^{(0)}_{\sigma\nu\rho}(t,r,q+p).
\eea 
The fully dressed version of this vertex satisfies instead a very complicated STI, which is of limited usefulness 
and will not be discussed here [see, {\it e.g.},~\cite{Binosi:2008qk}, Eq. (D.18)]. 

In addition, for reasons that will become apparent soon, we also consider 
a special, ghost-related two-point function (see~\fig{fig:H-Lambda-new}, bottom panel)
\bea
\Lambda_{\mu\nu}(q)&=&-i\gA\int_k\!\Delta_\mu^\sigma(k)D(q-k)H_{\nu\sigma}(-q,q-k,k)\nonumber\\
&=&g_{\mu\nu}G(q^2)+\frac{q_\mu q_\nu}{q^2}L(q^2),
\label{Lambda}
\eea
where $C_A$ represents the Casimir eigenvalue of the adjoint representation [$N$ for SU($N$)], $d=4-\epsilon$ is the space-time dimension, and we have introduced the integral measure
\be
\int_{k}\equiv\frac{\mu^{\epsilon}}{(2\pi)^{d}}\!\int\!\mathrm{d}^d k,
\label{dqd}
\ee
with $\mu$ being the 't Hooft mass.

\begin{figure}
	\includegraphics{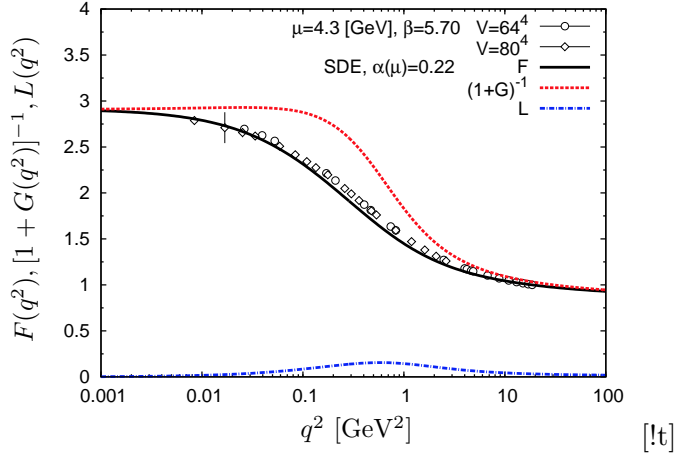}[!t]
	\caption{\label{1+GandL} Decomposition of the ghost dressing function $F$ into its $1+G$ and $L$ components. The renormalization point is $\mu=4.3$ GeV; lattice data are from~\cite{Bogolubsky:2009dc}.}
\end{figure}

Finally, note that 
the form factors $F(q^2)$, $G(q^2)$, and  $L(q^2)$ satisfy 
the exact relation~\cite{Grassi:2004yq,Aguilar:2009nf,Aguilar:2009pp,Aguilar:2010gm}  
\be
F^{-1}(q^2) = 1 + G(q^2) + L(q^2) ,
\label{funrel}
\ee
in the Landau gauge only. To facilitate the forthcoming analysis, we will use the approximate %(but rather accurate)  
relation 
\be
 1 + G(q^2) \approx F^{-1}(q^2),
\label{GFapp}
\ee
which becomes exact in the deep infrared region~\cite{Grassi:2004yq,Aguilar:2009nf,Aguilar:2009pp,Aguilar:2010gm}. 
We emphasize, however, that $L(q^2)$ is sizable at intermediate momenta, as shown in~\fig{1+GandL}. This, in turn, may induce appreciable contributions when calculating certain properties of phenomenological interest~\cite{Binosi:2014aea}.

\subsection{Gluon SDE in the PT-BFM framework}

The nonperturbative dynamics of the gluon propagator are governed by the corresponding SDE. In particular, 
within the conventional formulation~\cite{Roberts:1994dr,Maris:2003vk}, $\Pi_{\mu\nu}(q)$ 
is given by the fully dressed diagrams shown in \fig{glSDEQQ}.
This particular equation is known to be detrimentally affected by a serious complication, which in the vast 
majority of applications is tacitly ignored. Specifically, 
the SDE in \fig{glSDEQQ} cannot be truncated in any obvious way without 
compromising the validity of \1eq{trangen}. This is because the  
fully dressed vertices appearing in the diagrams of \fig{glSDEQQ} satisfy the complicated STIs
mentioned earlier, and it is only after the inclusion of {\it  all diagrams} that \1eq{trangen} may be enforced. 
This characteristic property constitutes a well-known textbook fact, as it 
has already manifested at the lowest order in perturbation theory. The one-loop version of ($a_1$) in~\fig{glSDEQQ}  is not transverse 
in isolation, it is only after the inclusion of the ($a_3$) ghost diagram
that the sum of both diagrams 
becomes transverse. As a result, the BRST symmetry of the theory [the most immediate consequence of which is \1eq{trangen}]
is bound to be compromised if one understands the term ``truncation'' as meaning 
the simple omission of diagrams.
Instead,
the formulation of this SDE in the context of the  PT-BFM scheme furnishes 
considerable advantages, because  it facilitates a systematic truncation 
that respects manifestly, and at every step, the crucial identity of \1eq{trangen}~\cite{Binosi:2007pi,Binosi:2008qk}.

To observe this mechanism in some detail,
let us employ the BFM terminology introduced above, and 
classify the gluon fields as either $B$ or $Q$.
Then, three types of gluon propagator may be defined: 
({\it i}) the conventional gluon propagator (with one $Q$ gluon entering and one exiting, $Q^2$), 
denoted (as above) by $\Delta(q^2)$; ({\it ii})
the background gluon propagator (with one $B$ gluon entering and one exiting, $B^2$), 
denoted by $\widehat\Delta(q^2)$; and ({\it iii})
the mixed  background-quantum gluon propagator (with the $Q$ gluon entering and the $B$ gluon exiting, $BQ$), 
denoted by  $\widetilde\Delta(q^2)$. 

%%%%%%%%%%%%%%%%%%%%%%
\begin{figure}[!t]
\begin{center}
\includegraphics[scale=1.0]{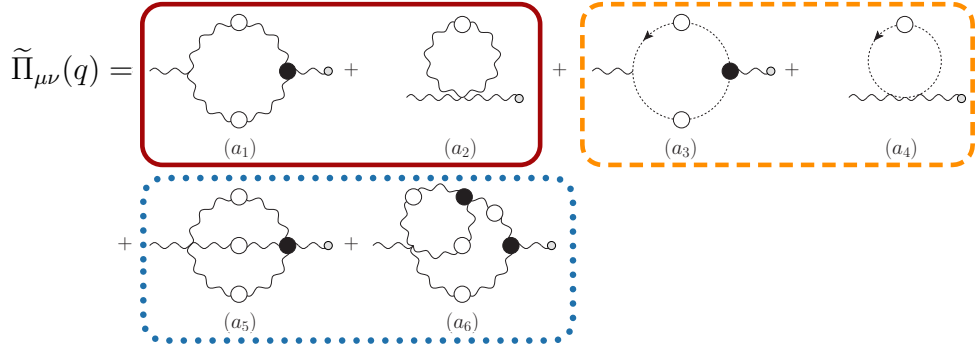}
\caption{\label{glSDE} The SDE obeyed by the $BQ$ gluon propagator. Black (white) blobs represents fully dressed 1-PI vertices (propagators);  
the small gray circles appearing on the external legs (entering from the right, only!) 
are used to indicate background gluons. The diagrams contained in each box form  
individually transverse subsets.}
\end{center}
\end{figure}
%%%%%%%%%%%%%%%%%%%%%%
%%%%%%%%%%%%%%%

We now consider the SDE that 
controls the self-energy of 
the mixed $BQ$ propagator $\widetilde\Pi_{\mu\nu}(q)$, which is shown in \fig{glSDE}.
The fully dressed vertices appearing in the corresponding diagrams,
namely the $BQ^2$, $B{\bar c}c$, and $BQ^3$ vertices, 
are denoted by $\widetilde{\Gamma}_{\alpha\mu\nu}$, $\widetilde{\Gamma}_\alpha$, and
$\widetilde{\Gamma}^{mnrs}_{\mu\nu\rho\sigma}$, respectively.
When contracted with the momentum carried by the $B$ gluon,
these vertices are known to satisfy Abelian STIs, specifically,
\be 
\label{WIBQ2}
q^\alpha \widetilde{\Gamma}_{\alpha\mu\nu}(q,r,p) = i\Delta_{\mu\nu}^{-1}(r) - i\Delta_{\mu\nu}^{-1}(p), 
\ee
\be 
\label{WIBcc}
q^\alpha \widetilde{\Gamma}_\alpha(q,r,-p) = D^{-1}(q+r) - D^{-1}(r),
\ee
and 
\bea 
q^\mu \widetilde{\Gamma}^{mnrs}_{\mu\nu\rho\sigma}(q,r,p,t) &=& f^{mse}f^{ern} \Gamma_{\nu\rho\sigma}(r,p,q+t) + f^{mne}f^{esr}\Gamma_{\rho\sigma\nu}(p,t,q+r) \nonumber \\
&+& f^{mre}f^{ens} \Gamma_{\sigma\nu\rho}(t,r,q+p).
\label{abSTIBQ3}
\eea
In particular, note that ~\1eq{abSTIBQ3} is the naive all-order generalization of~\1eq{tree-levelfour}, as stated, because the vertices appearing on the right-hand side (rhs) are the {\it fully dressed} $Q^3$ vertices.

We remind the reader that the tree-level expression for $\widetilde{\Gamma}_{\alpha\mu\nu}(q,r,p)$ depends explicitly on~$\xi$, such that 
\be 
\widetilde{\Gamma}^{(0)}_{\alpha\mu\nu}(q,r,p) = (r-p)_\alpha g_{\mu\nu} + (p - q + \xi^{-1}r)_\mu g_{\nu\alpha} + (q - r - \xi^{-1}p)_\nu g_{\alpha\mu},
\label{BQ2tree}
\ee
and satisfies the tree-level version of \1eq{WIBQ2}, where
\bea
q^\alpha\widetilde{\Gamma}^{(0)}_{\alpha\mu\nu}(q,r,p) &=&  \left\{ p^2 P_{\mu\nu}(p) + \xi^{-1} p_{\mu}p_{\nu}\right\} - 
\left\{ r^2 P_{\mu\nu}(r) + \xi^{-1} r_{\mu}r_{\nu} \right\},
\nonumber\\
&=& i\Delta_{(0)\,\mu\nu}^{-1}(r) - i\Delta_{(0)\,\mu\nu}^{-1}(p).
\eea 
An in-depth study of this vertex has been conducted in~\cite{Binosi:2011wi}.
On the other hand, the $BQ^3$ vertex coincides at tree level with the $Q^4$ conventional vertex,  {\it i.e.}, with the expression given in~\1eq{tree-levelfour}.

By virtue of the special Abelian STIs of \3eqs{WIBQ2}{WIBcc}{WIBQ3}, it is relatively straightforward to prove  
the block-wise transversality of $\widetilde\Pi_{\mu\nu}(q)$, where~\cite{Aguilar:2006gr}  
\be
q^{\nu} [(a_1) + (a_2)]_{\mu\nu} = 0;\qquad
q^{\nu} [(a_3) + (a_4)]_{\mu\nu} = 0; \qquad
q^{\nu} [(a_5) + (a_6)]_{\mu\nu} = 0.
\label{boxtr2}
\ee 
This is clearly an important property that has far-reaching practical implications for the treatment of the $\widetilde{\Delta}(q^2)$ SDE, as it furnishes a systematic, manifestly gauge-invariant truncation scheme~\cite{Aguilar:2006gr,Binosi:2007pi,Binosi:2008qk}. For instance, one can consider only the one-loop dressed gluon diagrams $(a_1)$ and $(a_2)$ and still find a transverse answer, despite the omission of the remaining graphs (most notably the ghost loops).

However, although it is evident that the diagrammatic representation of  $\widetilde\Pi_{\mu\nu}(q)$ is considerably better organized than that of the conventional  $\Pi_{\mu\nu}(q)$,  it is also clear that the SDE of $\widetilde\Delta(q^2)$ contains $\Delta(q^2)$ within its defining diagrams; therefore, in that sense, it cannot be considered as a {\it bona fide} dynamical equation for  $\widetilde\Delta(q^2)$ or $\Delta(q^2)$. At this point, a crucial identity (BQI) relating $\Delta(q^2)$ and   $\widetilde{\Delta}(q^2)$~\cite{Grassi:1999tp, Binosi:2002ez} enters the discussion.  Specifically, one has  
\be
\Delta(q^2) = [1 + G(q^2)] \widetilde{\Delta}(q^2), 
\label{BQIs}
\ee
with $G(q^2)$ having been defined in \1eq{Lambda}. 

The novel perspective put forth in~\cite{Aguilar:2006gr,Binosi:2007pi,Binosi:2008qk}
is that one may use the SDE for $\widetilde{\Delta}(q^2)$ expressed in terms of the 
BFM Feynman rules, take advantage of its improved truncation properties,  and then 
convert it to an equivalent equation for  $\Delta(q^2)$ (the propagator simulated on the lattice) 
by means of \1eq{BQIs}. Then, the SDE for the conventional gluon propagator within the PT-BFM formalism reads  
\be
\Delta^{-1}(q^2){ P}_{\mu\nu}(q) = 
\frac{q^2 {P}_{\mu\nu}(q) + i\,\sum_{i=1}^{6}(a_i)_{\mu\nu}}{1+G(q^2)}.
\label{sde}
\ee
The $(a_i)$ diagrams are shown in Fig.~\ref{glSDE}. 

\section{Demystifying the seagull graph}

In the context of non-Abelian gauge theories, the seagull graph [($a_2$) in Figs.~\ref{glSDEQQ} and~\ref{glSDE}] has traditionally been
considered quite controversial. At the perturbative level 
and within dimensional regularization, formulas such as 
\be
\int_{k} \frac{\ln^{n} (k^2) }{k^2} = 0 , \quad n =0, 1,2,\ldots,
\label{seapert}
\ee
cause this graph to vanish, a fact which enforces 
the masslessness of the gluon to all orders in perturbation theory.

Further complexity is found in relation to the nonperturbative case, 
because, in general, there is no mathematical justification whatsoever for setting
\be
\int_k \Delta(k^2) = 0 .
\label{seanopert}
\ee
Given that the seagull has dimensions of mass-squared, with 
no momentum for saturation, one might develop the impression that this graph alone 
(\ie without any concrete dynamical mechanism) might suffice for endowing the 
gluon with mass.
However, it eventually becomes apparent that there is a fundamental flaw in this conjecture. Indeed, this graph 
diverges ``quadratically'' as a $\Lambda^2$ term in cutoff language or 
as $\mu^2 (1/\epsilon)$ in dimensional regularization, if it does not vanish (which it is not required to do). 
The disposal of such divergences requires the inclusion in the original Lagrangian of a
counter-term of the form $\mu^2 A^2_{\mu}$, which is, however, 
forbidden by the local gauge invariance of the theory.

\subsection{Scalar QED: Enlightenment from the photon}

\begin{figure}[!t]
\begin{center}
\includegraphics[scale=0.6]{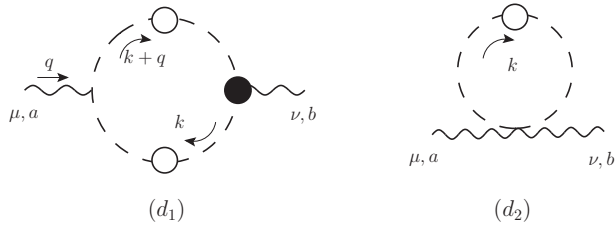}
\end{center}
\vspace{-1.0cm}
\caption{The ``one-loop dressed'' SDE for the photon self-energy.}
\label{p2}
\end{figure}

At this point, the question may be reversed. In a theory such as scalar QED, 
the seagull graph is generated by a definitely massive scalar propagator, and the 
corresponding seagull diagram is certainly non-zero [in fact, at one-loop level it 
can be computed exactly, see \1eq{dimint}]. However, on physical grounds, one cannot argue that the nonvanishing 
of the seagull graph would eventually endow the photon with a mass.
Therefore, the precise mechanism that prevents this from occurring must be determined.

At the one-loop dressed level, the SDE for the photon self-energy, $\Pi^{(1)}_{\mu\nu}(q)$, is given by 
the sum of the two diagrams shown in \fig{p2}, such that
\be 
\Pi^{(1)}_{\mu\nu}(q) = (d_1)_{\mu\nu} + (d_2)_{\mu\nu},
\label{projphotzeroPi}
\ee
with
\bea \label{d1photon}
(d_1)_{\mu\nu}  &=& e^2 \int_k (2k+q)_{\mu} {\cal D}(k){\cal D}(k+q)\Gamma_{\nu}(-q,k+q,-k),
\\
(d_2)_{\mu\nu} &=& -2 e^2 g_{\mu\nu} \int_k {\cal D}(k^2),
\label{d2photon}
\eea 
where  ${\cal D}(p^2)$ is the fully dressed propagator of the scalar field and 
$\Gamma_\mu(q,r,-p)$ the fully dressed photon-scalar vertex.
By virtue of the well-known Abelian STI relating these two quantities  
\be 
q^\mu \Gamma_\mu(q,r,-p) = {\cal D}^{-1}(p^2) - {\cal D}^{-1}(r^2) ,
\label{WIAphi2}
\ee
it is elementary to demonstrate the exact transversality of  $\Pi^{(1)}_{\mu\nu}(q)$, where 
\be
q^{\mu}\Pi^{(1)}_{\mu\nu}(q) = 0 ,
\label{transqed}
\ee
such that
\be
\Pi^{(1)}_{\mu\nu}(q) = \left( g_{\mu\nu}-\frac{q_\mu q_\nu}{q^2} \right){\Pi}^{(1)}(q^2) .
\label{transqed2}
\ee

It is clear that the seagull graph $(d_2)$ is independent of the momentum, and thus, 
proportional to $g_{\mu\nu}$ only. 
If we also set $q=0$ in $(d_1)$, its  
contribution is also proportional to $g_{\mu\nu}$; therefore, one immediately concludes that 
\be
{\Pi}^{(1)}(0) =0 ,
\label{Pi0}
\ee
because of \1eq{transqed2} and the fact that the $q_\mu q_\nu/q^2$ component vanishes.
Evidently, this is also true for the $g_{\mu\nu}$ component; the only question is 
how exactly this is enforced in the presence of the seagull graph.

Let us denote the corresponding co-factors of  $g_{\mu\nu}$ as $d_1$ and $d_2$;
then, we obtain 
\be
{\Pi}^{(1)}(0) = d_1 + d_2 ,
\ee
with 
\bea
d_1  &=& \frac{2 e^2}{d} \int_k k_{\mu} {\cal D}^2(k^2)\Gamma^\mu(0,k,-k) ,
\label{d1q0}\\
d_2  &=& -2 e^2 \int_k {\cal D}(k^2) .
\label{d2q0}
\eea
In order to proceed further, 
let us study \1eq{WIAphi2} in the limit $q\to 0$.
To that end, we perform a Taylor expansion of 
both sides around $q=0$ (and $p=-r$), such that
\be 
\label{TaylWIAphi2}
q^\mu \Gamma_\mu(q,r,-p) = q^\mu \Gamma_\mu(0,r,-r) + {\cal O}(q^2) = q^\mu \left.\frac{\partial}{\partial q^\mu}{\cal D}^{-1}(q+r)\right\vert_{q=0} + {\cal O}(q^2).
\ee
Then, equating the coefficients of the terms that are linear in $q^\mu$, one obtains the relation
\bea \label{IWIAphi2q0}
\Gamma_\mu(0,r,-r) &=& \left.\frac{\partial}{\partial q^\mu}{\cal D}^{-1}(q+r)\right\vert_{q=0}=\frac{\partial {\cal D}^{-1}(r^2)}{\partial r^\mu},
\eea
which is the exact analogue of the familiar textbook Ward identity (WI) of spinor QED.
Then, 
\be 
{\cal D}^2(k^2)\Gamma^\mu(0,k,-k) = -\frac{\partial {\cal D}(k^2)}{\partial k^\mu},
\label{D2IWIAphi2}
\ee
and so 
\be
d_1 = - \frac{4 e^2}{d} \int_k k^2 \frac{\partial {\cal D}(k^2)}{\partial k^2} ,
\ee
using 
\be
k^\mu \frac{\partial f(k^2)}{\partial k^\mu} = 2 k^2 \frac{\partial f(k^2)}{\partial k^2} .
\label{derf}
\ee
Then, summing $d_1$ and $d_2$,  we finally obtain 
\be
{\Pi}^{(1)}(0) = - \frac{4 e^2}{d}
\left[\int_k k^2 \frac{\partial {\cal D}(k^2)}{\partial k^2} + \frac{d}{2}\int_k {\cal D}(k^2)\right].
\label{derf1}
\ee
However, we know from \1eq{Pi0} that the rhs of \1eq{derf1} must vanish. 
Therefore, we must determine the mathematical mechanism that causes this to occur.

\subsection{The seagull identity}

%%%%%%%%%%%%%%%%%%%%%%%%%%%%%%%%%%%%%%%%%%%%%%%%%%%

Let us consider a function $f(k^2)$ that  
satisfies the conditions originally imposed by Wilson~\cite{Wilson:1972cf}, \ie 
as $k^2\rightarrow \infty$ it vanishes sufficiently rapidly that the integral 
$\int_k f(k^2)$
converges for all positive values of $d$ below a certain value $d^{*}$. Then, 
the integral is well-defined within the interval $(0,d^{*})$, and may be analytically continued 
outside this interval, following the standard rules of dimensional regularization~\cite{Collins:1984xc}. 
Then, one can show that~\cite{Aguilar:2009ke}
\be 
\int_k k^2\frac{\partial f(k^2)}{\partial k^2} + \frac{d}{2}\int_k f(k^2) = 0 .
\label{sea}
\ee
In order to properly interpret \1eq{sea}, note that, 
if the function $f(k^2)$ were such that the two integrals appearing in this equation would converge for $d=4$,  then its validity could be demonstrated through simple integration by parts (suppressing the angular contribution), such that
\be 
\int_0^\infty dyy^{\frac{d}{2}}\frac{\partial f(y)}{\partial y} = y^{\frac{d}{2}}f(y)\big\vert_0^\infty - \frac{d}{2}\int_0^\infty dyy^{\frac{d}{2}-1}f(y), 
\label{intparts}
\ee
and dropping the surface term.  

Let us instead consider   
$f(k^2)$ to be a massive tree-level propagator, \ie
\be \label{fmassprop}
f(k^2) = \frac{1}{k^2-m^2},
\ee
for which the assumption of individual convergence of each contribution for $d=4$ is invalid; on the other hand, 
Wilson's condition is indeed  satisfied  with    
$d^{*}=2$, such that  both integrals converge in the interval $(0,2)$.
Then, 
one may still interpret \1eq{sea} {\it via} an integration by parts, 
where the surface term given by $\frac{y^{\frac{d}{2}}}{y+m^2 }\big\vert_0^\infty$ may be dropped if $d < d^{*}=2$.

To confirm that the validity of \1eq{sea} is completely natural 
within the dimensional regularization formalism, it is simply necessary to compute the left-hand side (lhs) of \1eq{sea} explicitly, 
using textbook integration rules. One obtains 
\bea 
\int_k \frac{k^2}{(k^2-m^2)^2} &=& -i(4\pi)^{-\frac{d}{2}}\frac{d}{2}\,\Gamma\left(1-\frac{d}{2}\right)  (m^2)^{\frac{d}{2}-1}, \nonumber \\
\int_k \frac{1}{k^2-m^2} &=& -i(4\pi)^{-\frac{d}{2}} \, \Gamma\left(1-\frac{d}{2}\right) (m^2)^{\frac{d}{2}-1}, 
\label{dimint}
\eea
and substitution into the lhs of \1eq{sea} gives exactly zero.

\subsection{The seagull cancellation in the PT-BFM framework}

Let us now consider the gluon propagator and 
examine the diagrams contributing to the first 
block. We denote by $\widetilde{\Pi}^{(1)}_{\mu\nu}(q)$ the
corresponding self-energy. Following exactly the same reasoning as in the scalar QED case, 
we have 
\be 
\widetilde{\Pi}^{(1)}(0) = a_1 + a_2,
\label{projQCDzeroPi}
\ee 
with 
\bea
a_1 &=& \frac{g^2C_A}{2d} \int_k \Gamma_{\mu\alpha\beta}^{(0)}(0,k,-k)\Delta^{\alpha\rho}(k)\Delta^{\beta\sigma}(k)\widetilde{\Gamma}^\mu_{\sigma\rho}(0,k,-k),
\label{a1q0}\\
a_2 &=& g^2C_A \frac{(d-1)}{d}\int_k\Delta^{\alpha}_{\alpha}(k),
\label{a2q0}
\eea
and 
\be 
\Gamma_{\mu\alpha\beta}^{(0)}(0,k,-k) = 2k_{\mu}g_{\alpha\beta} - k_{\beta}g_{\alpha\mu} -k_{\alpha}g_{\beta\mu}.
\ee

In order to obtain $a_1$, we may 
begin from \1eq{WIBQ2} and follow the steps presented in \1eq{TaylWIAphi2}. Then, 
the fact that \1eq{WIBQ2} is Abelian-like gives rise to the very simple result
\be 
\label{IWIBQ2}
\widetilde{\Gamma}_{\alpha\mu\nu}(0,r,-r) = -i \frac{\partial \Delta^{-1}_{\mu\nu}(r) }{\partial r^\alpha},
\ee
which then furnishes the exact equivalent to \1eq{D2IWIAphi2} 
\be 
\label{2propsvertex}
\Delta^{\alpha\rho}(k)\Delta^{\beta\sigma}(k)\widetilde{\Gamma}^\mu_{\sigma\rho}(0,k,-k) = \frac{\partial \Delta^{\alpha\beta}(k)}{\partial k^\mu}.
\ee
Thus, 
\be
a_1 = \frac{g^2C_A}{2d} \int_k \Gamma_{\mu\alpha\beta}^{(0)}(0,k,-k) \frac{\partial \Delta^{\alpha\beta}(k)}{\partial k^\mu}.
\ee
The derivative above is evaluated by acting on the expression for $\Delta^{\alpha\beta}(k)$ given in \1eq{QQprop} and, again using \1eq{derf}, we obtain
\be
a_1 =\frac{2(d-1)}{d} g^2C_A \left[ \int_k k^2 \frac{\partial \Delta(k^2)}{\partial k^2} + \frac{1}{2} \int_k \Delta(k^2) \right].
\ee
For the $a_2$ term and using \1eq{QQprop}, we find 
\be
a_2 = g^2C_A \frac{(d-1)^2}{d}\int_k \Delta(k^2).
\ee
Note that all terms proportional to $\xi$, both in $a_1$ and $a_2$, 
vanish by virtue of the most elementary version of \1eq{seapert}, \ie for $n=0$. 

Then, we obtain 
\bea 
\widetilde{\Pi}(0) &=& \frac{2(d-1)}{d} g^2C_A 
\left[\int_k k^2 \frac{\partial \Delta(k^2)}{\partial k^2} + \frac{d}{2}\int_k \Delta(k^2)\right]
= 0,
\label{pi0}
\eea
using \1eq{sea} with $f(k^2) \to \Delta(k^2)$ in the final step.

\section{Dynamical gluon mass with exact BRST symmetry}

In this section, we review the field-theoretic mechanism that endows the 
gluon with a dynamical mass, while maintaining the BRST symmetry of the 
theory.

\subsection{\label{remind}The Schwinger mechanism in Yang-Mills theories}

The self-consistent generation of a gluon mass in the context of a Yang-Mills theory
proceeds through the implementation of the well-known Schwinger mechanism~\cite{Schwinger:1962tn,Schwinger:1962tp} at the level of the 
gluon SDE.
The general concept may be encapsulated more directly at the level of its inverse propagator, 
$\Delta^{-1}({q^2})=q^2 [1 + i {\bf \Pi}(q^2)]$, where ${\bf \Pi}(q)$ 
is the dimensionless vacuum polarization, \ie ${\Pi}(q^2) = q^2 {\bf \Pi}(q^2)$.
According to Schwinger's fundamental observation, 
if ${\bf \Pi}(q^2)$ 
develops a pole at zero momentum transfer ($q^2=0$), then the 
vector meson (gluon) acquires a mass, even if the gauge symmetry 
forbids a mass term at the level of the fundamental Lagrangian.
Indeed, if ${\bf \Pi}(q^2) = m^2/q^2$, then (in Euclidean space)
\mbox{$\Delta^{-1}(q^2) = q^2 + m^2$}; therefore, 
the vector meson becomes massive, with $\Delta^{-1}(0) = m^2$, 
even though it is massless in the absence of interactions 
($g=0$, ${\bf \Pi} =0$)~\cite{Jackiw:1973tr,Jackiw:1973ha}.

\begin{figure}[!t]
\includegraphics[scale=.6]{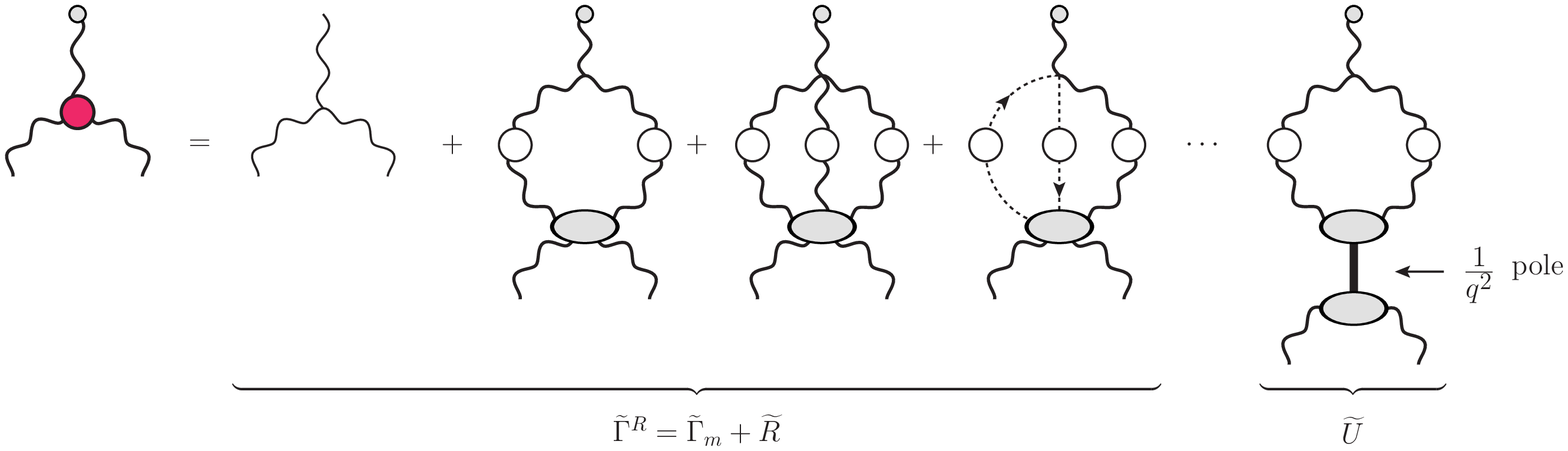}
\caption{\label{Gammaprime}The $\NV$ three-gluon vertex. Thick  internal gluon lines indicates massive propagators $\Delta_m$, as explained in the text.}
\end{figure}
%%%%%%%%%%%%%%%%%%%%%%%%%%%%%%%%%%%%%%%%%%%%%%%%%%%%

The dynamical realization of this concept at the level of a Yang-Mills theory 
requires the existence of a special type of nonperturbative vertex, which is generically denoted by $V$
(with appropriate Lorentz and color indexes). When added 
to the conventional fully dressed vertices, the $V$ vertices have a triple effect: 
\n{i} they evade the seagull cancellation and cause the SDE of the gluon propagator to yield $\Delta^{-1}(0)\neq 0$;
\n{ii} they guarantee that the Abelian and non-Abelian STIs of the theory 
remain intact, \ie they maintain exactly the same form before and after the mass generation; and 
\n{iii} they decouple from {\it on-shell} amplitudes. 
These crucial properties are possible because these special vertices \n{a}
contain massless poles 
and \n{b} are  
completely {\it longitudinally} coupled, \ie they satisfy 
conditions such as  
\be
P^{\alpha'\alpha}(q) P^{\mu'\mu}(r) P^{\nu'\nu}(p) \Vbqq_{\alpha'\mu'\nu'}(q,r,p)  = 0,
\label{totlon}
\ee 
(for a three-gluon vertex).
The origin of the aforementioned massless poles is due to purely non-perturbative dynamics: for sufficiently strong binding, 
the masses of certain (colored) bound states 
may be reduced to zero~\cite{Jackiw:1973tr,Jackiw:1973ha,Cornwall:1973ts,Eichten:1974et,Poggio:1974qs}.  
The actual dynamical realization of this scenario 
has been demonstrated in~\cite{Aguilar:2011xe}, 
where the homogeneous Bethe-Salpeter equation that controls the actual formation 
of these massless bound states was investigated.

From the kinematic perspective, 
we will describe the transition 
from a massless to a massive gluon propagator by performing the replacement  
(in Minkowski space)
\be
\Delta^{-1}(q^2) = q^2 J(q^2) \quad \longrightarrow\quad  \Deltam^{-1}(q^2)=q^2 \Jm(q^2)-m^2(q^2),
\label{massive}
\ee
where $m^2(q^2)$ is the (momentum-dependent) dynamically generated mass, and the subscript ``$m$'' in 
$\Jm$ indicates that, effectively, there is now a mass within the corresponding expressions 
(\ie in the SDE graphs).

Gauge invariance requires that the replacement described schematically in \1eq{massive} 
be accompanied by a simultaneous replacement of all relevant vertices by 
\be
\gfullb \quad \longrightarrow\quad   \NV = \bqqm + \Vbqq ,  
\label{nv}
\ee
where the vertex $\bqqm$ satisfies the STI originally satisfied by $\gfullb$, but now 
with $J(q^2) \to \Jm(q^2)$. Further, 
$\Vbqq$ must provide the missing components 
such that the full vertex $\NV$ satisfies the same STI as $\gfullb$. However, the 
gluon propagators appearing in this expression are now replaced by massive propagators 
[\ie the net effect is to obtain $\Deltam^{-1}(q^2)$ in place of $\Delta^{-1}(q^2)$].

To observe this concept explicitly, consider the example of $\widetilde{\Gamma}_{\alpha\mu\nu}$.
For a ``deactivated'' Schwinger mechanism and when this vertex is contracted with respect to 
the momentum of the $B$ gluon, it satisfies the WI
\be
q^\alpha\widetilde{\Gamma}_{\alpha\mu\nu}(q,r,p)=p^2\bcj(p^2)P_{\mu\nu}(p)-r^2\bcj(r^2)P_{\mu\nu}(r).
\label{STI}
\ee
The general replacement described in (\ref{nv}) amounts to introducing the vertex  
\be
\NV_{\alpha\mu\nu}(q,r,p) = \left[\widetilde{\Gamma}_{m}(q,r,p) + \NP(q,r,p)\right]_{\alpha\mu\nu},
\label{NV}
\ee
where 
\be
q_\alpha\widetilde{\Gamma}^{\alpha\mu\nu}_{m}(q,r,p)=p^2\bcj_{m}(p^2)P^{\mu\nu}(p)-r^2\bcj_{m}(r^2)P^{\mu\nu}(r),
\ee
[that is, \1eq{STI} with $J(q^2) \to \Jm(q^2)$] while 
\be
q^\alpha \NP_{\alpha\mu\nu}(q,r,p)= m^2(r^2)P_{\mu\nu}(r) - m^2(p^2)P_{\mu\nu}(p). 
\label{winp}
\ee
Thus, when the Schwinger mechanism is activated, the corresponding Abelian STI satisfied by $\NV$ reads  
\bea
q^{\alpha}\NV_{\alpha\mu\nu}(q,r,p) &=& 
q^{\alpha}\left[\widetilde{\Gamma}_{m}(q,r,p) + \NP(q,r,p)\right]_{\alpha\mu\nu},
\nonumber\\
&=& [p^2 \Jm (p^2) -m^2(p^2)]P_{\mu\nu}(p) - [r^2 \Jm (r^2) -m^2(r^2)]P_{\mu\nu}(r),
\nonumber\\
&=& \Deltam^{-1}({p^2})P_{\mu\nu}(p) - \Deltam^{-1}({r^2})P_{\mu\nu}(r) ,
\label{winpfull}
\eea
which is indeed the identity in Eq.~(\ref{STI}), with the aforementioned total replacement 
\mbox{$\Delta^{-1} \to \Delta_m^{-1}$} being enforced. 
The remaining STIs, which are triggered when $\NV_{\alpha\mu\nu}(q,r,p)$ is contracted 
with respect to the other two legs,  
are realized in exactly the same fashion.

A completely analogous procedure can be implemented for the four-gluon vertex 
$\widetilde{\Gamma}^{mnrs}_{\mu\nu\rho\sigma}(q,r,p,t)$; the details may be found in~\cite{Binosi:2012sj}. 
Finally, note that ``internal'' vertices, \ie vertices involving only $Q$ gluons,  
must also be supplemented by the corresponding $V$, such that their STIs remain unchanged 
in the presence of ``massive'' propagators. Clearly, these types of vertices 
do not contain $1/q^2$ poles, but rather poles in the virtual momenta;
therefore, they cannot contribute directly to the mass-generating mechanism. However, 
these poles must be included for the gauge invariance to remain intact.

Let us now return to the SDE of the gluon propagator.
By expressing the $\Deltam^{-1}(q^2)$ on the lhs of \1eq{sde}
in the form given in \1eq{massive}, one obtains 
\be
[q^2 \Jm(q^2)-m^2(q^2)]{P}_{\mu\nu}(q) = 
\frac{q^2 {P}_{\mu\nu}(q) + 
i\, \sum_{i=1}^{6}(a_i^{\prime})_{\mu\nu}}{1+G(q^2)},
\label{sdem}
\ee
where the ``prime''  indicates that the various fully dressed vertices 
appearing inside the corresponding diagrams 
must be replaced by their primed counterparts, as dictated by~\1eq{nv}. 
These modifications produce one of the primary desired effects, that is, that the blockwise transversality 
property of \1eq{boxtr2} also holds for the ``primed'' graphs, \ie when  
$(a_i)\to (a_i^{\prime})$. 

We next discuss the realization of the second desired effect, which is to evade the seagull cancellation and to enable the possibility of having $\Delta^{-1}(0)\neq 0$.

\subsection{Evading the seagull identity}

In the case of the $BQ^2$ vertex, the poles are included by setting 
\be 
\widetilde{V}_{\alpha\mu\nu}(q,r,p) = \widetilde{U}_{\alpha\mu\nu}(q,r,p) + \widetilde{R}_{\alpha\mu\nu}(q,r,p),
\label{BQ2pole}
\ee
where 
\be
\widetilde{U}_{\alpha\mu\nu}(q,r,p) = \frac{q_\alpha}{q^2}\widetilde{C}_{\mu\nu}(q,r,p),
\ee
contains $1/q^2$ explicitly. Further, $\widetilde{R}_{\alpha\mu\nu}$ has massless excitations in the other two channels, 
namely ${\cal O}(r^{-2})$ and/or ${\cal O}(p^{-2})$, but not ${\cal O}(q^{-2})$.
Note also that the explicit forms of $\widetilde{C}_{\mu\nu}$ and $\widetilde{R}_{\alpha\mu\nu}$ 
may be determined using the longitudinally coupled condition of \1eq{totlon}, 
as well as the known Abelian and non Abelian STIs satisfied by this vertex~\cite{Ibanez:2012zk}. 

We first focus on the vertex $\widetilde{\Gamma}'_{\alpha\mu\nu}(q,r,p)$ given by 
\be 
\widetilde{\Gamma}'_{\alpha\mu\nu}(q,r,p) = \left[\widetilde{\Gamma}_{\alpha\mu\nu}(q,r,p) + \widetilde{R}_{\alpha\mu\nu}(q,r,p)\right]
+ \frac{q_\alpha}{q^2}\widetilde{C}_{\mu\nu}(q,r,p),
\label{abc1}
\ee
where the two terms in the square brackets are both regular in $q$. Their combined contribution 
\be 
\widetilde{\Gamma}^{\rm R}_{\alpha\mu\nu}(q,r,p)  :=  \widetilde{\Gamma}_{\alpha\mu\nu}(q,r,p) + \widetilde{R}_{\alpha\mu\nu}(q,r,p),
\ee
is precisely the part of the total vertex $\widetilde{\Gamma}'$ 
that enters the calculation of $\widetilde \Pi(0) g_{\mu\nu}$,
and consequently participates in the seagull cancellation. On the other hand, the term 
with the massless pole in $q^2$ contributes to the  $\widetilde \Pi(0) q_{\mu}q_{\nu}/q^2$ term, which is not involved in the seagull cancellation. 
Of course, because of the exact transversality of the final answer, the total contribution of the $g_{\mu\nu}$ component 
(after the seagull cancellation) is exactly equal (and opposite in sign) to that proportional to $q_{\mu}q_{\nu}/q^2$.

The next task is to derive the Abelian STI satisfied by $\widetilde{\Gamma}^{\rm R}$.  
To that end, let us contract both sides of \1eq{abc1} by $q^\alpha$, such that 
\be
q^\alpha \widetilde{\Gamma}'_{\alpha\mu\nu}(q,r,p) = 
q^\alpha\widetilde{\Gamma}^{\rm R}_{\alpha\mu\nu}(q,r,p) + \widetilde{C}_{\mu\nu}(q,r,p). 
\label{qprimeBQ2}
\ee
Note that the massless pole $q_\alpha/q^2$ has been canceled by the contraction with $q^\alpha$,  and all quantities 
appearing on both sides of \1eq{qprimeBQ2} may be directly expanded around  $q=0$. 

To obtain the lhs of \1eq{qprimeBQ2} in this limit, consider the STI of \1eq{winpfull} satisfied by 
$\Gamma'$. It is clear that the Taylor expansion of both sides of that equation 
(neglecting terms of order ${\cal O}(q^2)$ and higher, as above) yields 
\be 
\widetilde{\Gamma}'_{\alpha\mu\nu}(0,r,-r) = -i \frac{\partial \Delta^{-1}_{m\,\mu\nu}(r) }{\partial r^\alpha},
\label{IWIBQ2prime}
\ee
which is simply \1eq{IWIBQ2} with $\Delta(q^2) \to \Delta_{m}(q^2)$.

On the other hand, the rhs of \1eq{qprimeBQ2}, expanded in the same limit, yields 
\be 
q^\alpha \widetilde{\Gamma}^{\rm R}_{\alpha\mu\nu}(0,r,-r) + 
\widetilde{C}_{\mu\nu}(0,r,-r)  + 
q^\alpha \left.\frac{\partial}{\partial q^\alpha}\widetilde{C}_{\mu\nu}(q,r,p)\right\vert_{q=0}.
\ee
Then, after equating the coefficients of the zeroth- and first-order terms in $q^\alpha$ on both sides, one obtains 
\be \label{CBQ2zero}
\widetilde{C}_{\mu\nu}(0,r,-r) = 0,
\ee
and 
\be 
\label{IWIBQ2pole}
\widetilde{\Gamma}^R_{\alpha\mu\nu}(0,r,-r) = -i\frac{\partial}{\partial r^\alpha}\Delta^{-1}_{m\,\mu\nu}(r) - \left.\frac{\partial}{\partial q^\alpha}\widetilde{C}_{\mu\nu}(q,r,p)\right\vert_{q=0}.
\ee

It is now clear that, if one were to repeat the calculation of subsection 3C, 
the seagull identity would again eliminate all contributions,
with the exception of the term that causes the deviation in the WI of \1eq{IWIBQ2pole}. 
The remaining term is given by 
\be 
\label{0a1a2mass}
\widetilde{\Pi}^{(1)}(0) = \frac{g^2C_A}{2d}g^2C_A \int_k \Gamma_{\mu\alpha\beta}^{(0)}\Delta^{\alpha\rho}(k)\Delta^{\beta\sigma}(k)\left.\frac{\partial}{\partial q^\mu}\widetilde{C}_{\sigma\rho}(-q,k+q,-k)\right\vert_{q=0}.
\ee

%%%%%%%%%%%%%%%%%%%%%%%%%
\section{\label{oneloop} The gluon gap equation}

The lhs  of \1eq{sdem} 
involves two unknown quantities, $J_m(q^2)$ and $m^{2}(q^2)$, which eventually 
satisfy two separate, but coupled, integral equations of the generic type
\bea
J_m(q^2) &=& 1+ \int_{k} {\cal K}_1 (q^{2},m^2,\Delta_m),
\nonumber\\
m^{2}(q^2) &=&  \int_{k} {\cal K}_2 (q^{2},m^2,\Delta_m),
\label{separ}
\eea
where $q^{2} {\cal K}_1 (q^{2},m^2,\Delta_m) \to 0$ as $q^{2}\to 0$. 
However, ${\cal K}_2 (q^{2},m^2,\Delta_m)\neq 0$ in the same limit,
precisely because it includes the  $1/q^2$ terms contained within the $\widetilde{V}$ terms.

%%%%%%%%%%%%%%%%%%%%%%%%%%%%%%%%%%%%%%%%%%%

Let us now derive the explicit form of the integral equation governing $m^{2}(q^2)$.
We perform this particular task in the Landau gauge, 
where the gluon propagator 
assumes the fully transverse form 
\be
i{\Delta}_{\mu\nu}(q)=-i{\Delta}(q^2)P_{\mu\nu}(q).
\label{prop-def}
\ee 
The primary reasons for this choice are the considerable simplifications 
that it introduces at the calculation level, and the fact that  
the vast majority of recent large-volume lattice simulations 
of Yang-Mills Green's functions have been performed in this special gauge.

As a gluon mass cannot be generated in the absence of ${\widetilde V}$, it is natural to expect that the  rhs of \1eq{separ}
is generated from the parts of the $(a_i^{\prime})_{\mu\nu}$ graphs that contain precisely ${\widetilde V}$, 
which we denote by $(a^{\s {\widetilde V}}_{i})_{\mu\nu}$. However, it may be less obvious that the 
$(a^{\s {\widetilde V}}_{i})_{\mu\nu}$ terms possess no $g_{\mu \nu}$ component in the Landau gauge, \ie
\be
(a^{\s {\widetilde V}}_{i})_{\mu\nu} = \frac{q_\mu q_\nu}{q^2} a^{\s {\widetilde V}}_{i}(q^2),
\ee
such that
\be
m^2(q^2) = \frac{ i \sum_{i}  a_{i}^{\s {\widetilde V}}(q^2)}{1+G(q^2)},
\label{masseq}
\ee
where the sum includes only the $i=1,5$, and $6$ graphs.

At first, this last statement may appear to contradict the earlier claim that the contribution from the mass must be completely transverse, that is, it must possess a $g_{\mu\nu}$ component that is equal in size and opposite in sign. The solution to this apparent paradox is intimately connected with the exact realization of the seagull cancellation, which operates exclusively in the $g_{\mu\nu}$ sector; for further detail, see the discussion following~\1eq{sdeJ}.

\begin{figure}
\mbox{}\hspace{-0.4cm}\includegraphics[scale=.64]{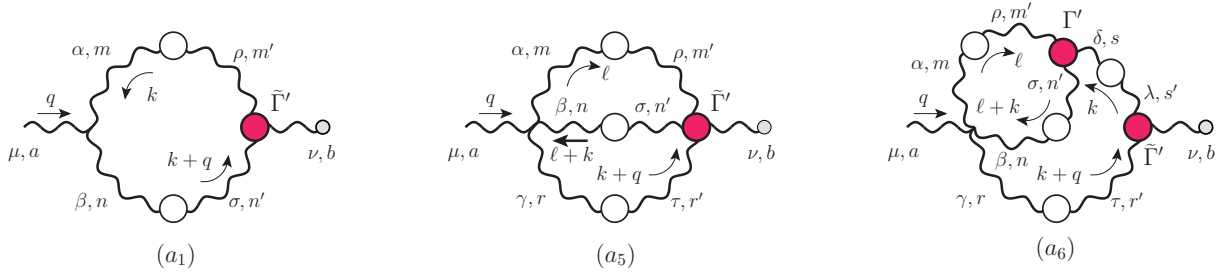} 
\caption{\label{a1-a2} (color online). The {\it one-} and {\it two-loop dressed} diagrams contributing to the gluon mass equation. Thick lines represent, as previously explained, gluon propagators endowed with a momentum-dependent mass. The fully dressed {\it primed} vertex, $\gfullb'$, enforces gauge invariance in the presence of such a mass.  The symmetry factors are $1/2$ ($a_1$ and $a_6$) and $1/6$ ($a_5$). We also show for the reader's convenience the color and Lorentz indexes, as well as  the momentum routing used in our calculations.}
\end{figure}

In order to observe all these features in some detail, 
we consider the contribution that originates from the $\NP$-part of the $(a_1^{\prime})_{\mu\nu}$ graph, 
which we denote by $(a^{\s {\widetilde V}}_{1})_{\mu\nu}$.
Then (see \fig{a1-a2}), 
\be
(a^{\s {\widetilde V}}_{1})_{\mu\nu} = \frac12\,\gA 
\int_k\! {\Gamma}^{(0)}_{\mu\alpha\beta}(q,k,-k-q)\Delta^{\alpha\rho}(k)
\Delta^{\beta\sigma}(k+q)\NP_{\nu\rho\sigma}(q,k,-k-q).
\label{gengau}
\ee

As explained in Sect.~\ref{remind}, 
the condition of gauge invariance requires that   
the vertex $\NP_{\nu\rho\sigma}(q,k,-k-q)$ satisfies the Abelian STI of Eq.~\noeq{winp} with $r=k$ and $p=-(k+q)$ when contracted by the momentum of the background leg. Thus, 
\be
q^\nu \NP_{\nu\rho\sigma}(q,k,-k-q)= m^2(k)P_{\rho\sigma}(k) - m^2(k+q)P_{\rho\sigma}(k+q).
\label{winp1}
\ee

It is relatively straightforward to determine that $(a^{\s {\widetilde V}}_{i})_{\mu\nu}$ 
is  proportional to $q_{\mu} q_{\nu}/q^2$ only.
Indeed, the condition of complete longitudinality of $\NP$, given in \1eq{totlon}, becomes 
\be 
P^{\nu\nu'}(q)P^{\alpha\rho}(k)P^{\beta\sigma}(k+q)\NP_{\nu'\rho\sigma}(q,k,-k-q)=0.
\label{transV}
\ee
Hence, it immediately follows that 
\be 
P^{\alpha\rho}(k)P^{\beta\sigma}(k+q)\NP^{\nu}_{\rho\sigma}(q,k,-k-q) = 
\frac{q^{\nu}}{q^2} \left[q^{\nu'}\NP_{\nu'\rho\sigma}(q,k,-k-q)\right]P^{\alpha\rho}(k)P^{\beta\sigma}(k+q), 
\label{sma}
\ee
and, thus, $(a^{\s {\widetilde V}}_{1})_{\mu\nu}$ is  proportional to $q_{\mu} q_{\nu}/q^2$ only, as stated.

It is interesting that the rhs of \1eq{sma} is {\it completely determined} from the Abelian STI of \1eq{winp}; 
specifically, using~\noeq{winp1}, we obtain
\be 
P^{\alpha\rho}(k)P^{\beta\sigma}(k+q)\NP^{\nu}_{\rho\sigma}(q,k,-k-q) = \frac{q^{\nu}}{q^2}
\left[m^2(k) - m^2(k+q)\right] P^{\alpha\rho}(k) P^{\beta}_{\rho}(k+q).
\label{smawi}
\ee
Then, using \1eq{Q3treeWI} and appropriate shifts of the integration variable, one can finally show that 
\be
a^{\s {\widetilde V}}_{1}(q^2) = \frac{\gA }{q^2}
\int_k\! m^2(k^2)\left[(k+q)^2-k^2\right] \Delta^{\alpha\rho}(k)\Delta_{\alpha\rho}(k+q).
\label{1l-dr-1}
\ee

We next turn to the $(a_6)$ graph and define the quantity 
\be
\Y_{\delta}^{\alpha\beta}(k)=\int_\ell\!\Delta^{\alpha\rho}(\ell)\Delta^{\beta\sigma}(\ell+k)\Gamma_{\sigma\rho\delta}(-\ell-k,\ell,k),
\label{defY}
\ee
which corresponds to the sub-diagram on the upper left corner of this graph.
Then, $(a_6^{\widetilde V})_{\mu\nu}$ is given by
\be
(a_6^{\widetilde V})_{\mu\nu} = \frac34i\gAsq\left(g_{\mu\alpha}g_{\beta\gamma}-g_{\mu\beta}g_{\alpha\gamma}\right)
\int_k\! \Y_{\delta}^{\alpha\beta}(k) \Delta^{\gamma\tau}(k+q)\Delta^{\delta\lambda}(k) {\widetilde V}_{\nu\tau\lambda}(-q,k+q,-k).
\ee
Using \3eqs{winp}{transV}{sma}, we obtain
\bea
(a_6^{\widetilde V})_{\mu\nu}
&=&\frac34i\gAsq\left(g_{\mu\alpha}g_{\beta\gamma}-g_{\mu\beta}g_{\alpha\gamma}\right)\frac{q^\nu}{q^2}\int_k\!\left[m^2(k)-m^2(k+q)\right]\Delta^{\delta}_{\lambda}(k)\Delta^{\gamma\lambda}(k+q)\Y_{\delta}^{\alpha\beta}(k), \nonumber \\
&=&\frac{q_\mu q_\nu}{q^2} a_6^{\widetilde V}(q^2),
\eea
and, therefore,
\be
a_6^{\widetilde V}(q^2)=\frac34i\gAsq\left(q_{\alpha}g_{\beta\gamma}-q_{\beta}g_{\alpha\gamma}\right)\frac1{q^2}\int_k\!\left[m^2(k)-m^2(k+q)\right]\Delta^\delta_\lambda(k)\Delta^{\gamma\lambda}(k+q)\Y_{\delta}^{\alpha\beta}(k).
\label{fnal}
\ee

At this point, it is easy to show that the integral $Y$ is antisymmetric under the $\alpha\leftrightarrow\beta$ exchange; 
thus, given also the antisymmetry of the $a_6^{\widetilde V}$ prefactor under the same exchange, one can state 
\be
\Y^{\alpha\beta}_{\delta}(k)=(k^\alpha g^\beta_\delta-k^\beta g^\alpha_\delta)\Y(k^2),
\label{Ydef}
\ee 
which gives the final result
\bea
a_6^{\widetilde V}(q^2) &=&\frac3{4}i\frac{\gAsq}{q^2}\int_k\!
m^2(k^2)[(k+q)^2-k^2][Y(k+q)+Y(k)]\Delta^\delta_\lambda(k)\Delta_\delta^\lambda(k+q)\nonumber \\
&+&\frac3{4}i\frac{\gAsq}{q^2}(q^2g_{\delta\gamma}-2q_\delta q_\gamma)\int_k\!
m^2(k^2)[Y(k+q)-Y(k)]\Delta^\delta_\lambda(k)\Delta^{\gamma\lambda}(k+q).
\label{a6Vtilde} 
\eea

Finally, a rather straightforward sequence of algebraic manipulations reveals a striking fact, \ie that the 
$(a_5)$ graph does not contribute to the mass equation in the Landau gauge~\cite{Binosi:2012sj}.

At this point, one may substitute the results of \1eq{1l-dr-1} and \1eq{a6Vtilde} into \1eq{masseq},  
in order to obtain the final form of the gluon gap equation.
Passing to Euclidean space by following standard rules, we find 
\begin{align}
	m^2(q^2)=-\frac{g^2 C_A}{1+G(q^2)}\frac1{q^2}\int_k m^2(k^2)\Delta_\alpha^\rho(k)\Delta_\beta^\rho(k+q){\cal K}^{\alpha\beta}(q,k),
	\label{meq}
\end{align} 
where the kernel ${\cal K}$ is given by
\begin{align}
	{\cal K}^{\alpha\beta}(q,k,-k-q)&=[(k+q)^2-k^2]S(q,k)g^{\alpha\beta}+q^2A(q,k)g^{\alpha\beta}+B(q,k)q^\alpha q^\beta,
	\label{Ker}
\end{align}
with
\begin{align}
	S(q,k)&=1 - \frac{3}{4} g^2 C_A[Y(k+q)+Y(k)];&\nonumber \\
	A(q,k)&= - \frac{1}{2} B(q,k)= \frac{3}{4} g^2 C_A [Y(k+q)-Y(k)].
\label{SAB}
\end{align}

%\begin{figure}
%\includegraphics[scale=1]{diagrammaticmass-1} 
%\caption{\label{diagrammaticmass} Diagrammatic representation of the condensed operations 
%leading to the all order gluon mass equation, where we have introduced the shorthand notation   
%${\widetilde m}^2 (q^2) = m^2(q^2)[1+G(q^2)]$. %All internal propagators are in the Landau %tttttttttttttttttttttttttttttttttttttttttttttttttttttttttttttttttttttttttttttttttttttttttttttttttttttttttttttttttttttttttttttttttttttttttttttttttttttttttttttttttttttttttttttttttttttttttttttttttttttttttttttttttttttttttttttttttttttttttttttttttttttttttttttttttttttttttttttttttttttttttttttttttttttttttttttttttttttttttttttttttttttttttttttttttttttttttttttttttttttttttttttttttttttttttttttttttttttttttttttttttttttttttttttttttttttttttttttttttttttttttttttttttttttttttttttttttttttttttttttttttttt	§gauge.}
%\end{figure} 

We next comment on the following additional important points: 

\begin{itemize}
	\item[\n{i}] The equation for $J_m(q^2)$ 
may be obtained from the  $q_{\mu} q_{\nu} /q^{2}$ component 
of the parts of the graphs that do not contain $\widetilde V$. These graphs 
are identical to the original set $(a_1)$--$(a_6)$, but now $\gfullb \longrightarrow\bqqm$, 
$\Delta \longrightarrow \Delta_m$, {\it etc.}, and their 
contributions may be separated into  $g_{\mu\nu}$ and $q_{\mu} q_{\nu} /q^{2}$ components, where 
\be
(a_i)_{\mu\nu} = g_{\mu\nu}\, {A_i}(q^2)  + \frac{q_{\mu} q_{\nu}}{q^{2}} {B_i}(q^2).
\label{AandB}
\ee
Note that $(a_2)$ and $(a_4)$
are proportional to $g_{\mu\nu}$ only; therefore, in the notation introduced above, $B_2(q^2)=B_4(q^2)=0$.
Then, the corresponding equation for $J_m(q^2)$ reads 
\be
- q^2 \Jm(q^2) =  \frac{-q^2 + i \sum_{i} B_i(q^2)}{1+G(q^2)},
\label{sdeJ}
\ee
with $i=1,3,5$, and $6$.

\item[\n{ii}] It is interesting to examine the case where the results obtained above  
are reproduced by considering the parts of \1eq{sdem} that are proportional to $g_{\mu\nu}$. The easiest way to 
disentangle and identify the contributions to $q^2 \Jm(q^2)$ and $m^2(q^2)$ is to first
provide $\{-a_{i}^{\s {\widetilde V}}(q^2)\} g_{\mu\nu}$ by hand, in order to manifest 
the transversality of the mass term, 
and then compensate by adding $a_{i}^{\s {\widetilde V}}(q^2) g_{\mu\nu}$
to the ${A_i}(q^2)$ defined in \1eq{AandB}. The sum of the combined contributions, ${A_i}(q^2) + a_{i}^{\s {\widetilde V}}(q^2)$, 
then determines the $q^2 \Jm(q^2) g_{\mu\nu}$ term. In fact, in order to demonstrate that  
${A_i}(0) + a_{i}^{\s {\widetilde V}}(0)$ vanishes 
(as it should, since it is to be identified with $q^2 \Jm(q^2)$, which vanishes as $q^2\to 0$) 
one must judiciously invoke the seagull cancellation of \1eq{sea}.

\item[\n{iii}] We emphasize once again that  
the Lagrangian of the Yang-Mills theory (or that of QCD) was  not altered throughout the entire mass-generating procedure. In addition, the 
crucial STIs that encapsulate the underlying BRST symmetry remained rigorously exact. 
Moreover, because of the validity of the seagull identity, along with the fact that the PT-BFM scheme permits 
this identity to manifest unambiguously, all would-be quadratic divergences were completely annihilated. This 
conclusively excludes the need for introduction of a symmetry-violating  ``bare gluon mass''. 

\item[\n{iv}] Although there is no ``bare gluon mass'' in the sense explained above, 
the momentum-dependent $m^2(q^2)$ undergoes renormalization. However, 
this is not associated with 
a new renormalization constant, but is rather implemented by the (already existing) 
wave-function renormalization constant of the gluon, namely,  $Z_{\chic A}$.
Specifically, from \1eq{massive} and given that $\Delta^{-1}(0) = m^2(0)$,
we find that the gluon masses before and after renormalization are related by~\cite{Aguilar:2014tka} 
\be
m^2_{\chic R}(q^2) = Z_{\chic A} m^2_0(q^2).
\label{glmassren}
\ee
Evidently, this particular ``renormalization'' is not associated with 
a counter-term of the type $\delta m^2 = m^2_{\chic R} -  m_0^2$, as is the case for hard boson masses [which is precisely the essence of point \n{iii}].

\item[\n{v}] In order to fully determine the nonperturbative $\Delta(q^2)$, one should, in principle, solve the coupled system of \1eq{separ}. However, 
the derivation of the all-order integral equation for $\Jm(q^2)$ is technically far more difficult, 
primarily because of the presence of the fully dressed vertex $BQ^3$ [see ($a_5$) in Fig.~\ref{glSDE}]. 
The latter is a practically unexplored quantity with an enormous number of form factors (for recent works on the subject see~\cite{Binosi:2014kka,Cyrol:2014kca}). 
Instead, we study \1eq{meq} in isolation, 
treating all full propagators appearing in this calculation as external quantities, the forms of which are determined 
by resorting to information beyond the SDEs, such as the large-volume lattice simulations. 
Therefore, \1eq{meq} is effectively converted into a homogeneous {\it linear} integral equation for 
the unknown  $m^2(q^2)$. 
\end{itemize}

\begin{figure}[!t]
\centerline{\includegraphics[scale=1]{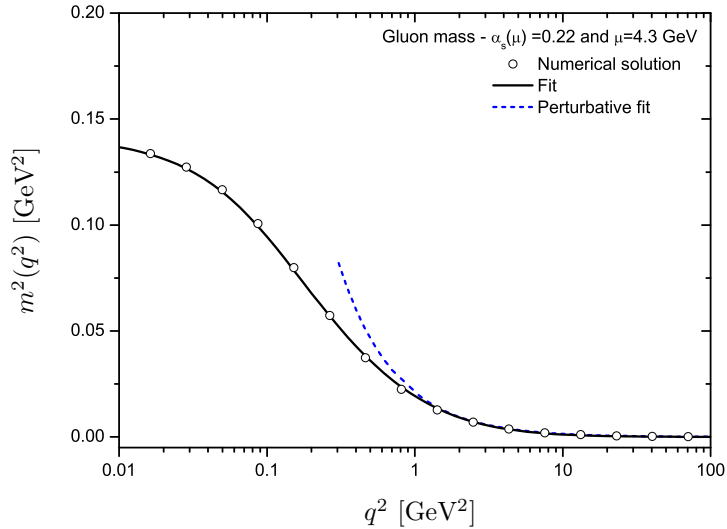}}
\caption{\label{fig:fit}(color online) The numerical solution for \mbox{$m^2(q^2)$} (black circles) 
compared with the corresponding fit~\noeq{fit} (black, continuous).  The (blue) dashed curve represents the asymptotic fit given by~\1eq{asfit}.}
\end{figure}

We now turn to the numerical analysis of the gluon gap equation. After its 
full renormalization has been carefully performed\footnote{This rather technical procedure, 
and the manner in which it affects the form of the 
renormalized kernel ${\cal K}^{\alpha\beta}$, has been presented in~\cite{Aguilar:2014tka}.}, \1eq{GFapp} has been utilized,
and the substitution of $\Delta(k^2)$ and $F(q^2)$ into \1eq{meq} using the lattice data of~\cite{Bogolubsky:2009dc,Bogolubsky:2007ud}
 has been implemented, one obtains positive-definite and monotonically 
decreasing solutions, as shown in~\fig{fig:fit}. 
This numerical solution can be accurately fit using the simple and physically motivated function
\begin{align}
	m^2(q^2)=\frac{m_0^2(q^2)}{1+(q^2/{\cal M}^2)^{1+p}}.
	\label{fit}
\end{align}
Specifically, the numerical solution shown in~\fig{fig:fit} is perfectly reproduced when the parameters~$({\cal M},p)$ assume the values (436 MeV, 0.15).

%%%%%%%%%%%%%%%%%%%%%%%%%%

%%%%%%%%%%%%%%%%%%%%%%%%

In addition, note that one can omit the 1 in the denominator of~\1eq{fit} for asymptotically large momentum values, yielding ``power-law'' behavior~\cite{Cornwall:1985bg,Lavelle:1991ve,Aguilar:2007ie}, where
\begin{align}
	m^2(q^2)\underset{q^2\gg {\cal M}^2}{\sim}\frac{m_0^2{\cal M}^2}{q^2}(q^2/{\cal M}^2)^{-p}.
	\label{asfit}
\end{align} 
This particular behavior reveals that 
condensates of dimension two do not contribute to the operator product expansion (OPE) of $m^2(q^2)$, given that    
their presence would have induced a logarithmic running of the solutions. 
Indeed, in the absence of quarks, the lowest-order condensates appearing in the OPE of the mass must be those of dimension four, namely, the (gauge-invariant) $\langle 0|\hspace{-0.125cm}:\hspace{-0.125cm}G_{\mu\nu}^{a}
G^{\mu\nu}_{a}\hspace{-0.125cm}:\hspace{-0.125cm}|0  \rangle$, and possibly the ghost condensate \mbox{$\langle 0|\hspace{-0.125cm}:\hspace{-0.125cm}{\overline c}^{a} \,\Box \, c^{a} \hspace{-0.125cm}:\hspace{-0.125cm}|0 \rangle$}~\cite{Lavelle:1988eg,Bagan:1989gt,Brodsky:2010xf}. 
As these condensates must be divided by $q^2$ on dimensional grounds, one obtains 
(up to logarithms) the observed power-law behavior.

We end this section by commenting that, as has been argued recently~\cite{Cloet:2013jya}, 
the nontrivial momentum dependence of the gluon mass shown in~\fig{fig:fit} may be considered responsible  
for the fact that, in contradistinction to a propagator with a constant mass, the $\Delta(q^2)$ 
of~\fig{fig:lattice-gluon} displays an inflection point. The presence of such  
a feature, in turn, is a sufficient condition for the spectral density of $\Delta(q^2)$, $\rho$, to be 
non-positive definite.  

%%%%%%%%%%%%%%%%%%%%%%%%%%%%%%%%%%%%%%%%%%%%%%%%%%%%%%%%%%%%%%%%%%%%%%%%%%
%Fig.  Inflection point - derivatives
%%%%%%%%%%%%%%%%%%%%%%%%%%%%%%%%%%%%%%%%%%%%%%%%%%%%%%%%%%%%%%%%%%%%%%%%%%%%
\begin{figure}[!t]
\centerline{\includegraphics[scale=1]{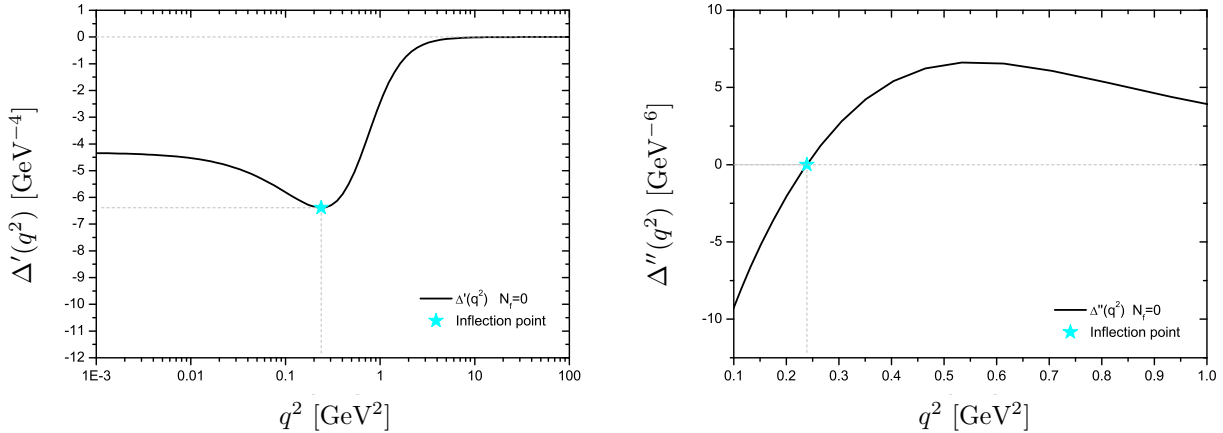}}
\vspace{-0.25cm}
\caption{\label{fig:derivatives} The first and second derivatives of the gluon propagator.}
\end{figure}
%%%%%%

Specifically, the K\"all\'en-Lehman representation of $\Delta(q^2)$ reads  
\be
\Delta(q^2)= \int^{\infty}_{0} d\sigma \frac{\rho(\sigma)}{q^2+ \sigma} ,
\label{spectral}
\ee
and if $\Delta(q^2)$  has an inflection point at $q^2_{\star}$, then its second 
derivative vanishes at that point (see~\fig{fig:derivatives}), such that~\cite{Tandy}
\be
\Delta^{\prime\prime}(q^2_{\star})= 2\int^{\infty}_{0} d\sigma \frac{\rho(\sigma)}{(q^2_{\star}+ \sigma)^3} = 0 .
\label{spectral2}
\ee
Given that $q^2_{\star}>0$, then the sign of $\rho(\sigma)$ is forced to reverse at least once. 
This non-positivity of  $\rho(\sigma)$ may be 
interpreted as an indication of confinement (see~\cite{Cloet:2013jya}, and references therein), 
because the resultant breeching of the axiom of reflection positivity excludes the gluon 
from the Hilbert space of observable states (for related works, see~\cite{DelDebbio:1996mh,Szczepaniak:2001rg,Langfeld:2001cz,Szczepaniak:2003ve,Greensite:2003bk,Gattnar:2004bf,Greensite:2011pj}).
As can be seen in~\fig{fig:derivatives}, the first derivative of $\Delta(q^2)$ exhibits a minimum at \mbox{$q^2_{\star} = 0.238 \, \mbox{GeV}^2$} 
and, consequently, the second derivative vanishes at the same point.

\section{Conclusions}

In this paper, we have considered the manner in which the 
 dynamical gluon mass is generated in pure Yang-Mills theories.
Lattice simulations reveal that this phenomenon 
also persists in the presence of light dynamical quarks, not 
only in ``quenched,'' but also in ``unquenched'' settings~\cite{Ayala:2012pb}. 
From the theoretical perspective, the generalization of the   
formalism outlined here to include the effects of a small 
number of families of light quarks has been developed in Refs.~\cite{Aguilar:2012rz,Aguilar:2013hoa}.
In addition, although we focused on the Landau gauge case throughout this discussion, recent lattice simulations~\cite{Bicudo:2015rma} and a variety of analytic studies~\cite{Aguilar:2015nqa,Huber:2015ria,Siringo:2015gia,Capri:2015ixa}
have indicated that gluon propagators continue to saturate in the infrared region 
for values of the gauge-fixing parameter that are at least within the $[0,0.5]$ interval.

A large number of profound implications are related to the generation of gluon mass~\cite{Roberts:2015dea}, 
such as the notion of a maximum gluon wavelength~\cite{Brodsky:2008be}, 
above which an effective decoupling (screening) of the gluonic modes occurs.
In addition, the crucial role of such a mass in overcoming the Gribov copy 
problem of Yang-Mills theories has also been noted. 
Moreover, the puzzling phenomenon of the saturation of the gluon parton distribution functions  
may also be a consequence of the emergence of such a mass~\cite{Roberts:2015dea}. 
We hope to examine some of these issues in more profound detail in the near future.

\acknowledgments

The research of J.~P. is supported by the Spanish MEYC under 
grant FPA2011-23596, FPA2014-53631-C2-1-P, and SEV-2014-0398 and the Generalitat Valenciana under grant “PrometeoII/2014/066”. The work of  A.~C.~A  is supported by the 
National Council for Scientific and Technological Development - CNPq
under the grant 306537/2012-5 and project 473260/2012-3,
and by S\~ao Paulo Research Foundation - FAPESP through the project 2012/15643-1.
We thank Craig Roberts for numerous stimulating discussions,  
and the organizers of the Workshop 
``Dyson-Schwinger Equations in Modern Mathematics and Physics'' for their hospitality.

%\bibliography{biblio}

\begin{thebibliography}{101}%
\makeatletter
\providecommand \@ifxundefined [1]{%
 \@ifx{#1\undefined}
}%
\providecommand \@ifnum [1]{%
 \ifnum #1\expandafter \@firstoftwo
 \else \expandafter \@secondoftwo
 \fi
}%
\providecommand \@ifx [1]{%
 \ifx #1\expandafter \@firstoftwo
 \else \expandafter \@secondoftwo
 \fi
}%
\providecommand \natexlab [1]{#1}%
\providecommand \enquote  [1]{``#1''}%
\providecommand \bibnamefont  [1]{#1}%
\providecommand \bibfnamefont [1]{#1}%
\providecommand \citenamefont [1]{#1}%
\providecommand \href@noop [0]{\@secondoftwo}%
\providecommand \href [0]{\begingroup \@sanitize@url \@href}%
\providecommand \@href[1]{\@@startlink{#1}\@@href}%
\providecommand \@@href[1]{\endgroup#1\@@endlink}%
\providecommand \@sanitize@url [0]{\catcode `\\12\catcode `\$12\catcode
  `\&12\catcode `\#12\catcode `\^12\catcode `\_12\catcode `\%12\relax}%
\providecommand \@@startlink[1]{}%
\providecommand \@@endlink[0]{}%
\providecommand \url  [0]{\begingroup\@sanitize@url \@url }%
\providecommand \@url [1]{\endgroup\@href {#1}{\urlprefix }}%
\providecommand \urlprefix  [0]{URL }%
\providecommand \Eprint [0]{\href }%
\providecommand \doibase [0]{http://dx.doi.org/}%
\providecommand \selectlanguage [0]{\@gobble}%
\providecommand \bibinfo  [0]{\@secondoftwo}%
\providecommand \bibfield  [0]{\@secondoftwo}%
\providecommand \translation [1]{[#1]}%
\providecommand \BibitemOpen [0]{}%
\providecommand \bibitemStop [0]{}%
\providecommand \bibitemNoStop [0]{.\EOS\space}%
\providecommand \EOS [0]{\spacefactor3000\relax}%
\providecommand \BibitemShut  [1]{\csname bibitem#1\endcsname}%
\let\auto@bib@innerbib\@empty
%</preamble>
\bibitem [{\citenamefont {Cornwall}(1982)}]{Cornwall:1981zr}%
  \BibitemOpen
  \bibfield  {author} {\bibinfo {author} {\bibfnamefont {John~M.}\ \bibnamefont
  {Cornwall}},\ }\bibfield  {title} {\enquote {\bibinfo {title} {{Dynamical
  Mass Generation in Continuum QCD}},}\ }\href@noop {} {\bibfield  {journal}
  {\bibinfo  {journal} {Phys. Rev.}\ }\textbf {\bibinfo {volume} {D26}},\
  \bibinfo {pages} {1453} (\bibinfo {year} {1982})}\BibitemShut {NoStop}%
%%CITATION = PHRVA,D26,1453;%%
\bibitem [{\citenamefont {Aguilar}\ \emph {et~al.}(2003)\citenamefont
  {Aguilar}, \citenamefont {Natale},\ and\ \citenamefont {Rodrigues~da
  Silva}}]{Aguilar:2002tc}%
  \BibitemOpen
  \bibfield  {author} {\bibinfo {author} {\bibfnamefont {A.~C.}\ \bibnamefont
  {Aguilar}}, \bibinfo {author} {\bibfnamefont {A.~A.}\ \bibnamefont {Natale}},
  \ and\ \bibinfo {author} {\bibfnamefont {P.~S.}\ \bibnamefont {Rodrigues~da
  Silva}},\ }\bibfield  {title} {\enquote {\bibinfo {title} {{Relating a gluon
  mass scale to an infrared fixed point in pure gauge QCD}},}\ }\href@noop {}
  {\bibfield  {journal} {\bibinfo  {journal} {Phys. Rev. Lett.}\ }\textbf
  {\bibinfo {volume} {90}},\ \bibinfo {pages} {152001} (\bibinfo {year}
  {2003})},\ \Eprint {http://arxiv.org/abs/hep-ph/0212105} {hep-ph/0212105}
  \BibitemShut {NoStop}%
%%CITATION = HEP-PH/0212105;%%
\bibitem [{\citenamefont {Aguilar}\ \emph {et~al.}(2004)\citenamefont
  {Aguilar}, \citenamefont {Mihara},\ and\ \citenamefont
  {Natale}}]{Aguilar:2004td}%
  \BibitemOpen
  \bibfield  {author} {\bibinfo {author} {\bibfnamefont {A.~C.}\ \bibnamefont
  {Aguilar}}, \bibinfo {author} {\bibfnamefont {A.}~\bibnamefont {Mihara}}, \
  and\ \bibinfo {author} {\bibfnamefont {A.~A.}\ \bibnamefont {Natale}},\
  }\bibfield  {title} {\enquote {\bibinfo {title} {{Phenomenological tests for
  the freezing of the QCD running coupling constant}},}\ }\href@noop {}
  {\bibfield  {journal} {\bibinfo  {journal} {Int. J. Mod. Phys.}\ }\textbf
  {\bibinfo {volume} {A19}},\ \bibinfo {pages} {249--269} (\bibinfo {year}
  {2004})}\BibitemShut {NoStop}%
%%CITATION = IMPAE,A19,249;%%
\bibitem [{\citenamefont {Binosi}\ \emph {et~al.}(2015)\citenamefont {Binosi},
  \citenamefont {Chang}, \citenamefont {Papavassiliou},\ and\ \citenamefont
  {Roberts}}]{Binosi:2014aea}%
  \BibitemOpen
  \bibfield  {author} {\bibinfo {author} {\bibfnamefont {Daniele}\ \bibnamefont
  {Binosi}}, \bibinfo {author} {\bibfnamefont {Lei}\ \bibnamefont {Chang}},
  \bibinfo {author} {\bibfnamefont {Joannis}\ \bibnamefont {Papavassiliou}}, \
  and\ \bibinfo {author} {\bibfnamefont {Craig~D.}\ \bibnamefont {Roberts}},\
  }\bibfield  {title} {\enquote {\bibinfo {title} {{Bridging a gap between
  continuum-QCD and ab initio predictions of hadron observables}},}\ }\href
  {\doibase 10.1016/j.physletb.2015.01.031} {\bibfield  {journal} {\bibinfo
  {journal} {Phys.Lett.}\ }\textbf {\bibinfo {volume} {B742}},\ \bibinfo
  {pages} {183--188} (\bibinfo {year} {2015})},\ \Eprint
  {http://arxiv.org/abs/1412.4782} {arXiv:1412.4782 [nucl-th]} \BibitemShut
  {NoStop}%
%%CITATION = ARXIV:1412.4782;%%
\bibitem [{\citenamefont {Cloet}\ and\ \citenamefont
  {Roberts}(2014)}]{Cloet:2013jya}%
  \BibitemOpen
  \bibfield  {author} {\bibinfo {author} {\bibfnamefont {Ian~C.}\ \bibnamefont
  {Cloet}}\ and\ \bibinfo {author} {\bibfnamefont {Craig~D.}\ \bibnamefont
  {Roberts}},\ }\bibfield  {title} {\enquote {\bibinfo {title} {{Explanation
  and Prediction of Observables using Continuum Strong QCD}},}\ }\href
  {\doibase 10.1016/j.ppnp.2014.02.001} {\bibfield  {journal} {\bibinfo
  {journal} {Prog. Part. Nucl. Phys.}\ }\textbf {\bibinfo {volume} {77}},\
  \bibinfo {pages} {1--69} (\bibinfo {year} {2014})},\ \Eprint
  {http://arxiv.org/abs/1310.2651} {arXiv:1310.2651 [nucl-th]} \BibitemShut
  {NoStop}%
%%CITATION = ARXIV:1310.2651;%%
\bibitem [{\citenamefont {Roberts}(2015)}]{Roberts:2015dea}%
  \BibitemOpen
  \bibfield  {author} {\bibinfo {author} {\bibfnamefont {Craig~D.}\
  \bibnamefont {Roberts}},\ }\bibfield  {title} {\enquote {\bibinfo {title}
  {{Hadron Physics and QCD: Just the Basic Facts}},}\ }in\ \href
  {http://inspirehep.net/record/1341291/files/arXiv:1501.06581.pdf} {\emph
  {\bibinfo {booktitle} {{37th Brazilian Workshop on Nuclear Physics Maresias,
  S{\~a}o Paulo, Brazil, September 8-12, 2014}}}}\ (\bibinfo {year} {2015})\
  \Eprint {http://arxiv.org/abs/1501.06581} {arXiv:1501.06581 [nucl-th]}
  \BibitemShut {NoStop}%
%%CITATION = ARXIV:1501.06581;%%
\bibitem [{\citenamefont {Jackiw}\ and\ \citenamefont
  {Johnson}(1973)}]{Jackiw:1973tr}%
  \BibitemOpen
  \bibfield  {author} {\bibinfo {author} {\bibfnamefont {R.}~\bibnamefont
  {Jackiw}}\ and\ \bibinfo {author} {\bibfnamefont {K.}~\bibnamefont
  {Johnson}},\ }\bibfield  {title} {\enquote {\bibinfo {title} {{Dynamical
  Model of Spontaneously Broken Gauge Symmetries}},}\ }\href@noop {} {\bibfield
   {journal} {\bibinfo  {journal} {Phys. Rev.}\ }\textbf {\bibinfo {volume}
  {D8}},\ \bibinfo {pages} {2386--2398} (\bibinfo {year} {1973})}\BibitemShut
  {NoStop}%
%%CITATION = PHRVA,D8,2386;%%
\bibitem [{\citenamefont {Jackiw}(1973)}]{Jackiw:1973ha}%
  \BibitemOpen
  \bibfield  {author} {\bibinfo {author} {\bibfnamefont {R.}~\bibnamefont
  {Jackiw}},\ }\bibfield  {title} {\enquote {\bibinfo {title} {{Dynamical
  Symmetry Breaking}},}\ }\href@noop {} {\bibfield  {journal} {\bibinfo
  {journal} {In *Erice 1973, Proceedings, Laws Of Hadronic Matter*, New York
  1975, 225-251 and M I T Cambridge - COO-3069-190 (73,REC.AUG 74) 23p}\ }
  (\bibinfo {year} {1973})}\BibitemShut {NoStop}%
\bibitem [{\citenamefont {Cornwall}\ and\ \citenamefont
  {Norton}(1973)}]{Cornwall:1973ts}%
  \BibitemOpen
  \bibfield  {author} {\bibinfo {author} {\bibfnamefont {J.~M.}\ \bibnamefont
  {Cornwall}}\ and\ \bibinfo {author} {\bibfnamefont {R.~E.}\ \bibnamefont
  {Norton}},\ }\bibfield  {title} {\enquote {\bibinfo {title} {{Spontaneous
  Symmetry Breaking Without Scalar Mesons}},}\ }\href@noop {} {\bibfield
  {journal} {\bibinfo  {journal} {Phys. Rev.}\ }\textbf {\bibinfo {volume}
  {D8}},\ \bibinfo {pages} {3338--3346} (\bibinfo {year} {1973})}\BibitemShut
  {NoStop}%
%%CITATION = PHRVA,D8,3338;%%
\bibitem [{\citenamefont {Eichten}\ and\ \citenamefont
  {Feinberg}(1974)}]{Eichten:1974et}%
  \BibitemOpen
  \bibfield  {author} {\bibinfo {author} {\bibfnamefont {E.}~\bibnamefont
  {Eichten}}\ and\ \bibinfo {author} {\bibfnamefont {F.}~\bibnamefont
  {Feinberg}},\ }\bibfield  {title} {\enquote {\bibinfo {title} {{Dynamical
  Symmetry Breaking of Nonabelian Gauge Symmetries}},}\ }\href@noop {}
  {\bibfield  {journal} {\bibinfo  {journal} {Phys. Rev.}\ }\textbf {\bibinfo
  {volume} {D10}},\ \bibinfo {pages} {3254--3279} (\bibinfo {year}
  {1974})}\BibitemShut {NoStop}%
%%CITATION = PHRVA,D10,3254;%%
\bibitem [{\citenamefont {Poggio}\ \emph {et~al.}(1975)\citenamefont {Poggio},
  \citenamefont {Tomboulis},\ and\ \citenamefont {Tye}}]{Poggio:1974qs}%
  \BibitemOpen
  \bibfield  {author} {\bibinfo {author} {\bibfnamefont {E.~C.}\ \bibnamefont
  {Poggio}}, \bibinfo {author} {\bibfnamefont {E.}~\bibnamefont {Tomboulis}}, \
  and\ \bibinfo {author} {\bibfnamefont {S.~H.~H.}\ \bibnamefont {Tye}},\
  }\bibfield  {title} {\enquote {\bibinfo {title} {{Dynamical Symmetry Breaking
  in Nonabelian Field Theories}},}\ }\href {\doibase 10.1103/PhysRevD.11.2839}
  {\bibfield  {journal} {\bibinfo  {journal} {Phys. Rev.}\ }\textbf {\bibinfo
  {volume} {D11}},\ \bibinfo {pages} {2839} (\bibinfo {year}
  {1975})}\BibitemShut {NoStop}%
%%CITATION = PHRVA,D11,2839;%%
\bibitem [{\citenamefont {Bernard}(1983)}]{Bernard:1982my}%
  \BibitemOpen
  \bibfield  {author} {\bibinfo {author} {\bibfnamefont {Claude~W.}\
  \bibnamefont {Bernard}},\ }\bibfield  {title} {\enquote {\bibinfo {title}
  {{Adjoint Wilson lines and the effective gluon mass}},}\ }\href@noop {}
  {\bibfield  {journal} {\bibinfo  {journal} {Nucl. Phys.}\ }\textbf {\bibinfo
  {volume} {B219}},\ \bibinfo {pages} {341} (\bibinfo {year}
  {1983})}\BibitemShut {NoStop}%
%%CITATION = NUPHA,B219,341;%%
\bibitem [{\citenamefont {Donoghue}(1984)}]{Donoghue:1983fy}%
  \BibitemOpen
  \bibfield  {author} {\bibinfo {author} {\bibfnamefont {John~F.}\ \bibnamefont
  {Donoghue}},\ }\bibfield  {title} {\enquote {\bibinfo {title} {{The gluon
  `mass' in the bag model}},}\ }\href@noop {} {\bibfield  {journal} {\bibinfo
  {journal} {Phys. Rev.}\ }\textbf {\bibinfo {volume} {D29}},\ \bibinfo {pages}
  {2559} (\bibinfo {year} {1984})}\BibitemShut {NoStop}%
%%CITATION = PHRVA,D29,2559;%%
\bibitem [{\citenamefont {Bogolubsky}\ \emph {et~al.}(2009)\citenamefont
  {Bogolubsky}, \citenamefont {Ilgenfritz}, \citenamefont {Muller-Preussker},\
  and\ \citenamefont {Sternbeck}}]{Bogolubsky:2009dc}%
  \BibitemOpen
  \bibfield  {author} {\bibinfo {author} {\bibfnamefont {I.L.}\ \bibnamefont
  {Bogolubsky}}, \bibinfo {author} {\bibfnamefont {E.M.}\ \bibnamefont
  {Ilgenfritz}}, \bibinfo {author} {\bibfnamefont {M.}~\bibnamefont
  {Muller-Preussker}}, \ and\ \bibinfo {author} {\bibfnamefont
  {A.}~\bibnamefont {Sternbeck}},\ }\bibfield  {title} {\enquote {\bibinfo
  {title} {{Lattice gluodynamics computation of Landau gauge Green's functions
  in the deep infrared}},}\ }\href {\doibase 10.1016/j.physletb.2009.04.076}
  {\bibfield  {journal} {\bibinfo  {journal} {Phys. Lett.}\ }\textbf {\bibinfo
  {volume} {B676}},\ \bibinfo {pages} {69--73} (\bibinfo {year} {2009})},\
  \Eprint {http://arxiv.org/abs/0901.0736} {arXiv:0901.0736 [hep-lat]}
  \BibitemShut {NoStop}%
%%CITATION = ARXIV:0901.0736;%%
\bibitem [{\citenamefont {Bogolubsky}\ \emph {et~al.}(2007)\citenamefont
  {Bogolubsky}, \citenamefont {Ilgenfritz}, \citenamefont {Muller-Preussker},\
  and\ \citenamefont {Sternbeck}}]{Bogolubsky:2007ud}%
  \BibitemOpen
  \bibfield  {author} {\bibinfo {author} {\bibfnamefont {I.~L.}\ \bibnamefont
  {Bogolubsky}}, \bibinfo {author} {\bibfnamefont {E.~M.}\ \bibnamefont
  {Ilgenfritz}}, \bibinfo {author} {\bibfnamefont {M.}~\bibnamefont
  {Muller-Preussker}}, \ and\ \bibinfo {author} {\bibfnamefont
  {A.}~\bibnamefont {Sternbeck}},\ }\bibfield  {title} {\enquote {\bibinfo
  {title} {{The Landau gauge gluon and ghost propagators in 4D SU(3)
  gluodynamics in large lattice volumes}},}\ }\href@noop {} {\  (\bibinfo
  {year} {2007})},\ \Eprint {http://arxiv.org/abs/arXiv:0710.1968 [hep-lat]}
  {arXiv:0710.1968 [hep-lat]} \BibitemShut {NoStop}%
%%CITATION = ARXIV:0710.1968;%%
\bibitem [{\citenamefont {Bowman}\ \emph {et~al.}(2007)\citenamefont {Bowman}
  \emph {et~al.}}]{Bowman:2007du}%
  \BibitemOpen
  \bibfield  {author} {\bibinfo {author} {\bibfnamefont {Patrick~O.}\
  \bibnamefont {Bowman}} \emph {et~al.},\ }\bibfield  {title} {\enquote
  {\bibinfo {title} {{Scaling behavior and positivity violation of the gluon
  propagator in full QCD}},}\ }\href@noop {} {\bibfield  {journal} {\bibinfo
  {journal} {Phys. Rev.}\ }\textbf {\bibinfo {volume} {D76}},\ \bibinfo {pages}
  {094505} (\bibinfo {year} {2007})},\ \Eprint
  {http://arxiv.org/abs/hep-lat/0703022} {hep-lat/0703022} \BibitemShut
  {NoStop}%
%%CITATION = HEP-LAT/0703022;%%
\bibitem [{\citenamefont {Oliveira}\ and\ \citenamefont
  {Silva}(2009)}]{Oliveira:2009eh}%
  \BibitemOpen
  \bibfield  {author} {\bibinfo {author} {\bibfnamefont {O.}~\bibnamefont
  {Oliveira}}\ and\ \bibinfo {author} {\bibfnamefont {P.J.}\ \bibnamefont
  {Silva}},\ }\bibfield  {title} {\enquote {\bibinfo {title} {{The Lattice
  infrared Landau gauge gluon propagator: The Infinite volume limit}},}\
  }\href@noop {} {\bibfield  {journal} {\bibinfo  {journal} {PoS}\ }\textbf
  {\bibinfo {volume} {LAT2009}},\ \bibinfo {pages} {226} (\bibinfo {year}
  {2009})},\ \Eprint {http://arxiv.org/abs/0910.2897} {arXiv:0910.2897
  [hep-lat]} \BibitemShut {NoStop}%
%%CITATION = ARXIV:0910.2897;%%
\bibitem [{\citenamefont {Cucchieri}\ and\ \citenamefont
  {Mendes}(2007)}]{Cucchieri:2007md}%
  \BibitemOpen
  \bibfield  {author} {\bibinfo {author} {\bibfnamefont {Attilio}\ \bibnamefont
  {Cucchieri}}\ and\ \bibinfo {author} {\bibfnamefont {Tereza}\ \bibnamefont
  {Mendes}},\ }\bibfield  {title} {\enquote {\bibinfo {title} {{What's up with
  IR gluon and ghost propagators in Landau gauge? A puzzling answer from huge
  lattices}},}\ }\href@noop {} {\bibfield  {journal} {\bibinfo  {journal}
  {PoS}\ }\textbf {\bibinfo {volume} {LAT2007}},\ \bibinfo {pages} {297}
  (\bibinfo {year} {2007})},\ \Eprint {http://arxiv.org/abs/0710.0412}
  {arXiv:0710.0412 [hep-lat]} \BibitemShut {NoStop}%
%%CITATION = ARXIV:0710.0412;%%
\bibitem [{\citenamefont {Cucchieri}\ and\ \citenamefont
  {Mendes}(2008)}]{Cucchieri:2007rg}%
  \BibitemOpen
  \bibfield  {author} {\bibinfo {author} {\bibfnamefont {A.}~\bibnamefont
  {Cucchieri}}\ and\ \bibinfo {author} {\bibfnamefont {T.}~\bibnamefont
  {Mendes}},\ }\bibfield  {title} {\enquote {\bibinfo {title} {{Constraints on
  the IR behavior of the gluon propagator in Yang-Mills theories}},}\ }\href
  {\doibase 10.1103/PhysRevLett.100.241601} {\bibfield  {journal} {\bibinfo
  {journal} {Phys.Rev.Lett.}\ }\textbf {\bibinfo {volume} {100}},\ \bibinfo
  {pages} {241601} (\bibinfo {year} {2008})},\ \Eprint
  {http://arxiv.org/abs/0712.3517} {arXiv:0712.3517 [hep-lat]} \BibitemShut
  {NoStop}%
%%CITATION = ARXIV:0712.3517;%%
\bibitem [{\citenamefont {Cucchieri}\ and\ \citenamefont
  {Mendes}(2010)}]{Cucchieri:2009zt}%
  \BibitemOpen
  \bibfield  {author} {\bibinfo {author} {\bibfnamefont {Attilio}\ \bibnamefont
  {Cucchieri}}\ and\ \bibinfo {author} {\bibfnamefont {Tereza}\ \bibnamefont
  {Mendes}},\ }\bibfield  {title} {\enquote {\bibinfo {title} {{Landau-gauge
  propagators in Yang-Mills theories at beta = 0: Massive solution versus
  conformal scaling}},}\ }\href {\doibase 10.1103/PhysRevD.81.016005}
  {\bibfield  {journal} {\bibinfo  {journal} {Phys.Rev.}\ }\textbf {\bibinfo
  {volume} {D81}},\ \bibinfo {pages} {016005} (\bibinfo {year} {2010})},\
  \Eprint {http://arxiv.org/abs/0904.4033} {arXiv:0904.4033 [hep-lat]}
  \BibitemShut {NoStop}%
%%CITATION = ARXIV:0904.4033;%%
\bibitem [{\citenamefont {Cucchieri}\ and\ \citenamefont
  {Mendes}(2009)}]{Cucchieri:2010xr}%
  \BibitemOpen
  \bibfield  {author} {\bibinfo {author} {\bibfnamefont {Attilio}\ \bibnamefont
  {Cucchieri}}\ and\ \bibinfo {author} {\bibfnamefont {Tereza}\ \bibnamefont
  {Mendes}},\ }\bibfield  {title} {\enquote {\bibinfo {title} {{Numerical test
  of the Gribov-Zwanziger scenario in Landau gauge}},}\ }\href@noop {}
  {\bibfield  {journal} {\bibinfo  {journal} {PoS}\ }\textbf {\bibinfo {volume}
  {QCD-TNT09}},\ \bibinfo {pages} {026} (\bibinfo {year} {2009})},\ \Eprint
  {http://arxiv.org/abs/1001.2584} {arXiv:1001.2584 [hep-lat]} \BibitemShut
  {NoStop}%
%%CITATION = ARXIV:1001.2584;%%
\bibitem [{\citenamefont {Aguilar}\ \emph {et~al.}(2008)\citenamefont
  {Aguilar}, \citenamefont {Binosi},\ and\ \citenamefont
  {Papavassiliou}}]{Aguilar:2008xm}%
  \BibitemOpen
  \bibfield  {author} {\bibinfo {author} {\bibfnamefont {A.~C.}\ \bibnamefont
  {Aguilar}}, \bibinfo {author} {\bibfnamefont {D.}~\bibnamefont {Binosi}}, \
  and\ \bibinfo {author} {\bibfnamefont {J.}~\bibnamefont {Papavassiliou}},\
  }\bibfield  {title} {\enquote {\bibinfo {title} {{Gluon and ghost propagators
  in the Landau gauge: Deriving lattice results from Schwinger-Dyson
  equations}},}\ }\href {\doibase 10.1103/PhysRevD.78.025010} {\bibfield
  {journal} {\bibinfo  {journal} {Phys. Rev.}\ }\textbf {\bibinfo {volume}
  {D78}},\ \bibinfo {pages} {025010} (\bibinfo {year} {2008})},\ \Eprint
  {http://arxiv.org/abs/0802.1870} {arXiv:0802.1870 [hep-ph]} \BibitemShut
  {NoStop}%
\bibitem [{\citenamefont {Szczepaniak}\ and\ \citenamefont
  {Swanson}(2002)}]{Szczepaniak:2001rg}%
  \BibitemOpen
  \bibfield  {author} {\bibinfo {author} {\bibfnamefont {Adam~P.}\ \bibnamefont
  {Szczepaniak}}\ and\ \bibinfo {author} {\bibfnamefont {Eric~S.}\ \bibnamefont
  {Swanson}},\ }\bibfield  {title} {\enquote {\bibinfo {title} {{Coulomb gauge
  QCD, confinement, and the constituent representation}},}\ }\href {\doibase
  10.1103/PhysRevD.65.025012} {\bibfield  {journal} {\bibinfo  {journal} {Phys.
  Rev.}\ }\textbf {\bibinfo {volume} {D65}},\ \bibinfo {pages} {025012}
  (\bibinfo {year} {2002})},\ \Eprint {http://arxiv.org/abs/hep-ph/0107078}
  {arXiv:hep-ph/0107078 [hep-ph]} \BibitemShut {NoStop}%
%%CITATION = HEP-PH/0107078;%%
\bibitem [{\citenamefont {Maris}\ and\ \citenamefont
  {Roberts}(2003)}]{Maris:2003vk}%
  \BibitemOpen
  \bibfield  {author} {\bibinfo {author} {\bibfnamefont {Pieter}\ \bibnamefont
  {Maris}}\ and\ \bibinfo {author} {\bibfnamefont {Craig~D.}\ \bibnamefont
  {Roberts}},\ }\bibfield  {title} {\enquote {\bibinfo {title}
  {{Dyson-Schwinger equations: A Tool for hadron physics}},}\ }\href {\doibase
  10.1142/S0218301303001326} {\bibfield  {journal} {\bibinfo  {journal}
  {Int.J.Mod.Phys.}\ }\textbf {\bibinfo {volume} {E12}},\ \bibinfo {pages}
  {297--365} (\bibinfo {year} {2003})},\ \Eprint
  {http://arxiv.org/abs/nucl-th/0301049} {arXiv:nucl-th/0301049 [nucl-th]}
  \BibitemShut {NoStop}%
%%CITATION = NUCL-TH/0301049;%%
\bibitem [{\citenamefont {Szczepaniak}(2004)}]{Szczepaniak:2003ve}%
  \BibitemOpen
  \bibfield  {author} {\bibinfo {author} {\bibfnamefont {Adam~P.}\ \bibnamefont
  {Szczepaniak}},\ }\bibfield  {title} {\enquote {\bibinfo {title}
  {{Confinement and gluon propagator in Coulomb gauge QCD}},}\ }\href {\doibase
  10.1103/PhysRevD.69.074031} {\bibfield  {journal} {\bibinfo  {journal} {Phys.
  Rev.}\ }\textbf {\bibinfo {volume} {D69}},\ \bibinfo {pages} {074031}
  (\bibinfo {year} {2004})},\ \Eprint {http://arxiv.org/abs/hep-ph/0306030}
  {arXiv:hep-ph/0306030 [hep-ph]} \BibitemShut {NoStop}%
%%CITATION = HEP-PH/0306030;%%
\bibitem [{\citenamefont {Aguilar}\ and\ \citenamefont
  {Natale}(2004)}]{Aguilar:2004sw}%
  \BibitemOpen
  \bibfield  {author} {\bibinfo {author} {\bibfnamefont {A.~C.}\ \bibnamefont
  {Aguilar}}\ and\ \bibinfo {author} {\bibfnamefont {A.~A.}\ \bibnamefont
  {Natale}},\ }\bibfield  {title} {\enquote {\bibinfo {title} {{A dynamical
  gluon mass solution in a coupled system of the Schwinger-Dyson equations}},}\
  }\href@noop {} {\bibfield  {journal} {\bibinfo  {journal} {JHEP}\ }\textbf
  {\bibinfo {volume} {08}},\ \bibinfo {pages} {057} (\bibinfo {year} {2004})},\
  \Eprint {http://arxiv.org/abs/hep-ph/0408254} {hep-ph/0408254} \BibitemShut
  {NoStop}%
%%CITATION = HEP-PH/0408254;%%
\bibitem [{\citenamefont {Kondo}(2006)}]{Kondo:2006ih}%
  \BibitemOpen
  \bibfield  {author} {\bibinfo {author} {\bibfnamefont {Kei-Ichi}\
  \bibnamefont {Kondo}},\ }\bibfield  {title} {\enquote {\bibinfo {title}
  {{Gauge-invariant gluon mass, infrared Abelian dominance and stability of
  magnetic vacuum}},}\ }\href@noop {} {\bibfield  {journal} {\bibinfo
  {journal} {Phys. Rev.}\ }\textbf {\bibinfo {volume} {D74}},\ \bibinfo {pages}
  {125003} (\bibinfo {year} {2006})},\ \Eprint
  {http://arxiv.org/abs/hep-th/0609166} {hep-th/0609166} \BibitemShut {NoStop}%
%%CITATION = HEP-TH/0609166;%%
\bibitem [{\citenamefont {Braun}\ \emph {et~al.}(2010)\citenamefont {Braun},
  \citenamefont {Gies},\ and\ \citenamefont {Pawlowski}}]{Braun:2007bx}%
  \BibitemOpen
  \bibfield  {author} {\bibinfo {author} {\bibfnamefont {Jens}\ \bibnamefont
  {Braun}}, \bibinfo {author} {\bibfnamefont {Holger}\ \bibnamefont {Gies}}, \
  and\ \bibinfo {author} {\bibfnamefont {Jan~M.}\ \bibnamefont {Pawlowski}},\
  }\bibfield  {title} {\enquote {\bibinfo {title} {{Quark Confinement from
  Color Confinement}},}\ }\href {\doibase 10.1016/j.physletb.2010.01.009}
  {\bibfield  {journal} {\bibinfo  {journal} {Phys.Lett.}\ }\textbf {\bibinfo
  {volume} {B684}},\ \bibinfo {pages} {262--267} (\bibinfo {year} {2010})},\
  \Eprint {http://arxiv.org/abs/0708.2413} {arXiv:0708.2413 [hep-th]}
  \BibitemShut {NoStop}%
%%CITATION = ARXIV:0708.2413;%%
\bibitem [{\citenamefont {Epple}\ \emph {et~al.}(2008)\citenamefont {Epple},
  \citenamefont {Reinhardt}, \citenamefont {Schleifenbaum},\ and\ \citenamefont
  {Szczepaniak}}]{Epple:2007ut}%
  \BibitemOpen
  \bibfield  {author} {\bibinfo {author} {\bibfnamefont {D.}~\bibnamefont
  {Epple}}, \bibinfo {author} {\bibfnamefont {H.}~\bibnamefont {Reinhardt}},
  \bibinfo {author} {\bibfnamefont {W.}~\bibnamefont {Schleifenbaum}}, \ and\
  \bibinfo {author} {\bibfnamefont {A.P.}\ \bibnamefont {Szczepaniak}},\
  }\bibfield  {title} {\enquote {\bibinfo {title} {{Subcritical solution of the
  Yang-Mills Schroedinger equation in the Coulomb gauge}},}\ }\href {\doibase
  10.1103/PhysRevD.77.085007} {\bibfield  {journal} {\bibinfo  {journal} {Phys.
  Rev.}\ }\textbf {\bibinfo {volume} {D77}},\ \bibinfo {pages} {085007}
  (\bibinfo {year} {2008})},\ \Eprint {http://arxiv.org/abs/0712.3694}
  {arXiv:0712.3694 [hep-th]} \BibitemShut {NoStop}%
%%CITATION = ARXIV:0712.3694;%%
\bibitem [{\citenamefont {Boucaud}\ \emph {et~al.}(2008)\citenamefont {Boucaud}
  \emph {et~al.}}]{Boucaud:2008ky}%
  \BibitemOpen
  \bibfield  {author} {\bibinfo {author} {\bibfnamefont {Ph.}\ \bibnamefont
  {Boucaud}} \emph {et~al.},\ }\bibfield  {title} {\enquote {\bibinfo {title}
  {{On the IR behaviour of the Landau-gauge ghost propagator}},}\ }\href
  {\doibase 10.1088/1126-6708/2008/06/099} {\bibfield  {journal} {\bibinfo
  {journal} {JHEP}\ }\textbf {\bibinfo {volume} {06}},\ \bibinfo {pages} {099}
  (\bibinfo {year} {2008})},\ \Eprint {http://arxiv.org/abs/0803.2161}
  {arXiv:0803.2161 [hep-ph]} \BibitemShut {NoStop}%
%%CITATION = 0803.2161;%%
\bibitem [{\citenamefont {Dudal}\ \emph {et~al.}(2008)\citenamefont {Dudal},
  \citenamefont {Gracey}, \citenamefont {Sorella}, \citenamefont
  {Vandersickel},\ and\ \citenamefont {Verschelde}}]{Dudal:2008sp}%
  \BibitemOpen
  \bibfield  {author} {\bibinfo {author} {\bibfnamefont {David}\ \bibnamefont
  {Dudal}}, \bibinfo {author} {\bibfnamefont {John~A.}\ \bibnamefont {Gracey}},
  \bibinfo {author} {\bibfnamefont {Silvio~Paolo}\ \bibnamefont {Sorella}},
  \bibinfo {author} {\bibfnamefont {Nele}\ \bibnamefont {Vandersickel}}, \ and\
  \bibinfo {author} {\bibfnamefont {Henri}\ \bibnamefont {Verschelde}},\
  }\bibfield  {title} {\enquote {\bibinfo {title} {{A refinement of the
  Gribov-Zwanziger approach in the Landau gauge: infrared propagators in
  harmony with the lattice results}},}\ }\href {\doibase
  10.1103/PhysRevD.78.065047} {\bibfield  {journal} {\bibinfo  {journal} {Phys.
  Rev.}\ }\textbf {\bibinfo {volume} {D78}},\ \bibinfo {pages} {065047}
  (\bibinfo {year} {2008})},\ \Eprint {http://arxiv.org/abs/0806.4348}
  {arXiv:0806.4348 [hep-th]} \BibitemShut {NoStop}%
%%CITATION = 0806.4348;%%
\bibitem [{\citenamefont {Fischer}\ \emph {et~al.}(2009)\citenamefont
  {Fischer}, \citenamefont {Maas},\ and\ \citenamefont
  {Pawlowski}}]{Fischer:2008uz}%
  \BibitemOpen
  \bibfield  {author} {\bibinfo {author} {\bibfnamefont {Christian~S.}\
  \bibnamefont {Fischer}}, \bibinfo {author} {\bibfnamefont {Axel}\
  \bibnamefont {Maas}}, \ and\ \bibinfo {author} {\bibfnamefont {Jan~M.}\
  \bibnamefont {Pawlowski}},\ }\bibfield  {title} {\enquote {\bibinfo {title}
  {{On the infrared behavior of Landau gauge Yang-Mills theory}},}\ }\href
  {\doibase 10.1016/j.aop.2009.07.009} {\bibfield  {journal} {\bibinfo
  {journal} {Annals Phys.}\ }\textbf {\bibinfo {volume} {324}},\ \bibinfo
  {pages} {2408--2437} (\bibinfo {year} {2009})},\ \Eprint
  {http://arxiv.org/abs/0810.1987} {arXiv:0810.1987 [hep-ph]} \BibitemShut
  {NoStop}%
%%CITATION = ARXIV:0810.1987;%%
\bibitem [{\citenamefont {Szczepaniak}\ and\ \citenamefont
  {Matevosyan}(2010)}]{Szczepaniak:2010fe}%
  \BibitemOpen
  \bibfield  {author} {\bibinfo {author} {\bibfnamefont {Adam~P.}\ \bibnamefont
  {Szczepaniak}}\ and\ \bibinfo {author} {\bibfnamefont {Hrayr~H.}\
  \bibnamefont {Matevosyan}},\ }\bibfield  {title} {\enquote {\bibinfo {title}
  {{A model for QCD ground state with magnetic disorder}},}\ }\href {\doibase
  10.1103/PhysRevD.81.094007} {\bibfield  {journal} {\bibinfo  {journal} {Phys.
  Rev.}\ }\textbf {\bibinfo {volume} {D81}},\ \bibinfo {pages} {094007}
  (\bibinfo {year} {2010})},\ \Eprint {http://arxiv.org/abs/1003.1901}
  {arXiv:1003.1901 [hep-ph]} \BibitemShut {NoStop}%
%%CITATION = ARXIV:1003.1901;%%
\bibitem [{\citenamefont {Watson}\ and\ \citenamefont
  {Reinhardt}(2010)}]{Watson:2010cn}%
  \BibitemOpen
  \bibfield  {author} {\bibinfo {author} {\bibfnamefont {Peter}\ \bibnamefont
  {Watson}}\ and\ \bibinfo {author} {\bibfnamefont {Hugo}\ \bibnamefont
  {Reinhardt}},\ }\bibfield  {title} {\enquote {\bibinfo {title} {{The Coulomb
  gauge ghost Dyson-Schwinger equation}},}\ }\href {\doibase
  10.1103/PhysRevD.82.125010} {\bibfield  {journal} {\bibinfo  {journal}
  {Phys.Rev.}\ }\textbf {\bibinfo {volume} {D82}},\ \bibinfo {pages} {125010}
  (\bibinfo {year} {2010})},\ \Eprint {http://arxiv.org/abs/1007.2583}
  {arXiv:1007.2583 [hep-th]} \BibitemShut {NoStop}%
%%CITATION = ARXIV:1007.2583;%%
\bibitem [{\citenamefont
  {Rodriguez-Quintero}(2011)}]{RodriguezQuintero:2010wy}%
  \BibitemOpen
  \bibfield  {author} {\bibinfo {author} {\bibfnamefont {J.}~\bibnamefont
  {Rodriguez-Quintero}},\ }\bibfield  {title} {\enquote {\bibinfo {title} {{On
  the massive gluon propagator, the PT-BFM scheme and the low-momentum
  behaviour of decoupling and scaling DSE solutions}},}\ }\href {\doibase
  10.1007/JHEP01(2011)105} {\bibfield  {journal} {\bibinfo  {journal} {JHEP}\
  }\textbf {\bibinfo {volume} {1101}},\ \bibinfo {pages} {105} (\bibinfo {year}
  {2011})},\ \Eprint {http://arxiv.org/abs/1005.4598} {arXiv:1005.4598
  [hep-ph]} \BibitemShut {NoStop}%
%%CITATION = ARXIV:1005.4598;%%
\bibitem [{\citenamefont {Campagnari}\ and\ \citenamefont
  {Reinhardt}(2010)}]{Campagnari:2010wc}%
  \BibitemOpen
  \bibfield  {author} {\bibinfo {author} {\bibfnamefont {Davide~R.}\
  \bibnamefont {Campagnari}}\ and\ \bibinfo {author} {\bibfnamefont {Hugo}\
  \bibnamefont {Reinhardt}},\ }\bibfield  {title} {\enquote {\bibinfo {title}
  {{Non-Gaussian wave functionals in Coulomb gauge Yang--Mills theory}},}\
  }\href {\doibase 10.1103/PhysRevD.82.105021} {\bibfield  {journal} {\bibinfo
  {journal} {Phys. Rev.}\ }\textbf {\bibinfo {volume} {D82}},\ \bibinfo {pages}
  {105021} (\bibinfo {year} {2010})},\ \Eprint {http://arxiv.org/abs/1009.4599}
  {arXiv:1009.4599 [hep-th]} \BibitemShut {NoStop}%
%%CITATION = ARXIV:1009.4599;%%
\bibitem [{\citenamefont {Tissier}\ and\ \citenamefont
  {Wschebor}(2010)}]{Tissier:2010ts}%
  \BibitemOpen
  \bibfield  {author} {\bibinfo {author} {\bibfnamefont {Matthieu}\
  \bibnamefont {Tissier}}\ and\ \bibinfo {author} {\bibfnamefont {Nicolas}\
  \bibnamefont {Wschebor}},\ }\bibfield  {title} {\enquote {\bibinfo {title}
  {{Infrared propagators of Yang-Mills theory from perturbation theory}},}\
  }\href {\doibase 10.1103/PhysRevD.82.101701} {\bibfield  {journal} {\bibinfo
  {journal} {Phys.Rev.}\ }\textbf {\bibinfo {volume} {D82}},\ \bibinfo {pages}
  {101701} (\bibinfo {year} {2010})},\ \Eprint {http://arxiv.org/abs/1004.1607}
  {arXiv:1004.1607 [hep-ph]} \BibitemShut {NoStop}%
%%CITATION = ARXIV:1004.1607;%%
\bibitem [{\citenamefont {Pennington}\ and\ \citenamefont
  {Wilson}(2011)}]{Pennington:2011xs}%
  \BibitemOpen
  \bibfield  {author} {\bibinfo {author} {\bibfnamefont {M.R.}\ \bibnamefont
  {Pennington}}\ and\ \bibinfo {author} {\bibfnamefont {D.J.}\ \bibnamefont
  {Wilson}},\ }\bibfield  {title} {\enquote {\bibinfo {title} {{Are the Dressed
  Gluon and Ghost Propagators in the Landau Gauge presently determined in the
  confinement regime of QCD?}}}\ }\href {\doibase 10.1103/PhysRevD.84.094028,
  10.1103/PhysRevD.84.119901} {\bibfield  {journal} {\bibinfo  {journal} {Phys.
  Rev.}\ }\textbf {\bibinfo {volume} {D84}},\ \bibinfo {pages} {119901}
  (\bibinfo {year} {2011})},\ \Eprint {http://arxiv.org/abs/1109.2117}
  {arXiv:1109.2117 [hep-ph]} \BibitemShut {NoStop}%
%%CITATION = ARXIV:1109.2117;%%
\bibitem [{\citenamefont {Watson}\ and\ \citenamefont
  {Reinhardt}(2012)}]{Watson:2011kv}%
  \BibitemOpen
  \bibfield  {author} {\bibinfo {author} {\bibfnamefont {Peter}\ \bibnamefont
  {Watson}}\ and\ \bibinfo {author} {\bibfnamefont {Hugo}\ \bibnamefont
  {Reinhardt}},\ }\bibfield  {title} {\enquote {\bibinfo {title} {{Leading
  order infrared quantum chromodynamics in Coulomb gauge}},}\ }\href {\doibase
  10.1103/PhysRevD.85.025014} {\bibfield  {journal} {\bibinfo  {journal}
  {Phys.Rev.}\ }\textbf {\bibinfo {volume} {D85}},\ \bibinfo {pages} {025014}
  (\bibinfo {year} {2012})},\ \Eprint {http://arxiv.org/abs/1111.6078}
  {arXiv:1111.6078 [hep-ph]} \BibitemShut {NoStop}%
%%CITATION = ARXIV:1111.6078;%%
\bibitem [{\citenamefont {Kondo}(2011)}]{Kondo:2011ab}%
  \BibitemOpen
  \bibfield  {author} {\bibinfo {author} {\bibfnamefont {Kei-Ichi}\
  \bibnamefont {Kondo}},\ }\bibfield  {title} {\enquote {\bibinfo {title} {{A
  low-energy effective Yang-Mills theory for quark and gluon confinement}},}\
  }\href {\doibase 10.1103/PhysRevD.84.061702} {\bibfield  {journal} {\bibinfo
  {journal} {Phys.Rev.}\ }\textbf {\bibinfo {volume} {D84}},\ \bibinfo {pages}
  {061702} (\bibinfo {year} {2011})},\ \Eprint {http://arxiv.org/abs/1103.3829}
  {arXiv:1103.3829 [hep-th]} \BibitemShut {NoStop}%
%%CITATION = ARXIV:1103.3829;%%
\bibitem [{\citenamefont {Siringo}(2014)}]{Siringo:2014lva}%
  \BibitemOpen
  \bibfield  {author} {\bibinfo {author} {\bibfnamefont {Fabio}\ \bibnamefont
  {Siringo}},\ }\bibfield  {title} {\enquote {\bibinfo {title} {{Gluon
  propagator in Feynman gauge by the method of stationary variance}},}\ }\href
  {\doibase 10.1103/PhysRevD.90.094021} {\bibfield  {journal} {\bibinfo
  {journal} {Phys. Rev.}\ }\textbf {\bibinfo {volume} {D90}},\ \bibinfo {pages}
  {094021} (\bibinfo {year} {2014})},\ \Eprint {http://arxiv.org/abs/1408.5313}
  {arXiv:1408.5313 [hep-ph]} \BibitemShut {NoStop}%
%%CITATION = ARXIV:1408.5313;%%
\bibitem [{\citenamefont {Schwinger}(1962{\natexlab{a}})}]{Schwinger:1962tn}%
  \BibitemOpen
  \bibfield  {author} {\bibinfo {author} {\bibfnamefont {Julian~S.}\
  \bibnamefont {Schwinger}},\ }\bibfield  {title} {\enquote {\bibinfo {title}
  {{Gauge Invariance and Mass}},}\ }\href@noop {} {\bibfield  {journal}
  {\bibinfo  {journal} {Phys. Rev.}\ }\textbf {\bibinfo {volume} {125}},\
  \bibinfo {pages} {397--398} (\bibinfo {year}
  {1962}{\natexlab{a}})}\BibitemShut {NoStop}%
%%CITATION = PHRVA,125,397;%%
\bibitem [{\citenamefont {Schwinger}(1962{\natexlab{b}})}]{Schwinger:1962tp}%
  \BibitemOpen
  \bibfield  {author} {\bibinfo {author} {\bibfnamefont {Julian~S.}\
  \bibnamefont {Schwinger}},\ }\bibfield  {title} {\enquote {\bibinfo {title}
  {{Gauge Invariance and Mass. 2}},}\ }\href@noop {} {\bibfield  {journal}
  {\bibinfo  {journal} {Phys. Rev.}\ }\textbf {\bibinfo {volume} {128}},\
  \bibinfo {pages} {2425--2429} (\bibinfo {year}
  {1962}{\natexlab{b}})}\BibitemShut {NoStop}%
%%CITATION = PHRVA,128,2425;%%
\bibitem{lectures} Lectures given by JP at the Workshop 
``Dyson-Schwinger Equations in Modern Mathematics and Physics'' (Trento, 22-26 September 2014).
\bibitem [{\citenamefont {Roberts}\ and\ \citenamefont
  {Williams}(1994)}]{Roberts:1994dr}%
  \BibitemOpen
  \bibfield  {author} {\bibinfo {author} {\bibfnamefont {Craig~D.}\
  \bibnamefont {Roberts}}\ and\ \bibinfo {author} {\bibfnamefont {Anthony~G.}\
  \bibnamefont {Williams}},\ }\bibfield  {title} {\enquote {\bibinfo {title}
  {{Dyson-Schwinger equations and their application to hadronic physics}},}\
  }\href@noop {} {\bibfield  {journal} {\bibinfo  {journal} {Prog. Part. Nucl.
  Phys.}\ }\textbf {\bibinfo {volume} {33}},\ \bibinfo {pages} {477--575}
  (\bibinfo {year} {1994})},\ \Eprint {http://arxiv.org/abs/hep-ph/9403224}
  {hep-ph/9403224} \BibitemShut {NoStop}%
%%CITATION = HEP-PH/9403224;%%
\bibitem [{\citenamefont {Cornwall}\ and\ \citenamefont
  {Papavassiliou}(1989)}]{Cornwall:1989gv}%
  \BibitemOpen
  \bibfield  {author} {\bibinfo {author} {\bibfnamefont {John~M.}\ \bibnamefont
  {Cornwall}}\ and\ \bibinfo {author} {\bibfnamefont {Joannis}\ \bibnamefont
  {Papavassiliou}},\ }\bibfield  {title} {\enquote {\bibinfo {title} {{Gauge
  Invariant Three Gluon Vertex in QCD}},}\ }\href@noop {} {\bibfield  {journal}
  {\bibinfo  {journal} {Phys. Rev.}\ }\textbf {\bibinfo {volume} {D40}},\
  \bibinfo {pages} {3474} (\bibinfo {year} {1989})}\BibitemShut {NoStop}%
%%CITATION = PHRVA,D40,3474;%%
\bibitem [{\citenamefont {Binosi}\ and\ \citenamefont
  {Papavassiliou}(2002{\natexlab{a}})}]{Binosi:2002ft}%
  \BibitemOpen
  \bibfield  {author} {\bibinfo {author} {\bibfnamefont {Daniele}\ \bibnamefont
  {Binosi}}\ and\ \bibinfo {author} {\bibfnamefont {Joannis}\ \bibnamefont
  {Papavassiliou}},\ }\bibfield  {title} {\enquote {\bibinfo {title} {{The
  pinch technique to all orders}},}\ }\href@noop {} {\bibfield  {journal}
  {\bibinfo  {journal} {Phys. Rev.}\ }\textbf {\bibinfo {volume} {D66}},\
  \bibinfo {pages} {111901(R)} (\bibinfo {year} {2002}{\natexlab{a}})},\
  \Eprint {http://arxiv.org/abs/hep-ph/0208189} {hep-ph/0208189} \BibitemShut
  {NoStop}%
%%CITATION = HEP-PH/0208189;%%
\bibitem [{\citenamefont {Binosi}\ and\ \citenamefont
  {Papavassiliou}(2004)}]{Binosi:2003rr}%
  \BibitemOpen
  \bibfield  {author} {\bibinfo {author} {\bibfnamefont {Daniele}\ \bibnamefont
  {Binosi}}\ and\ \bibinfo {author} {\bibfnamefont {Joannis}\ \bibnamefont
  {Papavassiliou}},\ }\bibfield  {title} {\enquote {\bibinfo {title} {{Pinch
  technique selfenergies and vertices to all orders in perturbation theory}},}\
  }\href {\doibase 10.1088/0954-3899/30/2/017} {\bibfield  {journal} {\bibinfo
  {journal} {J.Phys.G}\ }\textbf {\bibinfo {volume} {G30}},\ \bibinfo {pages}
  {203} (\bibinfo {year} {2004})},\ \Eprint
  {http://arxiv.org/abs/hep-ph/0301096} {arXiv:hep-ph/0301096 [hep-ph]}
  \BibitemShut {NoStop}%
\bibitem [{\citenamefont {Binosi}\ and\ \citenamefont
  {Papavassiliou}(2009)}]{Binosi:2009qm}%
  \BibitemOpen
  \bibfield  {author} {\bibinfo {author} {\bibfnamefont {D.}~\bibnamefont
  {Binosi}}\ and\ \bibinfo {author} {\bibfnamefont {J.}~\bibnamefont
  {Papavassiliou}},\ }\bibfield  {title} {\enquote {\bibinfo {title} {{Pinch
  Technique: Theory and Applications}},}\ }\href {\doibase
  10.1016/j.physrep.2009.05.001} {\bibfield  {journal} {\bibinfo  {journal}
  {Phys. Rept.}\ }\textbf {\bibinfo {volume} {479}},\ \bibinfo {pages} {1--152}
  (\bibinfo {year} {2009})},\ \Eprint {http://arxiv.org/abs/0909.2536}
  {arXiv:0909.2536 [hep-ph]} \BibitemShut {NoStop}%
%%CITATION = ARXIV:0909.2536;%%
\bibitem [{\citenamefont {Abbott}(1981)}]{Abbott:1980hw}%
  \BibitemOpen
  \bibfield  {author} {\bibinfo {author} {\bibfnamefont {L.~F.}\ \bibnamefont
  {Abbott}},\ }\bibfield  {title} {\enquote {\bibinfo {title} {{The Background
  Field Method Beyond One Loop}},}\ }\href@noop {} {\bibfield  {journal}
  {\bibinfo  {journal} {Nucl. Phys.}\ }\textbf {\bibinfo {volume} {B185}},\
  \bibinfo {pages} {189} (\bibinfo {year} {1981})}\BibitemShut {NoStop}%
%%CITATION = NUPHA,B185,189;%%
\bibitem [{\citenamefont {Abbott}(1982)}]{Abbott:1981ke}%
  \BibitemOpen
  \bibfield  {author} {\bibinfo {author} {\bibfnamefont {L.~F.}\ \bibnamefont
  {Abbott}},\ }\bibfield  {title} {\enquote {\bibinfo {title} {{Introduction to
  the Background Field Method}},}\ }\href@noop {} {\bibfield  {journal}
  {\bibinfo  {journal} {Acta Phys. Polon.}\ }\textbf {\bibinfo {volume}
  {B13}},\ \bibinfo {pages} {33} (\bibinfo {year} {1982})}\BibitemShut
  {NoStop}%
%%CITATION = APPOA,B13,33;%%
\bibitem [{\citenamefont {Aguilar}\ and\ \citenamefont
  {Papavassiliou}(2006)}]{Aguilar:2006gr}%
  \BibitemOpen
  \bibfield  {author} {\bibinfo {author} {\bibfnamefont {Arlene~C.}\
  \bibnamefont {Aguilar}}\ and\ \bibinfo {author} {\bibfnamefont {Joannis}\
  \bibnamefont {Papavassiliou}},\ }\bibfield  {title} {\enquote {\bibinfo
  {title} {{Gluon mass generation in the PT-BFM scheme}},}\ }\href@noop {}
  {\bibfield  {journal} {\bibinfo  {journal} {JHEP}\ }\textbf {\bibinfo
  {volume} {12}},\ \bibinfo {pages} {012} (\bibinfo {year} {2006})},\ \Eprint
  {http://arxiv.org/abs/hep-ph/0610040} {hep-ph/0610040} \BibitemShut {NoStop}%
%%CITATION = HEP-PH/0610040;%%
\bibitem [{\citenamefont {Binosi}\ and\ \citenamefont
  {Papavassiliou}(2008{\natexlab{a}})}]{Binosi:2007pi}%
  \BibitemOpen
  \bibfield  {author} {\bibinfo {author} {\bibfnamefont {Daniele}\ \bibnamefont
  {Binosi}}\ and\ \bibinfo {author} {\bibfnamefont {Joannis}\ \bibnamefont
  {Papavassiliou}},\ }\bibfield  {title} {\enquote {\bibinfo {title}
  {{Gauge-invariant truncation scheme for the Schwinger-Dyson equations of
  QCD}},}\ }\href {\doibase 10.1103/PhysRevD.77.061702} {\bibfield  {journal}
  {\bibinfo  {journal} {Phys.Rev.}\ }\textbf {\bibinfo {volume} {D77}},\
  \bibinfo {pages} {061702} (\bibinfo {year} {2008}{\natexlab{a}})},\ \Eprint
  {http://arxiv.org/abs/0712.2707} {arXiv:0712.2707 [hep-ph]} \BibitemShut
  {NoStop}%
\bibitem [{\citenamefont {Binosi}\ and\ \citenamefont
  {Papavassiliou}(2008{\natexlab{b}})}]{Binosi:2008qk}%
  \BibitemOpen
  \bibfield  {author} {\bibinfo {author} {\bibfnamefont {D.}~\bibnamefont
  {Binosi}}\ and\ \bibinfo {author} {\bibfnamefont {J.}~\bibnamefont
  {Papavassiliou}},\ }\bibfield  {title} {\enquote {\bibinfo {title} {{New
  Schwinger-Dyson equations for non-Abelian gauge theories}},}\ }\href
  {\doibase 10.1088/1126-6708/2008/11/063} {\bibfield  {journal} {\bibinfo
  {journal} {JHEP}\ }\textbf {\bibinfo {volume} {0811}},\ \bibinfo {pages}
  {063} (\bibinfo {year} {2008}{\natexlab{b}})},\ \Eprint
  {http://arxiv.org/abs/0805.3994} {arXiv:0805.3994 [hep-ph]} \BibitemShut
  {NoStop}%
\bibitem [{\citenamefont {Aguilar}\ and\ \citenamefont
  {Papavassiliou}(2010)}]{Aguilar:2009ke}%
  \BibitemOpen
  \bibfield  {author} {\bibinfo {author} {\bibfnamefont {Arlene~C.}\
  \bibnamefont {Aguilar}}\ and\ \bibinfo {author} {\bibfnamefont {Joannis}\
  \bibnamefont {Papavassiliou}},\ }\bibfield  {title} {\enquote {\bibinfo
  {title} {{Gluon mass generation without seagull divergences}},}\ }\href
  {\doibase 10.1103/PhysRevD.81.034003} {\bibfield  {journal} {\bibinfo
  {journal} {Phys.Rev.}\ }\textbf {\bibinfo {volume} {D81}},\ \bibinfo {pages}
  {034003} (\bibinfo {year} {2010})},\ \Eprint {http://arxiv.org/abs/0910.4142}
  {arXiv:0910.4142 [hep-ph]} \BibitemShut {NoStop}%
\bibitem [{\citenamefont {Aguilar}\ \emph {et~al.}(2011)\citenamefont
  {Aguilar}, \citenamefont {Binosi},\ and\ \citenamefont
  {Papavassiliou}}]{Aguilar:2011ux}%
  \BibitemOpen
  \bibfield  {author} {\bibinfo {author} {\bibfnamefont {A.~C.}\ \bibnamefont
  {Aguilar}}, \bibinfo {author} {\bibfnamefont {D.}~\bibnamefont {Binosi}}, \
  and\ \bibinfo {author} {\bibfnamefont {J.}~\bibnamefont {Papavassiliou}},\
  }\bibfield  {title} {\enquote {\bibinfo {title} {{The dynamical equation of
  the effective gluon mass}},}\ }\href {\doibase 10.1103/PhysRevD.84.085026}
  {\bibfield  {journal} {\bibinfo  {journal} {Phys. Rev.}\ }\textbf {\bibinfo
  {volume} {D84}},\ \bibinfo {pages} {085026} (\bibinfo {year} {2011})},\
  \Eprint {http://arxiv.org/abs/1107.3968} {arXiv:1107.3968 [hep-ph]}
  \BibitemShut {NoStop}%
%%CITATION = ARXIV:1107.3968;%%
\bibitem [{\citenamefont {Binosi}\ \emph {et~al.}(2012)\citenamefont {Binosi},
  \citenamefont {Iba\~nez},\ and\ \citenamefont
  {Papavassiliou}}]{Binosi:2012sj}%
  \BibitemOpen
  \bibfield  {author} {\bibinfo {author} {\bibfnamefont {D.}~\bibnamefont
  {Binosi}}, \bibinfo {author} {\bibfnamefont {D.}~\bibnamefont {Iba\~nez}}, \
  and\ \bibinfo {author} {\bibfnamefont {J.}~\bibnamefont {Papavassiliou}},\
  }\bibfield  {title} {\enquote {\bibinfo {title} {{The all-order equation of
  the effective gluon mass}},}\ }\href {\doibase 10.1103/PhysRevD.86.085033}
  {\bibfield  {journal} {\bibinfo  {journal} {Phys. Rev.}\ }\textbf {\bibinfo
  {volume} {D86}},\ \bibinfo {pages} {085033} (\bibinfo {year} {2012})},\
  \Eprint {http://arxiv.org/abs/1208.1451} {arXiv:1208.1451 [hep-ph]}
  \BibitemShut {NoStop}%
%%CITATION = ARXIV:1208.1451;%%
\bibitem [{\citenamefont {Nakanishi}(1966)}]{Nakanishi:1966cq}%
  \BibitemOpen
  \bibfield  {author} {\bibinfo {author} {\bibfnamefont {Noboru}\ \bibnamefont
  {Nakanishi}},\ }\bibfield  {title} {\enquote {\bibinfo {title} {Covariant
  quantization of the electromagnetic field in the landau gauge},}\ }\href@noop
  {} {\bibfield  {journal} {\bibinfo  {journal} {Prog. Theor. Phys.}\ }\textbf
  {\bibinfo {volume} {35}},\ \bibinfo {pages} {1111--1116} (\bibinfo {year}
  {1966})}\BibitemShut {NoStop}%
\bibitem [{\citenamefont {Lautrup}(1966)}]{Lautrup:1966cq}%
  \BibitemOpen
  \bibfield  {author} {\bibinfo {author} {\bibfnamefont {B.}~\bibnamefont
  {Lautrup}},\ }\bibfield  {title} {\enquote {\bibinfo {title} {Canonical
  quantum electrodynamics in covariant gauges},}\ }\href@noop {} {\bibfield
  {journal} {\bibinfo  {journal} {Mat. Fys. Medd. Dan. Vid. Selsk.}\ }\textbf
  {\bibinfo {volume} {35}},\ \bibinfo {pages} {1} (\bibinfo {year}
  {1966})}\BibitemShut {NoStop}%
\bibitem [{\citenamefont {Becchi}\ \emph {et~al.}(1975)\citenamefont {Becchi},
  \citenamefont {Rouet},\ and\ \citenamefont {Stora}}]{Becchi:1974md}%
  \BibitemOpen
  \bibfield  {author} {\bibinfo {author} {\bibfnamefont {C.}~\bibnamefont
  {Becchi}}, \bibinfo {author} {\bibfnamefont {A.}~\bibnamefont {Rouet}}, \
  and\ \bibinfo {author} {\bibfnamefont {R.}~\bibnamefont {Stora}},\ }\bibfield
   {title} {\enquote {\bibinfo {title} {{Renormalization of the Abelian
  Higgs-Kibble Model}},}\ }\href@noop {} {\bibfield  {journal} {\bibinfo
  {journal} {Commun. Math. Phys.}\ }\textbf {\bibinfo {volume} {42}},\ \bibinfo
  {pages} {127--162} (\bibinfo {year} {1975})}\BibitemShut {NoStop}%
%%CITATION = CMPHA,42,127;%%
\bibitem [{\citenamefont {Tyutin}()}]{Tyutin:1975qk}%
  \BibitemOpen
  \bibfield  {author} {\bibinfo {author} {\bibfnamefont {I.~V.}\ \bibnamefont
  {Tyutin}},\ }\bibfield  {title} {\enquote {\bibinfo {title} {Gauge invariance
  in field theory and statistical physics in operator formalism},}\ }\href@noop
  {} {\bibinfo  {journal} {LEBEDEV-75-39}\ }\BibitemShut {NoStop}%
\bibitem [{\citenamefont {Fujikawa}\ \emph {et~al.}(1972)\citenamefont
  {Fujikawa}, \citenamefont {Lee},\ and\ \citenamefont
  {Sanda}}]{Fujikawa:1972fe}%
  \BibitemOpen
\bibfield  {journal} {  }\bibfield  {author} {\bibinfo {author} {\bibfnamefont
  {K.}~\bibnamefont {Fujikawa}}, \bibinfo {author} {\bibfnamefont {B.~W.}\
  \bibnamefont {Lee}}, \ and\ \bibinfo {author} {\bibfnamefont {A.~I.}\
  \bibnamefont {Sanda}},\ }\bibfield  {title} {\enquote {\bibinfo {title}
  {{Generalized Renormalizable Gauge Formulation of Spontaneously Broken Gauge
  Theories}},}\ }\href@noop {} {\bibfield  {journal} {\bibinfo  {journal}
  {Phys. Rev.}\ }\textbf {\bibinfo {volume} {D6}},\ \bibinfo {pages}
  {2923--2943} (\bibinfo {year} {1972})}\BibitemShut {NoStop}%
%%CITATION = PHRVA,D6,2923;%%
\bibitem [{\citenamefont {Binosi}\ and\ \citenamefont
  {Quadri}(2013)}]{Binosi:2013cea}%
  \BibitemOpen
  \bibfield  {author} {\bibinfo {author} {\bibfnamefont {D.}~\bibnamefont
  {Binosi}}\ and\ \bibinfo {author} {\bibfnamefont {A.}~\bibnamefont
  {Quadri}},\ }\bibfield  {title} {\enquote {\bibinfo {title} {{AntiBRST
  symmetry and Background Field Method}},}\ }\href {\doibase
  10.1103/PhysRevD.88.085036} {\bibfield  {journal} {\bibinfo  {journal}
  {Phys.Rev.}\ }\textbf {\bibinfo {volume} {D88}},\ \bibinfo {pages} {085036}
  (\bibinfo {year} {2013})},\ \Eprint {http://arxiv.org/abs/1309.1021}
  {arXiv:1309.1021 [hep-th]} \BibitemShut {NoStop}%
%%CITATION = ARXIV:1309.1021;%%
\bibitem [{\citenamefont {Binosi}\ and\ \citenamefont
  {Papavassiliou}(2002{\natexlab{b}})}]{Binosi:2002ez}%
  \BibitemOpen
  \bibfield  {author} {\bibinfo {author} {\bibfnamefont {Daniele}\ \bibnamefont
  {Binosi}}\ and\ \bibinfo {author} {\bibfnamefont {Joannis}\ \bibnamefont
  {Papavassiliou}},\ }\bibfield  {title} {\enquote {\bibinfo {title} {{Pinch
  technique and the Batalin-Vilkovisky formalism}},}\ }\href {\doibase
  10.1103/PhysRevD.66.025024} {\bibfield  {journal} {\bibinfo  {journal}
  {Phys.Rev.}\ }\textbf {\bibinfo {volume} {D66}},\ \bibinfo {pages} {025024}
  (\bibinfo {year} {2002}{\natexlab{b}})},\ \Eprint
  {http://arxiv.org/abs/hep-ph/0204128} {arXiv:hep-ph/0204128 [hep-ph]}
  \BibitemShut {NoStop}%
\bibitem [{\citenamefont {Grassi}\ \emph {et~al.}(2001)\citenamefont {Grassi},
  \citenamefont {Hurth},\ and\ \citenamefont {Steinhauser}}]{Grassi:1999tp}%
  \BibitemOpen
  \bibfield  {author} {\bibinfo {author} {\bibfnamefont {Pietro~Antonio}\
  \bibnamefont {Grassi}}, \bibinfo {author} {\bibfnamefont {Tobias}\
  \bibnamefont {Hurth}}, \ and\ \bibinfo {author} {\bibfnamefont {Matthias}\
  \bibnamefont {Steinhauser}},\ }\bibfield  {title} {\enquote {\bibinfo {title}
  {{Practical algebraic renormalization}},}\ }\href@noop {} {\bibfield
  {journal} {\bibinfo  {journal} {Annals Phys.}\ }\textbf {\bibinfo {volume}
  {288}},\ \bibinfo {pages} {197--248} (\bibinfo {year} {2001})},\ \Eprint
  {http://arxiv.org/abs/hep-ph/9907426} {hep-ph/9907426} \BibitemShut {NoStop}%
%%CITATION = HEP-PH/9907426;%%
\bibitem [{\citenamefont {Ball}\ and\ \citenamefont
  {Chiu}(1980)}]{Ball:1980ax}%
  \BibitemOpen
  \bibfield  {author} {\bibinfo {author} {\bibfnamefont {James~S.}\
  \bibnamefont {Ball}}\ and\ \bibinfo {author} {\bibfnamefont {Ting-Wai}\
  \bibnamefont {Chiu}},\ }\bibfield  {title} {\enquote {\bibinfo {title}
  {{Analytic properties of the vertex function in gauge theories. 2}},}\
  }\href@noop {} {\bibfield  {journal} {\bibinfo  {journal} {Phys. Rev.}\
  }\textbf {\bibinfo {volume} {D22}},\ \bibinfo {pages} {2550} (\bibinfo {year}
  {1980})}\BibitemShut {NoStop}%
%%CITATION = PHRVA,D22,2550;%%
\bibitem [{\citenamefont {Pelaez}\ \emph {et~al.}(2013)\citenamefont {Pelaez},
  \citenamefont {Tissier},\ and\ \citenamefont {Wschebor}}]{Pelaez:2013cpa}%
  \BibitemOpen
  \bibfield  {author} {\bibinfo {author} {\bibfnamefont {Marcela}\ \bibnamefont
  {Pelaez}}, \bibinfo {author} {\bibfnamefont {Matthieu}\ \bibnamefont
  {Tissier}}, \ and\ \bibinfo {author} {\bibfnamefont {Nicolas}\ \bibnamefont
  {Wschebor}},\ }\bibfield  {title} {\enquote {\bibinfo {title} {{Three-point
  correlation functions in Yang-Mills theory}},}\ }\href {\doibase
  10.1103/PhysRevD.88.125003} {\bibfield  {journal} {\bibinfo  {journal}
  {Phys.Rev.}\ }\textbf {\bibinfo {volume} {D88}},\ \bibinfo {pages} {125003}
  (\bibinfo {year} {2013})},\ \Eprint {http://arxiv.org/abs/1310.2594}
  {arXiv:1310.2594 [hep-th]} \BibitemShut {NoStop}%
%%CITATION = ARXIV:1310.2594;%%
\bibitem [{\citenamefont {Aguilar}\ \emph
  {et~al.}(2014{\natexlab{a}})\citenamefont {Aguilar}, \citenamefont {Binosi},
  \citenamefont {Iba{\~n}ez},\ and\ \citenamefont
  {Papavassiliou}}]{Aguilar:2013vaa}%
  \BibitemOpen
  \bibfield  {author} {\bibinfo {author} {\bibfnamefont {A.~C.}\ \bibnamefont
  {Aguilar}}, \bibinfo {author} {\bibfnamefont {D.}~\bibnamefont {Binosi}},
  \bibinfo {author} {\bibfnamefont {D.}~\bibnamefont {Iba{\~n}ez}}, \ and\
  \bibinfo {author} {\bibfnamefont {J.}~\bibnamefont {Papavassiliou}},\
  }\bibfield  {title} {\enquote {\bibinfo {title} {{Effects of divergent ghost
  loops on the Green's functions of QCD}},}\ }\href {\doibase
  10.1103/PhysRevD.89.085008} {\bibfield  {journal} {\bibinfo  {journal} {Phys.
  Rev.}\ }\textbf {\bibinfo {volume} {D89}},\ \bibinfo {pages} {085008}
  (\bibinfo {year} {2014}{\natexlab{a}})},\ \Eprint
  {http://arxiv.org/abs/1312.1212} {arXiv:1312.1212 [hep-ph]} \BibitemShut
  {NoStop}%
%%CITATION = ARXIV:1312.1212;%%
\bibitem [{\citenamefont {Eichmann}\ \emph {et~al.}(2014)\citenamefont
  {Eichmann}, \citenamefont {Williams}, \citenamefont {Alkofer},\ and\
  \citenamefont {Vujinovic}}]{Eichmann:2014xya}%
  \BibitemOpen
  \bibfield  {author} {\bibinfo {author} {\bibfnamefont {Gernot}\ \bibnamefont
  {Eichmann}}, \bibinfo {author} {\bibfnamefont {Richard}\ \bibnamefont
  {Williams}}, \bibinfo {author} {\bibfnamefont {Reinhard}\ \bibnamefont
  {Alkofer}}, \ and\ \bibinfo {author} {\bibfnamefont {Milan}\ \bibnamefont
  {Vujinovic}},\ }\bibfield  {title} {\enquote {\bibinfo {title} {{The
  three-gluon vertex in Landau gauge}},}\ }\href {\doibase
  10.1103/PhysRevD.89.105014} {\bibfield  {journal} {\bibinfo  {journal}
  {Phys.Rev.}\ }\textbf {\bibinfo {volume} {D89}},\ \bibinfo {pages} {105014}
  (\bibinfo {year} {2014})},\ \Eprint {http://arxiv.org/abs/1402.1365}
  {arXiv:1402.1365 [hep-ph]} \BibitemShut {NoStop}%
%%CITATION = ARXIV:1402.1365;%%
\bibitem [{\citenamefont {Blum}\ \emph {et~al.}(2014)\citenamefont {Blum},
  \citenamefont {Huber}, \citenamefont {Mitter},\ and\ \citenamefont {von
  Smekal}}]{Blum:2014gna}%
  \BibitemOpen
  \bibfield  {author} {\bibinfo {author} {\bibfnamefont {Adrian}\ \bibnamefont
  {Blum}}, \bibinfo {author} {\bibfnamefont {Markus~Q.}\ \bibnamefont {Huber}},
  \bibinfo {author} {\bibfnamefont {Mario}\ \bibnamefont {Mitter}}, \ and\
  \bibinfo {author} {\bibfnamefont {Lorenz}\ \bibnamefont {von Smekal}},\
  }\bibfield  {title} {\enquote {\bibinfo {title} {{Gluonic three-point
  correlations in pure Landau gauge QCD}},}\ }\href {\doibase
  10.1103/PhysRevD.89.061703} {\bibfield  {journal} {\bibinfo  {journal}
  {Phys.Rev.}\ }\textbf {\bibinfo {volume} {D89}},\ \bibinfo {pages} {061703}
  (\bibinfo {year} {2014})},\ \Eprint {http://arxiv.org/abs/1401.0713}
  {arXiv:1401.0713 [hep-ph]} \BibitemShut {NoStop}%
%%CITATION = ARXIV:1401.0713;%%
\bibitem [{\citenamefont {Grassi}\ \emph {et~al.}(2004)\citenamefont {Grassi},
  \citenamefont {Hurth},\ and\ \citenamefont {Quadri}}]{Grassi:2004yq}%
  \BibitemOpen
  \bibfield  {author} {\bibinfo {author} {\bibfnamefont {Pietro~A.}\
  \bibnamefont {Grassi}}, \bibinfo {author} {\bibfnamefont {Tobias}\
  \bibnamefont {Hurth}}, \ and\ \bibinfo {author} {\bibfnamefont {Andrea}\
  \bibnamefont {Quadri}},\ }\bibfield  {title} {\enquote {\bibinfo {title} {{On
  the Landau background gauge fixing and the IR properties of YM Green
  functions}},}\ }\href@noop {} {\bibfield  {journal} {\bibinfo  {journal}
  {Phys. Rev.}\ }\textbf {\bibinfo {volume} {D70}},\ \bibinfo {pages} {105014}
  (\bibinfo {year} {2004})},\ \Eprint {http://arxiv.org/abs/hep-th/0405104}
  {hep-th/0405104} \BibitemShut {NoStop}%
%%CITATION = HEP-TH/0405104;%%
\bibitem [{\citenamefont {Aguilar}\ \emph
  {et~al.}(2009{\natexlab{a}})\citenamefont {Aguilar}, \citenamefont {Binosi},
  \citenamefont {Papavassiliou},\ and\ \citenamefont
  {Rodriguez-Quintero}}]{Aguilar:2009nf}%
  \BibitemOpen
  \bibfield  {author} {\bibinfo {author} {\bibfnamefont {A.~C.}\ \bibnamefont
  {Aguilar}}, \bibinfo {author} {\bibfnamefont {D.}~\bibnamefont {Binosi}},
  \bibinfo {author} {\bibfnamefont {J.}~\bibnamefont {Papavassiliou}}, \ and\
  \bibinfo {author} {\bibfnamefont {J.}~\bibnamefont {Rodriguez-Quintero}},\
  }\bibfield  {title} {\enquote {\bibinfo {title} {{Non-perturbative comparison
  of QCD effective charges}},}\ }\href {\doibase 10.1103/PhysRevD.80.085018}
  {\bibfield  {journal} {\bibinfo  {journal} {Phys. Rev.}\ }\textbf {\bibinfo
  {volume} {D80}},\ \bibinfo {pages} {085018} (\bibinfo {year}
  {2009}{\natexlab{a}})},\ \Eprint {http://arxiv.org/abs/0906.2633}
  {arXiv:0906.2633 [hep-ph]} \BibitemShut {NoStop}%
\bibitem [{\citenamefont {Aguilar}\ \emph
  {et~al.}(2009{\natexlab{b}})\citenamefont {Aguilar}, \citenamefont {Binosi},\
  and\ \citenamefont {Papavassiliou}}]{Aguilar:2009pp}%
  \BibitemOpen
  \bibfield  {author} {\bibinfo {author} {\bibfnamefont {A.C.}\ \bibnamefont
  {Aguilar}}, \bibinfo {author} {\bibfnamefont {D.}~\bibnamefont {Binosi}}, \
  and\ \bibinfo {author} {\bibfnamefont {J.}~\bibnamefont {Papavassiliou}},\
  }\bibfield  {title} {\enquote {\bibinfo {title} {{Indirect determination of
  the Kugo-Ojima function from lattice data}},}\ }\href {\doibase
  10.1088/1126-6708/2009/11/066} {\bibfield  {journal} {\bibinfo  {journal}
  {JHEP}\ }\textbf {\bibinfo {volume} {0911}},\ \bibinfo {pages} {066}
  (\bibinfo {year} {2009}{\natexlab{b}})},\ \Eprint
  {http://arxiv.org/abs/0907.0153} {arXiv:0907.0153 [hep-ph]} \BibitemShut
  {NoStop}%
\bibitem [{\citenamefont {Aguilar}\ \emph {et~al.}(2010)\citenamefont
  {Aguilar}, \citenamefont {Binosi},\ and\ \citenamefont
  {Papavassiliou}}]{Aguilar:2010gm}%
  \BibitemOpen
  \bibfield  {author} {\bibinfo {author} {\bibfnamefont {A.~C.}\ \bibnamefont
  {Aguilar}}, \bibinfo {author} {\bibfnamefont {D.}~\bibnamefont {Binosi}}, \
  and\ \bibinfo {author} {\bibfnamefont {J.}~\bibnamefont {Papavassiliou}},\
  }\bibfield  {title} {\enquote {\bibinfo {title} {{QCD effective charges from
  lattice data}},}\ }\href {\doibase 10.1007/JHEP07(2010)002} {\bibfield
  {journal} {\bibinfo  {journal} {JHEP}\ }\textbf {\bibinfo {volume} {1007}},\
  \bibinfo {pages} {002} (\bibinfo {year} {2010})},\ \Eprint
  {http://arxiv.org/abs/1004.1105} {arXiv:1004.1105 [hep-ph]} \BibitemShut
  {NoStop}%
\bibitem [{\citenamefont {Binosi}\ and\ \citenamefont
  {Papavassiliou}(2011)}]{Binosi:2011wi}%
  \BibitemOpen
  \bibfield  {author} {\bibinfo {author} {\bibfnamefont {D.}~\bibnamefont
  {Binosi}}\ and\ \bibinfo {author} {\bibfnamefont {J.}~\bibnamefont
  {Papavassiliou}},\ }\bibfield  {title} {\enquote {\bibinfo {title} {{Gauge
  invariant Ansatz for a special three-gluon vertex}},}\ }\href {\doibase
  10.1007/JHEP03(2011)121} {\bibfield  {journal} {\bibinfo  {journal} {JHEP}\
  }\textbf {\bibinfo {volume} {1103}},\ \bibinfo {pages} {121} (\bibinfo {year}
  {2011})},\ \Eprint {http://arxiv.org/abs/1102.5662} {arXiv:1102.5662
  [hep-ph]} \BibitemShut {NoStop}%
\bibitem [{\citenamefont {Wilson}(1973)}]{Wilson:1972cf}%
  \BibitemOpen
  \bibfield  {author} {\bibinfo {author} {\bibfnamefont {Kenneth~G.}\
  \bibnamefont {Wilson}},\ }\bibfield  {title} {\enquote {\bibinfo {title}
  {{Quantum field theory models in less than four-dimensions}},}\ }\href
  {\doibase 10.1103/PhysRevD.7.2911} {\bibfield  {journal} {\bibinfo  {journal}
  {Phys.Rev.}\ }\textbf {\bibinfo {volume} {D7}},\ \bibinfo {pages}
  {2911--2926} (\bibinfo {year} {1973})}\BibitemShut {NoStop}%
%%CITATION = PHRVA,D7,2911;%%
\bibitem [{\citenamefont {Collins}(1984)}]{Collins:1984xc}%
  \BibitemOpen
  \bibfield  {author} {\bibinfo {author} {\bibfnamefont {John~C.}\ \bibnamefont
  {Collins}},\ }\bibfield  {title} {\enquote {\bibinfo {title}
  {{Renormalization. An introduction to renormalization, the renormalization
  group, and the operator product expansion}},}\ }\href@noop {} {\  (\bibinfo
  {year} {1984})}\BibitemShut {NoStop}%
%%CITATION = INSPIRE-209810;%%
\bibitem [{\citenamefont {Aguilar}\ \emph
  {et~al.}(2012{\natexlab{a}})\citenamefont {Aguilar}, \citenamefont {Ibanez},
  \citenamefont {Mathieu},\ and\ \citenamefont
  {Papavassiliou}}]{Aguilar:2011xe}%
  \BibitemOpen
  \bibfield  {author} {\bibinfo {author} {\bibfnamefont {A.C.}\ \bibnamefont
  {Aguilar}}, \bibinfo {author} {\bibfnamefont {D.}~\bibnamefont {Ibanez}},
  \bibinfo {author} {\bibfnamefont {V.}~\bibnamefont {Mathieu}}, \ and\
  \bibinfo {author} {\bibfnamefont {J.}~\bibnamefont {Papavassiliou}},\
  }\bibfield  {title} {\enquote {\bibinfo {title} {{Massless bound-state
  excitations and the Schwinger mechanism in QCD}},}\ }\href {\doibase
  10.1103/PhysRevD.85.014018} {\bibfield  {journal} {\bibinfo  {journal}
  {Phys.Rev.}\ }\textbf {\bibinfo {volume} {D85}},\ \bibinfo {pages} {014018}
  (\bibinfo {year} {2012}{\natexlab{a}})},\ \Eprint
  {http://arxiv.org/abs/1110.2633} {arXiv:1110.2633 [hep-ph]} \BibitemShut
  {NoStop}%
%%CITATION = ARXIV:1110.2633;%%
\bibitem [{\citenamefont {Iba{\~n}ez}\ and\ \citenamefont
  {Papavassiliou}(2013)}]{Ibanez:2012zk}%
  \BibitemOpen
  \bibfield  {author} {\bibinfo {author} {\bibfnamefont {D.}~\bibnamefont
  {Iba{\~n}ez}}\ and\ \bibinfo {author} {\bibfnamefont {J.}~\bibnamefont
  {Papavassiliou}},\ }\bibfield  {title} {\enquote {\bibinfo {title} {{Gluon
  mass generation in the massless bound-state formalism}},}\ }\href {\doibase
  10.1103/PhysRevD.87.034008} {\bibfield  {journal} {\bibinfo  {journal}
  {Phys.Rev.}\ }\textbf {\bibinfo {volume} {D87}},\ \bibinfo {pages} {034008}
  (\bibinfo {year} {2013})},\ \Eprint {http://arxiv.org/abs/1211.5314}
  {arXiv:1211.5314 [hep-ph]} \BibitemShut {NoStop}%
%%CITATION = ARXIV:1211.5314;%%
\bibitem [{\citenamefont {Aguilar}\ \emph
  {et~al.}(2014{\natexlab{b}})\citenamefont {Aguilar}, \citenamefont {Binosi},\
  and\ \citenamefont {Papavassiliou}}]{Aguilar:2014tka}%
  \BibitemOpen
  \bibfield  {author} {\bibinfo {author} {\bibfnamefont {A.~C.}\ \bibnamefont
  {Aguilar}}, \bibinfo {author} {\bibfnamefont {D.}~\bibnamefont {Binosi}}, \
  and\ \bibinfo {author} {\bibfnamefont {J.}~\bibnamefont {Papavassiliou}},\
  }\bibfield  {title} {\enquote {\bibinfo {title} {{Renormalization group
  analysis of the gluon mass equation}},}\ }\href {\doibase
  10.1103/PhysRevD.89.085032} {\bibfield  {journal} {\bibinfo  {journal} {Phys.
  Rev.}\ }\textbf {\bibinfo {volume} {D89}},\ \bibinfo {pages} {085032}
  (\bibinfo {year} {2014}{\natexlab{b}})},\ \Eprint
  {http://arxiv.org/abs/1401.3631} {arXiv:1401.3631 [hep-ph]} \BibitemShut
  {NoStop}%
%%CITATION = ARXIV:1401.3631;%%
\bibitem [{\citenamefont {Binosi}\ \emph {et~al.}(2014)\citenamefont {Binosi},
  \citenamefont {Iba{\~n}ez},\ and\ \citenamefont
  {Papavassiliou}}]{Binosi:2014kka}%
  \BibitemOpen
  \bibfield  {author} {\bibinfo {author} {\bibfnamefont {D.}~\bibnamefont
  {Binosi}}, \bibinfo {author} {\bibfnamefont {D.}~\bibnamefont {Iba{\~n}ez}},
  \ and\ \bibinfo {author} {\bibfnamefont {J.}~\bibnamefont {Papavassiliou}},\
  }\bibfield  {title} {\enquote {\bibinfo {title} {{Nonperturbative study of
  the four gluon vertex}},}\ }\href {\doibase 10.1007/JHEP09(2014)059}
  {\bibfield  {journal} {\bibinfo  {journal} {JHEP}\ }\textbf {\bibinfo
  {volume} {1409}},\ \bibinfo {pages} {059} (\bibinfo {year} {2014})},\ \Eprint
  {http://arxiv.org/abs/1407.3677} {arXiv:1407.3677 [hep-ph]} \BibitemShut
  {NoStop}%
%%CITATION = ARXIV:1407.3677;%%
\bibitem [{\citenamefont {Cyrol}\ \emph {et~al.}(2015)\citenamefont {Cyrol},
  \citenamefont {Huber},\ and\ \citenamefont {von Smekal}}]{Cyrol:2014kca}%
  \BibitemOpen
  \bibfield  {author} {\bibinfo {author} {\bibfnamefont {Anton~K.}\
  \bibnamefont {Cyrol}}, \bibinfo {author} {\bibfnamefont {Markus~Q.}\
  \bibnamefont {Huber}}, \ and\ \bibinfo {author} {\bibfnamefont {Lorenz}\
  \bibnamefont {von Smekal}},\ }\bibfield  {title} {\enquote {\bibinfo {title}
  {{A Dyson--Schwinger study of the four-gluon vertex}},}\ }\href {\doibase
  10.1140/epjc/s10052-015-3312-1} {\bibfield  {journal} {\bibinfo  {journal}
  {Eur. Phys. J.}\ }\textbf {\bibinfo {volume} {C75}},\ \bibinfo {pages} {102}
  (\bibinfo {year} {2015})},\ \Eprint {http://arxiv.org/abs/1408.5409}
  {arXiv:1408.5409 [hep-ph]} \BibitemShut {NoStop}%
%%CITATION = ARXIV:1408.5409;%%
\bibitem [{\citenamefont {Cornwall}\ and\ \citenamefont
  {Hou}(1986)}]{Cornwall:1985bg}%
  \BibitemOpen
  \bibfield  {author} {\bibinfo {author} {\bibfnamefont {John~M.}\ \bibnamefont
  {Cornwall}}\ and\ \bibinfo {author} {\bibfnamefont {Wei-Shu}\ \bibnamefont
  {Hou}},\ }\bibfield  {title} {\enquote {\bibinfo {title} {{Extension of the
  gauge technique to broken symmetry and finite temperature}},}\ }\href@noop {}
  {\bibfield  {journal} {\bibinfo  {journal} {Phys. Rev.}\ }\textbf {\bibinfo
  {volume} {D34}},\ \bibinfo {pages} {585} (\bibinfo {year}
  {1986})}\BibitemShut {NoStop}%
%%CITATION = PHRVA,D34,585;%%
\bibitem [{\citenamefont {Lavelle}(1991)}]{Lavelle:1991ve}%
  \BibitemOpen
  \bibfield  {author} {\bibinfo {author} {\bibfnamefont {Martin}\ \bibnamefont
  {Lavelle}},\ }\bibfield  {title} {\enquote {\bibinfo {title} {{Gauge
  invariant effective gluon mass from the operator product expansion}},}\
  }\href@noop {} {\bibfield  {journal} {\bibinfo  {journal} {Phys. Rev.}\
  }\textbf {\bibinfo {volume} {D44}},\ \bibinfo {pages} {26--28} (\bibinfo
  {year} {1991})}\BibitemShut {NoStop}%
%%CITATION = PHRVA,D44,26;%%
\bibitem [{\citenamefont {Aguilar}\ and\ \citenamefont
  {Papavassiliou}(2008)}]{Aguilar:2007ie}%
  \BibitemOpen
  \bibfield  {author} {\bibinfo {author} {\bibfnamefont {Arlene~C.}\
  \bibnamefont {Aguilar}}\ and\ \bibinfo {author} {\bibfnamefont {Joannis}\
  \bibnamefont {Papavassiliou}},\ }\bibfield  {title} {\enquote {\bibinfo
  {title} {{Power-law running of the effective gluon mass}},}\ }\href {\doibase
  10.1140/epja/i2008-10535-4} {\bibfield  {journal} {\bibinfo  {journal}
  {Eur.Phys.J.}\ }\textbf {\bibinfo {volume} {A35}},\ \bibinfo {pages}
  {189--205} (\bibinfo {year} {2008})},\ \Eprint
  {http://arxiv.org/abs/0708.4320} {arXiv:0708.4320 [hep-ph]} \BibitemShut
  {NoStop}%
\bibitem [{\citenamefont {Lavelle}\ and\ \citenamefont
  {Schaden}(1988)}]{Lavelle:1988eg}%
  \BibitemOpen
  \bibfield  {author} {\bibinfo {author} {\bibfnamefont {M.~J.}\ \bibnamefont
  {Lavelle}}\ and\ \bibinfo {author} {\bibfnamefont {M.}~\bibnamefont
  {Schaden}},\ }\bibfield  {title} {\enquote {\bibinfo {title} {{Propagators
  and condensates in QCD}},}\ }\href@noop {} {\bibfield  {journal} {\bibinfo
  {journal} {Phys. Lett.}\ }\textbf {\bibinfo {volume} {B208}},\ \bibinfo
  {pages} {297} (\bibinfo {year} {1988})}\BibitemShut {NoStop}%
%%CITATION = PHLTA,B208,297;%%
\bibitem [{\citenamefont {Bagan}\ and\ \citenamefont
  {Steele}(1989)}]{Bagan:1989gt}%
  \BibitemOpen
  \bibfield  {author} {\bibinfo {author} {\bibfnamefont {E.}~\bibnamefont
  {Bagan}}\ and\ \bibinfo {author} {\bibfnamefont {Thomas~G.}\ \bibnamefont
  {Steele}},\ }\bibfield  {title} {\enquote {\bibinfo {title} {{QCD condensates
  and the Slavnov-Taylor identities}},}\ }\href@noop {} {\bibfield  {journal}
  {\bibinfo  {journal} {Phys. Lett.}\ }\textbf {\bibinfo {volume} {B219}},\
  \bibinfo {pages} {497} (\bibinfo {year} {1989})}\BibitemShut {NoStop}%
%%CITATION = PHLTA,B219,497;%%
\bibitem [{\citenamefont {Brodsky}\ \emph {et~al.}(2010)\citenamefont
  {Brodsky}, \citenamefont {Roberts}, \citenamefont {Shrock},\ and\
  \citenamefont {Tandy}}]{Brodsky:2010xf}%
  \BibitemOpen
  \bibfield  {author} {\bibinfo {author} {\bibfnamefont {Stanley~J.}\
  \bibnamefont {Brodsky}}, \bibinfo {author} {\bibfnamefont {Craig~D.}\
  \bibnamefont {Roberts}}, \bibinfo {author} {\bibfnamefont {Robert}\
  \bibnamefont {Shrock}}, \ and\ \bibinfo {author} {\bibfnamefont {Peter~C.}\
  \bibnamefont {Tandy}},\ }\bibfield  {title} {\enquote {\bibinfo {title}
  {{Essence of the vacuum quark condensate}},}\ }\href {\doibase
  10.1103/PhysRevC.82.022201} {\bibfield  {journal} {\bibinfo  {journal} {Phys.
  Rev.}\ }\textbf {\bibinfo {volume} {C82}},\ \bibinfo {pages} {022201}
  (\bibinfo {year} {2010})},\ \Eprint {http://arxiv.org/abs/1005.4610}
  {arXiv:1005.4610 [nucl-th]} \BibitemShut {NoStop}%
%%CITATION = ARXIV:1005.4610;%%
\bibitem{Tandy} This demonstration is taken from Peter C. Tandy's talk 
``Some Chapters from the Do-It-Yourself Hadron Theory Manual''.
%%
\bibitem [{\citenamefont {Del~Debbio}\ \emph {et~al.}(1997)\citenamefont
  {Del~Debbio}, \citenamefont {Faber}, \citenamefont {Greensite},\ and\
  \citenamefont {Olejnik}}]{DelDebbio:1996mh}%
  \BibitemOpen
  \bibfield  {author} {\bibinfo {author} {\bibfnamefont {L.}~\bibnamefont
  {Del~Debbio}}, \bibinfo {author} {\bibfnamefont {Manfried}\ \bibnamefont
  {Faber}}, \bibinfo {author} {\bibfnamefont {J.}~\bibnamefont {Greensite}}, \
  and\ \bibinfo {author} {\bibfnamefont {S.}~\bibnamefont {Olejnik}},\
  }\bibfield  {title} {\enquote {\bibinfo {title} {{Center dominance and Z(2)
  vortices in SU(2) lattice gauge theory}},}\ }\href {\doibase
  10.1103/PhysRevD.55.2298} {\bibfield  {journal} {\bibinfo  {journal}
  {Phys.Rev.}\ }\textbf {\bibinfo {volume} {D55}},\ \bibinfo {pages}
  {2298--2306} (\bibinfo {year} {1997})},\ \Eprint
  {http://arxiv.org/abs/hep-lat/9610005} {arXiv:hep-lat/9610005 [hep-lat]}
  \BibitemShut {NoStop}%
%%CITATION = HEP-LAT/9610005;%%
\bibitem [{\citenamefont {Langfeld}\ \emph {et~al.}(2002)\citenamefont
  {Langfeld}, \citenamefont {Reinhardt},\ and\ \citenamefont
  {Gattnar}}]{Langfeld:2001cz}%
  \BibitemOpen
  \bibfield  {author} {\bibinfo {author} {\bibfnamefont {K.}~\bibnamefont
  {Langfeld}}, \bibinfo {author} {\bibfnamefont {H.}~\bibnamefont {Reinhardt}},
  \ and\ \bibinfo {author} {\bibfnamefont {J.}~\bibnamefont {Gattnar}},\
  }\bibfield  {title} {\enquote {\bibinfo {title} {{Gluon propagators and quark
  confinement}},}\ }\href {\doibase 10.1016/S0550-3213(01)00574-0} {\bibfield
  {journal} {\bibinfo  {journal} {Nucl.Phys.}\ }\textbf {\bibinfo {volume}
  {B621}},\ \bibinfo {pages} {131--156} (\bibinfo {year} {2002})},\ \Eprint
  {http://arxiv.org/abs/hep-ph/0107141} {arXiv:hep-ph/0107141 [hep-ph]}
  \BibitemShut {NoStop}%
%%CITATION = HEP-PH/0107141;%%
\bibitem [{\citenamefont {Greensite}(2003)}]{Greensite:2003bk}%
  \BibitemOpen
  \bibfield  {author} {\bibinfo {author} {\bibfnamefont {See, for example, J.}~\bibnamefont
  {Greensite}},\ }\bibfield  {title} {\enquote {\bibinfo {title} {{The
  confinement problem in lattice gauge theory}},}\ }\href@noop {} {\bibfield
  {journal} {\bibinfo  {journal} {Prog. Theor. Phys. Suppl.}\ ,\ \bibinfo
  {pages} {1}} (\bibinfo {year} {2003}), and references therein}\BibitemShut {NoStop}%
%%CITATION = PPNPD,51,1;%%
\bibitem [{\citenamefont {Gattnar}\ \emph {et~al.}(2004)\citenamefont
  {Gattnar}, \citenamefont {Langfeld},\ and\ \citenamefont
  {Reinhardt}}]{Gattnar:2004bf}%
  \BibitemOpen
  \bibfield  {author} {\bibinfo {author} {\bibfnamefont {Jochen}\ \bibnamefont
  {Gattnar}}, \bibinfo {author} {\bibfnamefont {Kurt}\ \bibnamefont
  {Langfeld}}, \ and\ \bibinfo {author} {\bibfnamefont {Hugo}\ \bibnamefont
  {Reinhardt}},\ }\bibfield  {title} {\enquote {\bibinfo {title} {{Signals of
  confinement in Green functions of SU(2) Yang- Mills theory}},}\ }\href@noop
  {} {\bibfield  {journal} {\bibinfo  {journal} {Phys. Rev. Lett.}\ }\textbf
  {\bibinfo {volume} {93}},\ \bibinfo {pages} {061601} (\bibinfo {year}
  {2004})},\ \Eprint {http://arxiv.org/abs/hep-lat/0403011} {hep-lat/0403011}
  \BibitemShut {NoStop}%
%%CITATION = HEP-LAT/0403011;%%
\bibitem [{\citenamefont {Greensite}\ \emph {et~al.}(2011)\citenamefont
  {Greensite}, \citenamefont {Matevosyan}, \citenamefont {Olejnik},
  \citenamefont {Quandt}, \citenamefont {Reinhardt} \emph
  {et~al.}}]{Greensite:2011pj}%
  \BibitemOpen
  \bibfield  {author} {\bibinfo {author} {\bibfnamefont {J.}~\bibnamefont
  {Greensite}}, \bibinfo {author} {\bibfnamefont {H.}~\bibnamefont
  {Matevosyan}}, \bibinfo {author} {\bibfnamefont {S.}~\bibnamefont {Olejnik}},
  \bibinfo {author} {\bibfnamefont {M.}~\bibnamefont {Quandt}}, \bibinfo
  {author} {\bibfnamefont {H.}~\bibnamefont {Reinhardt}},  \emph {et~al.},\
  }\bibfield  {title} {\enquote {\bibinfo {title} {{Testing Proposals for the
  Yang-Mills Vacuum Wavefunctional by Measurement of the Vacuum}},}\ }\href
  {\doibase 10.1103/PhysRevD.83.114509} {\bibfield  {journal} {\bibinfo
  {journal} {Phys.Rev.}\ }\textbf {\bibinfo {volume} {D83}},\ \bibinfo {pages}
  {114509} (\bibinfo {year} {2011})},\ \Eprint {http://arxiv.org/abs/1102.3941}
  {arXiv:1102.3941 [hep-lat]} \BibitemShut {NoStop}%
%%CITATION = ARXIV:1102.3941;%%
\bibitem [{\citenamefont {Ayala}\ \emph {et~al.}(2012)\citenamefont {Ayala},
  \citenamefont {Bashir}, \citenamefont {Binosi}, \citenamefont
  {Cristoforetti},\ and\ \citenamefont {Rodriguez-Quintero}}]{Ayala:2012pb}%
  \BibitemOpen
  \bibfield  {author} {\bibinfo {author} {\bibfnamefont {A.}~\bibnamefont
  {Ayala}}, \bibinfo {author} {\bibfnamefont {A.}~\bibnamefont {Bashir}},
  \bibinfo {author} {\bibfnamefont {D.}~\bibnamefont {Binosi}}, \bibinfo
  {author} {\bibfnamefont {M.}~\bibnamefont {Cristoforetti}}, \ and\ \bibinfo
  {author} {\bibfnamefont {J.}~\bibnamefont {Rodriguez-Quintero}},\ }\bibfield
  {title} {\enquote {\bibinfo {title} {{Quark flavour effects on gluon and
  ghost propagators}},}\ }\href {\doibase 10.1103/PhysRevD.86.074512}
  {\bibfield  {journal} {\bibinfo  {journal} {Phys. Rev.}\ }\textbf {\bibinfo
  {volume} {D86}},\ \bibinfo {pages} {074512} (\bibinfo {year} {2012})},\
  \Eprint {http://arxiv.org/abs/1208.0795} {arXiv:1208.0795 [hep-ph]}
  \BibitemShut {NoStop}%
%%CITATION = ARXIV:1208.0795;%%
\bibitem [{\citenamefont {Aguilar}\ \emph
  {et~al.}(2012{\natexlab{b}})\citenamefont {Aguilar}, \citenamefont {Binosi},\
  and\ \citenamefont {Papavassiliou}}]{Aguilar:2012rz}%
  \BibitemOpen
  \bibfield  {author} {\bibinfo {author} {\bibfnamefont {A.~C.}\ \bibnamefont
  {Aguilar}}, \bibinfo {author} {\bibfnamefont {D.}~\bibnamefont {Binosi}}, \
  and\ \bibinfo {author} {\bibfnamefont {J.}~\bibnamefont {Papavassiliou}},\
  }\bibfield  {title} {\enquote {\bibinfo {title} {{Unquenching the gluon
  propagator with Schwinger-Dyson equations}},}\ }\href@noop {} {\bibfield
  {journal} {\bibinfo  {journal} {Phys. Rev.}\ }\textbf {\bibinfo {volume}
  {D86}},\ \bibinfo {pages} {014032} (\bibinfo {year} {2012}{\natexlab{b}})},\
  \Eprint {http://arxiv.org/abs/1204.3868} {arXiv:1204.3868 [hep-ph]}
  \BibitemShut {NoStop}%
%%CITATION = ARXIV:1204.3868;%%
\bibitem [{\citenamefont {Aguilar}\ \emph {et~al.}(2013)\citenamefont
  {Aguilar}, \citenamefont {Binosi},\ and\ \citenamefont
  {Papavassiliou}}]{Aguilar:2013hoa}%
  \BibitemOpen
  \bibfield  {author} {\bibinfo {author} {\bibfnamefont {A.~C.}\ \bibnamefont
  {Aguilar}}, \bibinfo {author} {\bibfnamefont {D.}~\bibnamefont {Binosi}}, \
  and\ \bibinfo {author} {\bibfnamefont {J.}~\bibnamefont {Papavassiliou}},\
  }\bibfield  {title} {\enquote {\bibinfo {title} {{Gluon mass generation in
  the presence of dynamical quarks}},}\ }\href {\doibase
  10.1103/PhysRevD.88.074010} {\bibfield  {journal} {\bibinfo  {journal} {Phys.
  Rev.}\ }\textbf {\bibinfo {volume} {D88}},\ \bibinfo {pages} {074010}
  (\bibinfo {year} {2013})},\ \Eprint {http://arxiv.org/abs/1304.5936}
  {arXiv:1304.5936 [hep-ph]} \BibitemShut {NoStop}%
%%CITATION = ARXIV:1304.5936;%%
\bibitem [{\citenamefont {Bicudo}\ \emph {et~al.}(2015)\citenamefont {Bicudo},
  \citenamefont {Binosi}, \citenamefont {Cardoso}, \citenamefont {Oliveira},\
  and\ \citenamefont {Silva}}]{Bicudo:2015rma}%
  \BibitemOpen
  \bibfield  {author} {\bibinfo {author} {\bibfnamefont {P.}~\bibnamefont
  {Bicudo}}, \bibinfo {author} {\bibfnamefont {D.}~\bibnamefont {Binosi}},
  \bibinfo {author} {\bibfnamefont {N.}~\bibnamefont {Cardoso}}, \bibinfo
  {author} {\bibfnamefont {O.}~\bibnamefont {Oliveira}}, \ and\ \bibinfo
  {author} {\bibfnamefont {P.~J.}\ \bibnamefont {Silva}},\ }\bibfield  {title}
  {\enquote {\bibinfo {title} {{The lattice gluon propagator in renormalizable
  $\xi$ gauges}},}\ }\href@noop {} {\  (\bibinfo {year} {2015})},\ \Eprint
  {http://arxiv.org/abs/1505.05897} {arXiv:1505.05897 [hep-lat]} \BibitemShut
  {NoStop}%
%%CITATION = ARXIV:1505.05897;%%
\bibitem [{\citenamefont {Aguilar}\ \emph {et~al.}(2015)\citenamefont
  {Aguilar}, \citenamefont {Binosi},\ and\ \citenamefont
  {Papavassiliou}}]{Aguilar:2015nqa}%
  \BibitemOpen
  \bibfield  {author} {\bibinfo {author} {\bibfnamefont {A.C.}\ \bibnamefont
  {Aguilar}}, \bibinfo {author} {\bibfnamefont {D.}~\bibnamefont {Binosi}}, \
  and\ \bibinfo {author} {\bibfnamefont {J.}~\bibnamefont {Papavassiliou}},\
  }\bibfield  {title} {\enquote {\bibinfo {title} {{Yang-Mills two-point
  functions in linear covariant gauges}},}\ }\href {\doibase
  10.1103/PhysRevD.91.085014} {\bibfield  {journal} {\bibinfo  {journal}
  {Phys.Rev.}\ }\textbf {\bibinfo {volume} {D91}},\ \bibinfo {pages} {085014}
  (\bibinfo {year} {2015})},\ \Eprint {http://arxiv.org/abs/1501.07150}
  {arXiv:1501.07150 [hep-ph]} \BibitemShut {NoStop}%
%%CITATION = ARXIV:1501.07150;%%
\bibitem [{\citenamefont {Huber}(2015)}]{Huber:2015ria}%
  \BibitemOpen
  \bibfield  {author} {\bibinfo {author} {\bibfnamefont {Markus~Q.}\
  \bibnamefont {Huber}},\ }\bibfield  {title} {\enquote {\bibinfo {title}
  {{Gluon and ghost propagators in linear covariant gauges}},}\ }\href
  {\doibase 10.1103/PhysRevD.91.085018} {\bibfield  {journal} {\bibinfo
  {journal} {Phys.Rev.}\ }\textbf {\bibinfo {volume} {D91}},\ \bibinfo {pages}
  {085018} (\bibinfo {year} {2015})},\ \Eprint
  {http://arxiv.org/abs/1502.04057} {arXiv:1502.04057 [hep-ph]} \BibitemShut
  {NoStop}%
%%CITATION = ARXIV:1502.04057;%%
\bibitem [{\citenamefont {Siringo}(2015)}]{Siringo:2015gia}%
  \BibitemOpen
  \bibfield  {author} {\bibinfo {author} {\bibfnamefont {Fabio}\ \bibnamefont
  {Siringo}},\ }\bibfield  {title} {\enquote {\bibinfo {title} {{Second order
  gluon polarization for SU(N) theory in linear covariant gauge}},}\
  }\href@noop {} {\  (\bibinfo {year} {2015})},\ \Eprint
  {http://arxiv.org/abs/1507.00122} {arXiv:1507.00122 [hep-ph]} \BibitemShut
  {NoStop}%
%%CITATION = ARXIV:1507.00122;%%
\bibitem [{\citenamefont {Capri}\ \emph {et~al.}(2015)\citenamefont {Capri},
  \citenamefont {Dudal}, \citenamefont {Fiorentini}, \citenamefont {Guimaraes},
  \citenamefont {Justo}, \citenamefont {Pereira}, \citenamefont {Mintz},
  \citenamefont {Palhares}, \citenamefont {Sobreiro},\ and\ \citenamefont
  {Sorella}}]{Capri:2015ixa}%
  \BibitemOpen
  \bibfield  {author} {\bibinfo {author} {\bibfnamefont {M.~A.~L.}\
  \bibnamefont {Capri}}, \bibinfo {author} {\bibfnamefont {D.}~\bibnamefont
  {Dudal}}, \bibinfo {author} {\bibfnamefont {D.}~\bibnamefont {Fiorentini}},
  \bibinfo {author} {\bibfnamefont {M.~S.}\ \bibnamefont {Guimaraes}}, \bibinfo
  {author} {\bibfnamefont {I.~F.}\ \bibnamefont {Justo}}, \bibinfo {author}
  {\bibfnamefont {A.~D.}\ \bibnamefont {Pereira}}, \bibinfo {author}
  {\bibfnamefont {B.~W.}\ \bibnamefont {Mintz}}, \bibinfo {author}
  {\bibfnamefont {L.~F.}\ \bibnamefont {Palhares}}, \bibinfo {author}
  {\bibfnamefont {R.~F.}\ \bibnamefont {Sobreiro}}, \ and\ \bibinfo {author}
  {\bibfnamefont {S.~P.}\ \bibnamefont {Sorella}},\ }\bibfield  {title}
  {\enquote {\bibinfo {title} {{Exact nilpotent nonperturbative BRST symmetry
  for the Gribov-Zwanziger action in the linear covariant gauge}},}\ }\href
  {\doibase 10.1103/PhysRevD.92.045039} {\bibfield  {journal} {\bibinfo
  {journal} {Phys. Rev.}\ }\textbf {\bibinfo {volume} {D92}},\ \bibinfo {pages}
  {045039} (\bibinfo {year} {2015})},\ \Eprint
  {http://arxiv.org/abs/1506.06995} {arXiv:1506.06995 [hep-th]} \BibitemShut
  {NoStop}%
%%CITATION = ARXIV:1506.06995;%%
\bibitem [{\citenamefont {Brodsky}\ and\ \citenamefont
  {Shrock}(2008)}]{Brodsky:2008be}%
  \BibitemOpen
  \bibfield  {author} {\bibinfo {author} {\bibfnamefont {Stanley~J.}\
  \bibnamefont {Brodsky}}\ and\ \bibinfo {author} {\bibfnamefont {Robert}\
  \bibnamefont {Shrock}},\ }\bibfield  {title} {\enquote {\bibinfo {title}
  {{Maximum Wavelength of Confined Quarks and Gluons and Properties of Quantum
  Chromodynamics}},}\ }\href {\doibase 10.1016/j.physletb.2008.06.054}
  {\bibfield  {journal} {\bibinfo  {journal} {Phys. Lett.}\ }\textbf {\bibinfo
  {volume} {B666}},\ \bibinfo {pages} {95--99} (\bibinfo {year} {2008})},\
  \Eprint {http://arxiv.org/abs/0806.1535} {arXiv:0806.1535 [hep-th]}
  \BibitemShut {NoStop}%
%%CITATION = 0806.1535;%%
\end{thebibliography}


\end{document}